\documentclass[sn-basic]{sn-jnl}

\usepackage{graphicx}%
\usepackage{multirow}%
\usepackage{amsmath,amssymb,amsfonts}%
\usepackage{amsthm}%
\usepackage{mathrsfs}%
\usepackage[title]{appendix}%
\usepackage{xcolor}%
\usepackage{textcomp}%
\usepackage{manyfoot}%
\usepackage{booktabs}%
\usepackage{algorithm}%
\usepackage{algorithmicx}%
\usepackage{algpseudocode}%
\usepackage{listings}%
\usepackage{xspace}       %

\newcommand{\about}{$\simeq$}

\newcommand{\Msol}{M\ensuremath{_\odot}\xspace}

\newcommand\rd[1]{{#1}}

\newcommand{\araa}{Ann.Rev.Astron.\&Astroph. }%
\newcommand{\apj}{Astrophys. J. }%
\newcommand{\apjl}{Astrophys. J. Lett. }%
\newcommand{\apjs}{Astrophys. J. Suppl. Ser. }%
\newcommand{\apss}{Astroph.J.\&Sp.Sci. }%
\newcommand{\aap}{Astron. Astrophys. }%
\newcommand{\aapr}{Astron. Astrophys.~Rev. }%
\newcommand{\aaps}{Astron. Astrophys. Suppl. }%
\newcommand{\aj}{Astron. J. }%
\newcommand{\gca}{Geoch. Cosmoch. Acta}%
\newcommand{\icarus}{Icarus }%
\newcommand{\jcap}{J. Cosmol. Astropart.Phys. }%
\newcommand{\mnras}{Mon. Notices Royal Astron. Soc. }%
\newcommand{\nar}{New Astron. Rev. }%
\newcommand{\prc}{Phys.~Rev.~C }%
\newcommand{\prd}{Phys.~Rev.~D }%
\newcommand{\prl}{Phys.~Rev.~Lett. }%
\newcommand{\pasa}{PASA }%
\newcommand{\pasp}{Proc.Astr.Soc.Pac. }%
\newcommand{\pasj}{Proc.Astr.Soc.Jap. }%
\newcommand{\ssr}{Space~Sci.~Rev. }%
\newcommand{\nat}{Nature }%
\newcommand{\nphysa}{Nucl.~Phys.~A }%
\newcommand{\physrep}{Phys.~Rep. }%
\begin{document}

\title[Nuclear Astrophysics]{Nuclear Astrophysics}


\author*[1]{\fnm{Roland} \sur{Diehl}}\email{rod@mpe.mpg.de}
\author[2]{\fnm{Michael} \sur{Wiescher}}

\affil*[1]{\orgname{Max Planck Institut f\"ur extraterrestrische Physik and Technical University (TUM)}, \orgaddress{\postcode{85748} \city{Garching}, \country{Germany}}}
\affil[2]{\orgname{Department of Physics and Astronomy, University of Notre Dame, and Joint Institute for Nuclear Astrophysics}, \orgaddress{\city{Notre Dame}, \postcode{IN 46556} \country{USA}}}


\abstract{Reactions between atomic nuclei are measured in great detail in terrestrial laboratory experiments; transferring and extrapolating this knowledge to how the same reactions act within cosmic environments presents major challenges. Cross-disciplinary efforts are needed in view of the many nuclear reactions that govern the chemical evolution of the universe, and occur in a broad range of stellar plasma conditions that require astrophysical exploration.
The variety of quiescent and explosive astrophysical environments for nuclear processes reaches from Big Bang conditions through stellar interiors to a multitude of explosive processes of and near compact stars. Since the early identification of 'processes' associated with the buildup of elements or nucleosynthesis, new insights have been obtained on the complexity of nuclear reaction mechanisms. 
This article will provide an overview in nuclear processing shaped by reactions during near equilibrium conditions, cooling and freeze out times. 
The emergence of molecular-like nucleon configurations within nuclei incurs important features at the low energies given in stellar interiors. 
Multiple capture and fusion reactions are key in the overall nucleosynthesis patterns. Here we use $^{12}$C induced capture and fusion processes to illustrate the challenge of low-energy measurements and the challenges of using theoretical methods to extrapolate measurements towards energy regimes within cosmic sources.
Slow and rapid neutron captures processes facilitate the gradual buildup of heavy elements.
Particle beam experiments at accelerator facilities above and deep underground simulate stellar reactions, and new experimental facilities and methods complement these by providing short-lived isotope-separated beams and high-flux photon and neutron sources in a new generation of laboratories, with laser driven plasma facilities and particle storage rings as the latest tools for the experimenters. 
This is complemented by improved theoretical tools to calculate the quantum effects of nuclear reactions at the various cosmic conditions.

Astronomical signatures of nuclear reactions from within cosmic sources are deduced through a growing range of observational tools, from the determination of rapidly changing light curves {characterizing} cosmic explosions from supernova, novae, and kilonovae, through gamma-ray lines and presolar grains, to the detection of rare neutrino particles from our Sun and distant cosmic events. High resolution spectroscopy of distant stars has been expanded to objects and transient events measured in the X-ray and the gamma energy range of the electromagnetic spectrum. 
The analysis of vibrational behavior of stars in astro-seismology provides new tools in probing stellar interiors. The isotopic analysis of meteoritic inclusions provides detailed information about various nucleosynthesis sources, which are important tools for the understanding of complex dynamic convection and mixing processes in the interior of stars. While requiring care in interpreting observational data to account for various biases and systematics, this variety of tools provides new opportunities and synergies. 
Chemical-evolution models provide a bridge through stellar-abundance archeology and have recently developed to also describe the complex dynamics during the evolution of galaxies.

This article seeks to summarize the experimental and theoretical efforts for a better understanding of the complex mechanisms that lead to the compositional evolution of our universe, supplemented by an overview of the broad range of observational tools that have been developed to test the experimental data and the theoretical interpretation of nuclear processes in the cosmos.}




\maketitle

\section{Introduction}\label{sec1}
The field of \emph{Nuclear Astrophysics} is situated between the two disciplines of \emph{nuclear physics} and of  \emph{astrophysics}. Each of these includes experimental/observational and theoretical branches. 
\rd{ The foundations of this field have been laid out in pioneering papers by \citet{Burbidge:1957} and \citet{Cameron:1957,Cameron:1957a}, and later written up in textbooks such as Donald Clayton's ``Principles of stellar evolution and nucleosynthesis'' \citep{Clayton:1968}, then ``Cauldrons in the Cosmos'' by Claus Rolfs and William Rodney \citep{Rolfs:1988}, and the more-recent ``Nuclear physics of stars'' by Christian Iliadis \citep{Iliadis:2007} and ``The synthesis of the elements'' by Giora Shaviv \citep{Shaviv:2012}.
Each of these take a different look at the complex interplay that }occurs between the microscopic quantum processes that govern the nuclear reaction processes and the macroscopic processes of transporting energy and materials that characterize the conditions in stellar interiors. This results in considerable  uncertainty for the analysis and determination of nuclear reaction processes in cosmic sites. 
{ On the other hand we understand} enough about the origin of chemical elements to describe and teach this astrophysical and nuclear-physics problem \citep[see above-cited textbooks and, for example,][and other review articles \rd{that have been written} over seven decades]{Burbidge:1957,Clayton:1969b,Wallerstein:1997,Heger:2003,Arnould:2020,Kobayashi:2020a, Cowan:2021,Arcones:2023}).
{Often herein the descriptions of cosmic nucleosynthesis processes are inter-woven with our plausible understanding of its cosmic sources, thus mixing better consolidated with more-uncertain aspects.}
For astrophysicists, a description of cosmic sources which includes nuclear transformations {remains a} challenge, while nuclear physicists {are challenged to determine} nuclear reaction processes at conditions {that are} typically not available in the laboratory. 
The data base for this science are the measured outcomes of nuclear-reaction experiments, and the observations of isotopic variety and abundances in material samples or objects throughout the universe. 
Combining the characteristics of nuclear reactions with those of cosmic sites and their dynamics, nuclear astrophysics exploits astronomical data {in attempts to} rule out the invalid components of the complex and nested modeling that represents our understanding of the compositional evolution of matter in the universe.

Conferences and workshops {are of great importance in such a diverse field of} nuclear astrophysics. They are exemplified by the series of  conferences of  "Nuclei in the Cosmos" (since 1990, no. XVIII held recently\footnote{see https://indico.icc.ub.edu/event/341/ for the most-recent NIC in 2025}) and of  "Nuclear Physics in Astrophysics" (since 2003, no. XI held recently\footnote{see https://events.hifis.net/event/540/ for the most recent NPA in 2024}), \rd{typically held every 2 years}.  Gathering 200 people or more, these conferences may appear to novices as either overloaded with specialist discussions in sub-fields, or too shallow to address physics issues of interest with the required depth. Yet, they are essential to mediate among sub-fields, and are complemented by workshops addressing specific topics and questions in the field.
International community networking projects such as IReNA (USA and international), JINA/CENAM (USA), ChETEC/ChETEC-Infra (EU/UK), UKAKUREN/JaFNA (Japan) and INAC (China) provide an umbrella for the diverse projects and for exchanges of tools; these have an important role to maintain the exchanges among diverse research groups. 
The experimental challenges ahead for this field are addressed by \emph{white papers} that lobby for this science with the various funding agencies, see for example \citet{Lewitowicz:2024} and  \citet{Schatz:2022}.
Nuclear astrophysics as a whole appears too broad a field to be addressed in a single review paper, due to this interdisciplinary nature. Partial reviews \citep[e.g.][]{Wiescher:2023} are useful to help appreciating the interplay between achievements, uncertainties, and tacit assumptions in each of the aspects of nuclear aspects.

In this article we follow a different approach, presenting separately the concepts of nuclear reactions and \rd{those of the } cosmic sources, while pointing out the overlapping aspects. Thus we characterize the field along the aspects of \emph{nuclear physics aimed at cosmic environments}, then \emph{nucleosynthesis reactions} within \emph{their cosmic environments} as we understand these from astrophysics. 
\rd{Thus, unlike other reviews, we focus on the underlying concepts of nuclear physics and astro-physics, respectively, aiming to remind experienced scientists of these, and to guide scientists entering the field accordingly. In the same spirit, }
 we present  \emph{astronomy} and how it could be focused towards the specific research questions of nuclear astrophysics in a dedicated \emph{nuclear astronomy}. 
We thus aim to emphasize the foundations of the research field, i.e. the puzzles that arise when the cosmos is seen with the eyes of a nuclear physicist, and the research pathways as we attempt to combine the potentials of astronomy and astrophysics modeling with those of nuclear theory and experiment. 
 \rd{At the same time, our review also presents latest developments and how they address current challenges. Although this aims to represent the current status of nuclear astrophysics, our choices of key concepts and current topics are subjective, and represent our own view of the complex and broad field of nuclear astrophysics. We hope that this adds a usefully-different aspect and complements the literature in this field.
 Our review of} the characteristics of the field, and our comments on the challenges, achievements, and prospects of its different disciplines, \rd{date as of late} 2025. 

\section{Nuclear physics in cosmic environments}\label{sec2} 
\subsection{General considerations}\label{sec2.1}
\begin{figure}[ht]  
\centering
\includegraphics[width=1.0\textwidth]{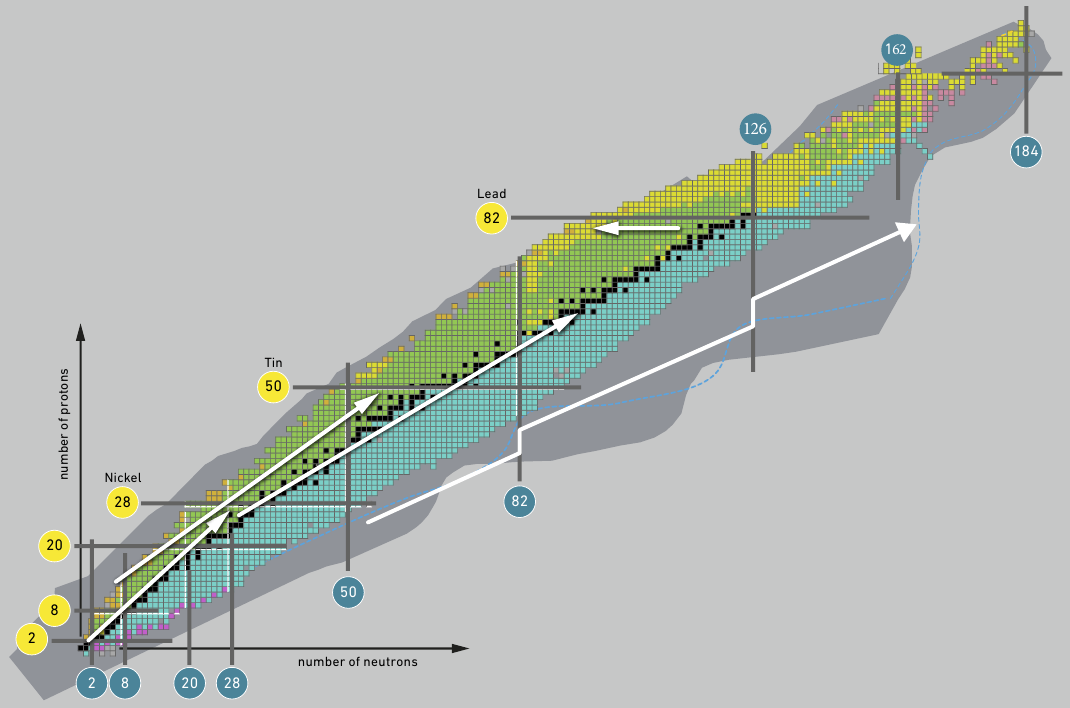}
\caption{Table of known isotopes with their proton (vertical axis) and neutron (horizontal axis) numbers, showing the stable isotopes (black squares) as well as unstable isotopes (colored squares). The total number of isotopes that are expected to exist is more than 7000, with\about 3000 of them known, and \about~300 of them being stable. Characteristic nuclear-reaction networks in cosmic environments can be approximated through reaction paths, or \emph{processes}. These are discussed in Section 2.6 and illustrated in this figure of the table of isotopes by (white) arrows, reflecting the \rd{charged-particle induced reactions and the s~process} along the valley of stable isotopes, the r-process path on the neutron rich side, the rp~process on the proton-rich side, and the p (or $\gamma$-)~process pointing horizontally away from the higher end of the stability valley. Magic numbers, which characterize enhanced nuclear stability, are indicated for proton and neutron numbers, characteristically aligning with kinks in the r-process path. }\label{fig_ToI-reactionpaths}
\end{figure}

The variety of cosmic environments covers a large dynamic range of densities and timescales, characteristically evolving within a cosmic source. 
If one conservatively assumes that any exothermic nuclear reaction between two existing isotopes (including ones with short radioactive lifetimes) can occur and contribute to compositional evolution, one needs to develop a complex reaction network for following the energy production and nucleosynthesis in the environment. 
{The nuclear chart of known and predicted isotopes is shown in Figure~\ref{fig_ToI-reactionpaths}. It }includes at least 7,000 different isotopes, and since every nucleus is potentially coupled by on average 10 reactions with other nuclei one faces about 70,000 reactions in total. 
{The coloring of isotope boxes encodes the level of nuclear-physics knowledge about the isotopes, from black (stable) through colored (unstable, with more or less knowledge) to grey (unknown territory, except that nuclei are expected to exist).}
While stellar explosions \rd{are launched} within a second, \rd{and may proceed for minutes to hours if occurring on a compact-star surface,} stellar evolution extends over \rd{much longer} periods of time, which depend sensitively on the stellar mass, \rd{ranging} from 8--10 Gy for the Sun to just a few My for stars more massive than 25--30~\Msol \citep{Schaller:1992}. 

Only a subset of reactions, anticipated for a specific environment, are typically considered in astrophysical models. 
{ For hydrogen burning, the different branches of the \emph{pp chains} \citep{Salpeter:1952a} and the cyclic pattern of the four \emph{CNO cycles}  \citep{Wiescher:2018a} are an early example of identifying reaction sequences that plausibly operate in stars.
Similarly, the nuclear-reaction paths that may occur in cosmic sites are commonly organized in \emph{processes}, or reaction patterns describing a reaction flow.
These share a common environmental characteristic such as temperatures, densities, and abundance of reaction agents, and link a specific set of reactions. 
Regimes of reactions that are expected to occur under such characteristic conditions are illustrated in Figure~\ref{fig_ToI-reactionpaths} by white arrows, the \emph{processes of nucleosynthesis} that are most-commonly discussed.}
These have been presented in the early period of nuclear astrophysics \citep{Burbidge:1957}, and later \rd{a variety of} other such \emph{processes and paths}  were proposed \rd{from different reasonings and scenarios} \citep[e.g.][\rd{(these will be discussed below in their respective context)}]{Hainebach:1974,Wallace:1981,Woosley:1990,Howard:1991,Woosley:1992,Hoffman:1996,Schatz:1998,Rapp:2006,Pignatari:2010,Wanajo:2011,Pignatari:2018,Choplin:2021,Choplin:2022}.
Shown are the sequences of $\alpha$~capture reactions (\emph{$\alpha$~process}) that \rd{may illustrate production of} $^{56}$Ni and iron group elements from light nuclei, the continuation of heavier-nuclei production through slow \rd{neutron capture~/~$\beta$~decay sequences} (\emph{s~process}) along the valley of stable nuclei (black squares), the alternate neutron capture path of rapid neutron captures (\emph{r~process}) far on the neutron-rich side, a rapid proton capture path (\emph{rp~process}) on the proton-rich side, and a horizontal production of proton-rich nuclei from stability through a \emph{p or $\gamma$~process}, \rd{i.e. triggered by proton releases from ($\alpha$,p)  or ($\gamma$,p) reactions on neutron-deficient oxygen or neon isotopes}.
{Not shown are the generic $\beta$-decay reaction flows that drive unstable isotopes towards the valley of stability (i.e., forming diagonal lines in this Figure), driven only by the $\beta$-decay lifetimes of each isotope and thus (largely) independent from environmental characteristics.}
More such \emph{processes} are being discussed in nuclear astrophysics as one aims to characterize a coherent nuclear-reaction network that may operate in specific cosmic sources.
{So, an intermittent \emph{n process}  \citep{Blake:1976,Pignatari:2018} and the intermediate \emph{i~process} \citep{Cameron:1983,Choplin:2021} are discussed to describe reactions at neutron intensities between the extremes characterizing the r and s process. 
Then, a \emph{$\nu$~process} \citep{Woosley:1990,Sieverding:2017} with its variants is used to characterize neutrino-driven nuclear reactions at high neutrino fluxes as they occur in the cores of gravitational collapse supernovae, 
and may characteristically alter the nucleosynthesis to produce specific rare isotopes that cannot be understood from the neutron capture nor $\gamma$~processes, such as $^{180}$Ta, $^{138}$La, or $^{19}$F.
Also, reaction flows with electron captures and successive $\beta$~decays or vice versa through \emph{URCA pairs} of nuclei define an \emph{URCA process} \citep{Barkat:1967};  not altering the composition, the URCA process establishes an energy leak through neutrino production, which affects burning-site conditions.
These latter processes illustrate that details of nuclear reactions in cosmic sites are often more complex, and cannot all be shown in this Figure for clarity.}
{This overview Figure~\ref{fig_ToI-reactionpaths} provides a valuable tool, to first order, for such explorations in order to find the most-relevant regions where nuclear-reaction properties matter for cosmic nucleosynthesis.
It therefore is a key figure for the communication between nuclear physicists and astrophysicists.}

{Stellar burning sites and the associated nuclear-reaction patterns are typically defined by their specific fuel component, such as hydrogen, helium available for capture reactions, of species of heavy ions causing fusion or photo-disintegration processes. Depending on the environmental conditions, such processes can be characterized as quiescent burning in stellar environments over extensive periods of time or as explosive burning within very short periods. Nuclear structure and reaction strength characterize the overall nuclear burning pattern or reaction flow within a burning site through the nuclear reaction rates of the associated capture of fusion processes. This is also the case for neutron-induced reaction processes at limited neutron flux conditions. In that case the reaction rates for neutron capture is slower than the $\beta$-decay for radioactive nuclei along the reaction path. This are the conditions that determine the \emph{slow-neutron capture} or \emph{s-process}. For increased neutron flux, such as anticipated for the \emph{i-process} (i for intermediate), the reaction path includes long-lived radioactive nuclei, but the reaction rates are important to know for determining the pattern of the reaction path. At conditions of very high neutron flux, one expects the \emph{rapid neutron capture} or \emph{r-process} because the neutron induced reactions are much faster, driving the reaction path far away from stability to very short-lived isotopes. If the path is defined by a balance between neutron capture and $\beta^-$-decay, the process is labeled as the cold r-process. In the case the reaction path is defined by a $(n,\gamma)$-$(\gamma,n)$ equilibrium between neutron capture and photo-dissociation one speaks about the hot r-process. The reaction path is different for the two cases, in the first case it depends on the nuclear structure far off stability determining the decay and capture properties, while in the second case the reaction path is primarily determined by the binding energies of the associated nuclei.}

{At higher temperatures as expected for extreme astrophysical events like supernovae and neutron-star mergers, nuclear burning processes are frequently in equilibrium, with forward and backward reactions in balance. This is described as \emph{nuclear statistic equilibrium (NSE)} The distribution of elements produced under NSE is highly dependent on the electron fraction ($Y_{e}$), which is the ratio of electrons to baryons (protons and neutrons) in the matter. For a plasma with equal numbers of protons and neutrons and absence of weak reactions, $^{56}$Ni would thus be produced. In events involving the collapse of a massive star or the merger of neutron stars, the core conditions are different leading to considerably lower $Y_{e}$ values and causing very different nucleosynthesis pattern.
Key reactions in a quiescent of explosive burning environment are typically associated with bottlenecks or waiting points. 
Bottlenecks result from a reaction flow that effectively only can proceed through a specific link, when bypass reactions are much less favored. Waiting points delay the reaction flow due to a reaction or decay rate for a specific nucleus that is small compared to other nuclei along the flow, causing an abundance enhancement of the waiting point nucleus.}

{\emph{Sensitivity studies} are frequently employed to identify key reactions by manually modifying rates and reaction rates to study possible changes in the reaction pattern. These methods often seem like a numerical demonstration of scientific intuition, while more variations of rates simultaneously and randomized through the Monte-Carlo method pay tribute to the complexity of large reaction networks and their feedback} (e.g., \citet{Iliadis:2020,Downen:2022,Martinet:2025}). 
The identification of key reactions also requires a reliable model treatment of the astrophysical site and its environment, \rd{including variations of environmental situations}, in which the impact of reactions can be simulated in so-called post-processing mode \citep{Rapp:2006,Parikh:2013,Mumpower:2016,Psaltis:2025}.
\rd{As illustrated by these examples, one should be aware of the choices made in such an analysis, which may limit broader interpretation of results: sensitivity studies are meant to guide the search for most-relevant reactions in complex reaction networks.}

Conditions in cosmic sites are very different from the situation at terrestrial experimental facilities of nuclear physics. 
The reaction partners in all cosmic sites are part of a multi-component plasma, with evolving composition and thus densities of specific reaction partners, and with broad energy distributions that mostly are far below the Coulomb barrier to reactions, a repulsion that represents the electromagnetic interaction of the nuclear charges, which is also at a scale of several MeV. Quantum tunneling through the Coulomb barrier is important \citep{Gamow:1928,Wiescher:2025}. 
The convolution of a Maxwellian (thermal) energy distribution of reaction partners and the exponential probability for penetrating the the Coulomb barrier leads to the \emph{Gamow peak energies} or \emph{Gamow window} for charged-particle reactions, which depends on the temperature of the environment and the charge of the interacting ionized particles. While for quiescent burning environments the Gamow window covers a range of typically 10 to 100~keV, for explosive burning it can extend up to the 2~MeV range. This corresponds to plasma temperatures of several MK (10$^6$K) to a few GK (10$^9$K). 
Correcting the reaction cross section for the impact of the Coulomb barrier leads to the \emph{astrophysical S-factor}, which in first approximation reflects the quantum mechanical transition probability for the reaction process~\citep{Gamow:1938,Bethe:1937,Salpeter:1952,Salpeter:1957}. 

Direct heavy-ion fusion reactions are prevented by the Coulomb repulsion for nuclei heavier than the Ne-Mg group\footnote{We will discuss the heavy ion fusion reactions between $^{12}$C and $^{16}$O nuclei below, which are important in the context of explosive sites.}, and \rd{in most cosmic environments} $p$ and $\alpha$~capture reactions are the only charged-particle reactions that occur. With increasing energy and hence temperature also photon reactions come into play, as they can lead to ejections of nucleons. 

Uncertainties on reduced Coulomb repulsion in nuclear reactions under cosmic environments result from screening of the nuclear charge because of free electrons clustering around the positive charges. 
Particle beams in accelerator experiments are fundamentally different in their electron environments of the reaction compared to mixtures of high-temperature plasma in cosmic sites. 
Screening corrections are required for each of these two cases, and are different \citep[see review by][]{Aliotta:2022}. 
In the laboratory, potential extremes related to the screening towards the lowest energies motivate experiments studying reactions below keV energies between light nuclei such as $^3$He and deuterium towards $^4$He fusion \citep{Casey:2023,Vesic:2024,Wiescher:2025}.


Nuclear physicists aim to experimentally measure the nuclear reaction rates at the energies relevant in cosmic environments.
Typical particle beam energies in terrestrial beam facilities are, however, 100~keV to about 1~MeV \citep{Rolfs:1988}. These can be extended down into the few-ten-keV region through special efforts, \rd{yet the 10~keV domain of stellar burning remains out of reach}. 
But the direct measurement of charged-particle nuclear reactions at astrophysically-relevant energies remains experimentally challenging, and limited to very few, compared to the variety in cosmic nucleosynthesis. 

With absence of Coulomb repulsion, this is different for neutron and photon reactions. 
Neutron reactions of the s-wave type, i.e. at zero angular momentum of the collision, have reaction rates independent of energy as the cross section dependency is compensated by the time spent during the collision. 
Therefore,  measurements at a convenient energy can be converted to astrophysical reaction rates for a thermal distribution of neutron energies. 
An orbital momentum of the collision geometry, however, translates into an angular-momentum barrier of the neutron reaction, presenting a similar case to the Gamow window for charged-particle reactions. 
Here, suitable beams with a near-Maxwellian energy distribution as obtained from certain nuclear reactions can be set up for integrated reaction-rate measurements that should be representative for the stellar environment. These are complemented by measurements at specific neutron energies that can be determined through the time-of-flight measurement of neutrons between production and reaction targets. 
For the study of photon reactions, continuum-spectrum sources such as Bremsstrahlung from electron beams can be used. 
Photon sources for calibration purposes of $\gamma$-ray detectors have been the radioactive decays of sources such as $^{60}$Co, complemented in recent decades backscattering a charged-particle beam on a laser photon source.

Experimental studies of nuclear physics for astrophysics include many indirect techniques, such as breaking up nuclei in heavy-ion collisions, or measuring surrogate reactions with similarities to the reactions and nuclear characteristics in question \citep[see][for details]{Wiescher:2023}.

The results from nuclear-physics experiments need to be combined with mathematical and theoretical methods to properly interpret and extrapolate experiments, or to complement these in various realms of the astrophysical parameter space where experiments remain unfeasible. 
The experimental data points can be fitted in the framework of R-matrix theory \citep{Wigner:1947,Lane:1958,Descouvemont:2010}, including all reaction and scattering channels that involve the same compound nucleus, for a reliable extrapolation toward the stellar energy range \citep{Azuma:2010,deBoer:2017,Wiescher:2025}. 
In the case of multiple resonance contributions, the system is often analyzed through a statistical approach, labeled a \emph{Hauser-Feshbach} method \citep{Weisskopf:1937,Hauser:1952,Friedman:1985,Rauscher:1997,Goriely:2008}. 
Through these constrained fitting functions it is possible to extrapolate downwards beyond measured and towards the relevant energies.
Different experiments and datasets with different systematic and statistical uncertainties require special care; beyond selection, Bayesian analysis and combining this with Monte Carlo variation methods promises to reduce systematics for such extrapolation \citep{Odell:2022}. 
Using such modern statistical methods, the results from various experiments and theoretical modelings can then be exploited as 'typical reaction parameters'  to estimate  the relevant astrophysical reaction rates and their uncertainties  \citep{Iliadis:2016}. 

The available reaction rates continue to be implemented in \emph{reaction rate libraries} \citep{Fowler:1967} such as NACRE \citep{Angulo:1999}, the JINA REACLIB \citep{Cyburt:2010}, STARLIB  \citep{Sallaska:2013}, {and KADoNiS \citep{Dillmann:2006,Dillmann:2008}}. Such libraries provide the basis of nuclear reaction networks for nuclear burning in stellar environments or specific processes of nucleosynthesis \citep{Thielemann:2023}. 
{Based on these libraries, versatile nuclear-reaction codes have been developed which allow evaluation of nucleosynthesis in various astrophysical environments, with publicly-available codes such as WINNET \citep{Reichert:2023} or SkyNet \citep{Lippuner:2017a} supplementing the specialized codes developed by many groups that model nucleosynthesis in cosmic sources.}
\subsection{Nuclear Reaction Studies}\label{sec2.2}

Nuclear reaction networks can contain up to several thousand reaction and decay processes linking the different isotopes. Reaction processes are characterized by resonant and non-resonant processes as summarized in the previous chapter. In this section we will provide an overview of the various experimental techniques utilized to determine reliable reaction rates. This will be followed by a discussion of a number of key reactions associated with the formation and depletion of $^{12}$C as one of the key elements in our universe. This will help to illustrate the approach in determining the reaction cross sections and the associated reaction rates.

\subsubsection{Experimental Techniques and Methods}
Direct reaction studies of charged-particle processes at the low energies corresponding to the temperature range in stellar quiescent burning are becoming increasingly difficult if not impossible, as the cross sections fall off exponentially with lower energy. While traditionally such measurements have been performed at low-energy high-intensity accelerators, the background radiation caused by cosmic rays (`cosmogenic' background) prevented a reliable study of many of the reactions important for stellar hydrogen, helium and carbon burning due to the background dominating the reaction yield. An increase in beam intensity and event identification using electronic coincidence techniques offer an improvement in the peak to background ratio, which makes above-ground developments still competitive \citep{Iliadis:2022}, in particular when utilizing inverse kinematics techniques of heavy-ion beams impinging on light-ion target materials~\citep{Ruiz:2025}. The approach promises a much high efficiency for the recoil detection compared to the detection of photons or light-ion reaction products with low-efficiency detector material in a small angle range. 

To reduce the cosmogenic background, an increasing number of light-isotope reaction experiments are conducted in underground laboratories, where overlying rock shields the experiments from cosmic-ray induced background, with about 3 orders of magnitude background suppression compared to surface laboratories. A price to pay is an increase of radiogenic background from radioactive isotopes within the Earth crust by about a factor 3 when going underground. This depends on the type of surrounding rock and its uranium and thorium content, and needs to be shielded locally. 
The \rd{LUNA deep underground accelerator at the Gran Sasso laboratory (LNGS)} \citep{Costantini:2009,Broggini:2010,Ananna:2024}, located in an annex to the traffic tunnel through the Gran Sasso mountain in Italy pioneered this field. CASPAR \citep{Robertson:2016} in South Dakota/USA, and JUNA \citep{Liu:2024} in the Jinping mountain in China are the two other main and currently-active deep-underground accelerator laboratories. 
Their overburden rock layer thickness and their water equivalents compare as 1400~m (3800~m.w.e., \rd{LNGS}) / 1600~m (6700~m.w.e., CASPAR) / 2400~m (6720~m.w.e., JUNA). 
Although depth is important (JUNA's cosmic-ray background is 1/100 of the one at \rd{LNGS}), also at only 45~m of overburden rock (140~m.w.e.) in the Felsenkeller facility in Germany useful experiments can be made at its 3~MeV accelerator \citep{Masha:2024}, as (beam-induced, and depth-independent) background from target impurities often also presents a limit in such experiments. This modest rock shielding already achieves a factor \about~200 background suppression with respect to surface laboratories, and using ease of access for prototype experiments before entering the deep laboratories. The remaining and eventually dominant background component comes from the beam-induced background on low-Z impurities in the target material, which, to a certain extent, can be removed by coincidence techniques. 
The particle beam facilities have matured over the years, with a new MeV accelerator as part of the Bellotti facility \citep{Junker:2023} supplementing its original 400~kV accelerator \rd{employed by the LUNA collaboration at LNGS}, and CASPAR with a 1~MV VdG accelerator for proton and $\alpha$~beams; JUNA's highlight is a significantly higher beam current of up to 10 equivalent mA for H$^+$ and He$^+$ ions from its 400~kV accelerator. 
After a first measurement phase ('Run~1') of several prominent reactions in 2024 \citep{Liu:2024} and subsequent disassembly of facilities, JUNA is being re-assembled during 2025 with further improved beam intensity and also featuring a windowless gas target.
In 2026, a modified accelerator design with higher voltage is foreseen to further advance beam intensities ('Super-JUNA').
Typically reactions relevant for Big Bang nucleosynthesis, for solar hydrogen burning, and with lighter elements up to C, N, O, F, Al, Ne, and Mg are being investigated at these underground facilities. 

{An alternative to the direct measurement of nuclear reactions is the inverse kinematic technique using \emph {recoil separators}. The method relies on the use of heavy ion beams impinging on light target particles. The advantage is that the secondary reaction products are moving into forward direction and can be collected with higher efficiency. The disadvantage is that they need to be separated from the primary beam particles with a rejection ratio of primary/secondary $\le 10^{-15}$. This can be achieved with recoil separators, an arrangement of magnetic and electric field for charge, and velocity separation of primary and secondary reaction products. This is often a challenge and requires a high demand on the beam optics for low energy particles. Successful separators have been developed with ERNA at Caserta~\citep{Schuermann:2004}, DRAGON at TRIUMF~\citep{Hutcheon:2005}, and St. GEORGE  at Notre Dame~\citep{Couder:2008}. Inverse kinematics measurements with recoil separators are very important for the study of proton and alpha capture reactions on radioactive isotopes. Besides DRAGON, a new separator SECAR has been developed at the FRIB radioactive beam facility for radiative reaction studies far off stability~\citep{Tsintari:2022}.}

{For small samples or low reaction cross sections, \emph{storage rings} offer a suitable technique with high sensitivity for future experiments \citep{Bruno:2023a}.  Monitoring the flux of a specific short-lived nucleus over multiple revolutions in the storage ring provides better sensitivity to low-rate proton captures on light hydrogen through multiple use of the specific short-lived particles. The newly-produced isotopes would recoil from the circular trajectory and focused into a mass spectrometer for detection and analysis. Using the ring technique can achieve effective amplification by 10$^5$ over direct separator experiments.}  

The essential parameters of storage rings are the injection of preselected specific isotopes from a facility that allows production and filtering of the isotope in question; then precision beam optics with acceleration sections to compensate for radiative losses, and electron cooler sections to narrow down the energy distribution of nuclei in the ring; finally target sections where neutron captures can be initiated, with suitable detectors for reaction products leaving the ring orbit.

One example facility is the CRYRING at GSI/FAIR in Germany, where suitably-selected nuclei could be decelerated to low and astrophysically-relevant energies, for measurements of their reaction and decay properties \citep{Glorius:2023}.  

{All these experiments provide information about the reaction cross section, over the accessible energy range of the accelerator facility. The cross section needs to be converted into astrophysically-relevant reaction rates by folding the cross-section data over the Maxwell-Boltzmann distribution of interacting nuclei for the temperatures of interest. Complementing such reaction-rate modeling, the development of laser plasma facilities makes a direct measurement of the reaction rate possible as long as stellar temperature and density conditions can be achieved.~\citep{Casey:2017} So far this effort is limited to very short-period studies of fusion reactions between very light isotopes as anticipated for the pp-chains, but future efforts aim at proton capture studies in the CNO mass range.}

Neutron capture reaction rates as measured in laboratory experiments may not be representative for neutron captures in stars: thermal excitations of nuclei in stars populate states above the ground state, and neutron captures thus may occur through a greater variety of states than in the laboratory case \citep{Lugaro:2023}. This is accounted for by a \emph{stellar enhancement factor}, determined from estimates based on the nuclear excitation levels \citep{Rauscher:2012,Rauscher:2015}.
Similarly, $\beta$~decay lifetimes determined, e.g., by activation experiments, require corrections for the stellar environment \citep{Langanke:2003,Langanke:2023}, as lifetimes can be significantly shorter when occurring from excited levels above the ground state, where decay is measured in laboratory experiments.

{The study of the reactions for simulating reaction patterns such as the $rp-$, the $p-$, or the $r-$process for explosive nucleosynthesis events requires large-scale experimental facilities for the production of neutron deficient and neutron rich heavy nuclei. Two techniques have been developed to ensure sufficient production of short-lived isotopes near the anticipated reaction path. 
Acceleration of a proton or heavy ion provides the particle beam leading to decomposition of a heavy target nucleus by spallation or fragmentation, respectively. Suitable analysis and separation of the fragments extracts nuclei of interest, which then can be post-accelerated to provide a beam of rare nuclear species of interest for specific experiments (\emph{radioactive beam facilities}).} 

\begin{figure}[h]  
\centering
\includegraphics[width=0.8\textwidth]{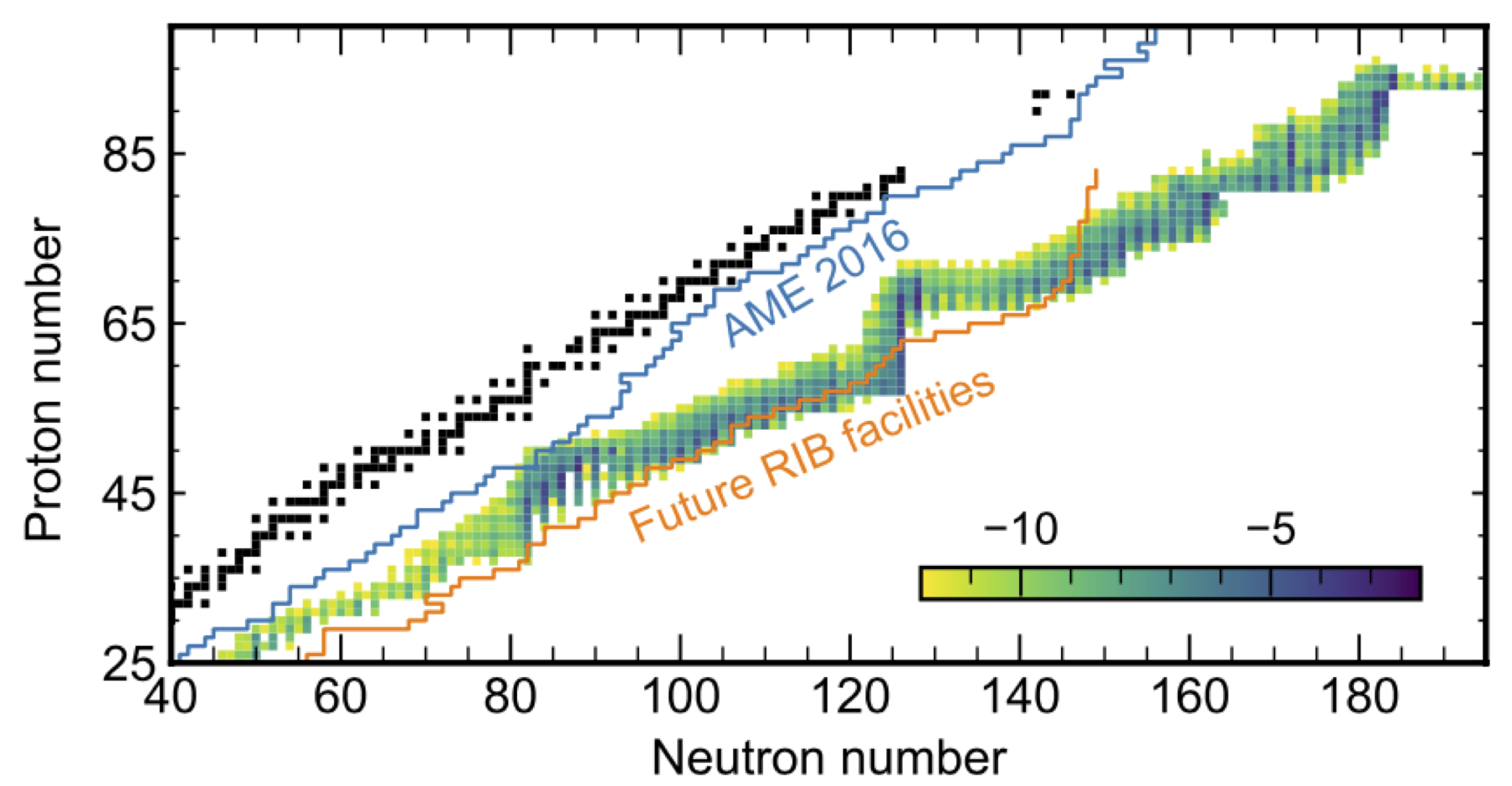}
\caption{{Illustration of proceeding knowledge about isotopes on the neutron-rich side of the stability valley. The table of known isotopes with their proton (vertical axis) and neutron (horizontal axis) numbers shows the stable isotopes (black squares), where knowledge about nuclear properties is best as experiments can provide the required masses and energy levels.} `AME 2016' indicates how far the knowledge of nuclear masses extended beyond stable isotopes in the regularly-updated tables  \citep{Audi:1996,Kondev:2021,Wang:2021a} of atomic mass evaluations (AME) in 2016. For nuclei further away from the valley of stability, capabilities of future radioactive-ion beam (RIB) facilities are indicated. The colored squares reflect abundances of an r-process calculation after freeze-out of neutron captures, {and illustrate the region of interest}. (From \citet{Cowan:2021})}\label{fig_ToI-knowledge}
\end{figure}

Separators for isotopes produced by spallation are slowly extracted via the \emph{ISOL} technique, while fragment separators are used to separate the fast reaction products of fragmentation processes. 
Fragment separation facilities connected to heavy-ion accelerators have been built recently in Germany (GSI/FAIR), in Korea (RAON), and in the USA (ANL/CARIBU) and (MSU/FRIB).  These begin to deliver experimental data beyond the pioneering results obtained in Japan at RIKEN/RIBF. Existing facilities using the ISOL method to produce radioactive beams (such as ISOLDE at CERN, IGISOL at Jyv\"askyl\"a/Finland, and ISAC at TRIUMF/Canada) have seen improvements, as have fission-based facilities such as CARIBU/ANL, US, and SPIRAL2/GANIL, France. All these facilities still contribute their share in exploration of the nuclear landscape. Smaller-scale, reaction based facilities such as \emph{TriSol} at Notre Dame and \emph{Resolut} at Florida State University (US) provide opportunities for independent complementary experiments with radioactive species in the light mass region.
{An illustration of the experimental reach is shown in Figure~\ref{fig_ToI-knowledge} for the neutron-rich side of unstable isotopes beyond the valley of stability}.

{While the progress in developing these facilities is impressive, the experimental programs are primarily focused on mass and decay measurements for isotopes far off stability. This provides important information about the nuclear structure effects influencing large scale reaction patterns, namely in terms of nuclear masses for the $rp-$ and the $r$-process as well as for $\beta$-decay lifetimes, which are important for considering waiting points along the reaction path. However, the number of direct reaction studies is limited, mainly due to the limitations in the intensity of radioactive beams. 
A number of charged particle reactions have been studied in inverse kinematics at high energies, well outside the Gamow range. The experimental results were extrapolated using statistical model reaction theory towards low energies. That is a promising first step, but it is far from clear that this statistical model is a reliable tool, in particular in far off stability systems near the proton or neutron threshold~\citep{Wiescher:2025}. More intensity is necessary to expand the measurements into the Gamow range of the associated explosive events.}

{The situation is worse for determining neutron capture rates far off stability, which would require the development of a high intensity neutron beam on short-lived radioactive targets or short-lived radioactive beam impinging on some "neutron target" produced by nuclear reactions and thermalized for longer-term containment. While such efforts are presently under development~\citep{Wiescher:2023} the determination of neutron capture rates so far relies on indirect reactions or so-called surrogate reactions that provide reaction information via neutron transfer studies~\citep{Nunes:2020}. This approach is limited to nuclear reactions with limited level density in the compound system~\citep{Kozub:2012}.} 

In the case of high-level density in compound systems, statistical methods become a suitable practical approach. This is typically the case for heavy nuclei at high excitation energies. 
Statistical methods can be applied whenever the number of resonances of a compound nucleus in the relevant energy range will be large \citep{Thielemann:2023}.
The reaction probabilities can be expressed as \emph{transmission coefficients} from the production or entrance channel of the reaction as well as the decay channel.
Reaction-rate calculations \rd{based on statistical methods such as  the 'SMOKER', 'non-SMOKER' \citep{Rauscher:2015}} and TALYS \citep{Koning:2023} nuclear-reaction codes also include treatments of direct versus compound-nucleus reactions and an optical model, evaluating nuclear mass models in a statistical model interpretation. 
Neutron capture predictions based on inferences from measurements have been made through the OSLO method \citep{Spyrou:2014,Larsen:2019,Larsen:2020,Chalil:2024}: here the determination of the gamma strength function provides a statistical summary of potential reactions. 
Reaction rates thus have been estimated in the domain of heavy nuclei beyond the iron group for situations relevant in nuclear astrophysics  \citep{Thielemann:2023}.

An interesting alternative approach is to constrain the reaction rates of interest from astrophysical considerations, using stellar evolution models and their outcomes as constrained by astronomical data. 
{A first such attempt to constrain the reaction rate of $^{12}$C($\alpha,\gamma$)$^{16}$O was made by \citet{Weaver:1993}, by analyzing a grid of massive stars ranging from 12 to 40 $M_{\odot}$ through all stages of nuclear burning up to the point of iron core collapse. Varying the $^{12}$C+$\alpha$ reaction rate and comparing the predicted abundances for a number of isotopes with solar abundance distribution, they predicted a nuclear-reaction rate that is in fair agreement with the present R-matrix analysis of nuclear reaction data~\citep{deBoer:2017}.}
Similarly, core-collapse supernovae explosions in the mass range 10--40~\Msol have recently been used to constrain the $^{12}$C($\alpha,\gamma$)$^{16}$O rate to rather lower values \citep{Xin:2025}
Degeneracies with other model parameters such as stellar rotation still weaken such constraints.
The uncertainties in mixing within stellar cores during stellar evolution (see chapter 3 on Cosmic Environments), thus also weaken interpretations of measured C/O ratios in white dwarfs, as used for $^{12}$C($\alpha,\gamma$)$^{16}$O rate constraints \citep{Straniero:2003}. 
Observations of cosmic radioisotopes  $^{26}$Al, $^{60}$Fe, and $^{44}$Ti  also had been used earlier to derive constraints on their yields from core-collapse supernovae, thus exploring possible rate variations for both the 3$\alpha$ and  $^{12}$C($\alpha,\gamma$)$^{16}$O reactions \citep{Tur:2010}. 

Another opportunity arises from black-hole masses as measured from gravitational wave events and their comparison with stellar evolution predictions of such masses {(see section 4.4 on Constraining Stellar Matter in Compact Stars)}. 
Models of massive-star evolution predict the pair instability to occur above a certain mass of the stellar core, which would result in supernovae that disrupt the star and would not produce a compact remnant; this results in an expected \emph{pair-instability mass gap}  \citep{Ozel:2010} for black holes {(see section 3.1 on Stellar Evolution). 
Rate variations for the 3$\alpha$ reaction and $^{12}$C($\alpha,\gamma$)$^{16}$O alter the limiting masses for this mass gap, so that observed black-hole masses constrain such rate variations \citep{Farmer:2020,van-Son:2020,Woosley:2021,Metha:2022,Shen:2023,Antonini:2025}.}

Astronomical measurements of pulsations of a white dwarf reflect its internal composition, and specifically the central oxygen content \citep{Giammichele:2018}. This oxygen content turned out to be significantly higher than the values predicted by stellar evolution models. 
This may be the result of the $^{12}$C($\alpha,\gamma$)$^{16}$O rate in modeling the composition. 
But on the other hand, also the Carbon-burning $^{12}$C+$^{12}$C reaction may occur in stellar cores that have been depleted in He following He burning in the giant phase of their evolution, or in white dwarf stars that result from stellar evolution through their giant phase in the mass range up to \about~8~\Msol, and are mainly composed of C, O, and small admixtures of heavier elements (Ne, Mg) \citep{De-Geronimo:2024,Giammichele:2022}.
Again, astronomical observations, here constraining internal composition of white dwarfs, can be exploited towards the Carbon-destructing nuclear rates but are still hampered by the inherent uncertainties of the astrophysics models.

\subsubsection{$^{12}$C induced reaction processes}
There are a multitude of nuclear reactions that have influenced the chemical evolution of the universe. In this review we only focus an a very limited subset of reactions, namely those associated with the production and depletion of $^{12}$C as the key element for the formation of organic matter and life. $^{12}$C plays a key role in most of the quiescent stellar burning scenarios, but it is also of great importance in explosive events such as novae and thermonuclear supernovae as will be shown in the following sections.

Hydrogen burning in stars along their main sequence (see section 3.1. on Stellar Interiors) occurs via the pp reaction chain or through the CNO cycle \citep{Bethe:1939,Salpeter:1952a}. The $^{12}$C nucleus plays a key role as seed material in the CNO cycle \citep{Wiescher:2018a}, which is characterized by the  cyclic reaction sequence of
$^{12}$C($p,\gamma$)$^{13}$N($\beta\nu^+$)$^{13}$C($p,\gamma$)$^{14}$N$p,\gamma$)$^{15}$O($\beta\nu^+$)$^{15}$N($p,\alpha$)$^{12}$C.
The reactions of the CNO cycle present a perfect
example for the different contributing reaction mechanisms in
stellar reactions, the compound mechanism and the direct reaction
mechanism. The compound reaction is a two step process in which the
particle, in this case the proton, is captured forming an 
excited unbound state in the compound nucleus, which then decays into
energetically allowed gamma or particle channels. The direct
reaction is a one step process, in which the initial two particle
system transits through interaction with the Coulomb field into a
final bound nuclear configuration. 

The initial CNO cycle reaction of $^{12}$C($p,\gamma$)$^{13}$N is characterized 
by both reaction components, the two-step compound mechanism through 
an excited unbound state in $^{13}$N, which subsequently decays by 
$\gamma$ emission to the ground state of $^{13}$N and 
the one-step direct capture mechanisms populating directly the 
ground state of $^{13}$N. The compound mechanism is reflected in a
broad pronounced resonance at 457 keV, whose low energy tail 
impacts the reaction rate. The E1 direct capture
contribution to the ground state adds at low energies to the 
reaction rate. As in many other cases for stellar hydrogen and helium 
burning, both compound and direct reaction components play a critical 
role in determining the reaction cross section at very low energies.

The reaction $^{12}$C($p,\gamma$)$^{13}$N was the first experimentally investigated CNO reaction
~\citep{Hall:1950}. This
early experiment already demonstrated that the CNO cycle was not 
the energy source of the sun as previously
claimed~\citep{Bethe:1939}. However, the reaction 
remained a prime target for the experimentalists over the decades
because of the interference effects between resonance tail and direct capture that influence the reaction mechanism. This was only 
completely elucidated on the basis of recent deep underground reaction
studies~\citep{Skowronski:2023} and the associated
R-matrix analysis of the data~\citep{Kettner:2023}.

The production of $^{12}$C in stars and the Big Bang has been a long-standing puzzle because it requires bridging the
gap in existing stable-mass nuclei at nuclear masses 5 and 8 in a stellar helium burning environment. It was shown by Fred Hoyle~\citep{Hoyle:1954} that this may occur directly through the triple-$\alpha$~reaction of $^4$He+$^4$He+$^4$He by first building an equilibrium abundance of $^8$Be between the fusion of two $\alpha$ particles and the rapid decay of the unbound $^8$Be. Subsequent $\alpha$ capture on the $^8$Be equilibrium populates the $0^{+}$ \emph{Hoyle state}, which is a compound state of $J^{\pi}=0^+$ in $^{12}$C at 7.65~MeV which has been modeled as a three-alpha-cluster configuration \citep{Epelbaum:2011,Freer:2014,Otsuka:2022}. The Hoyle state then rapidly decays via a $\gamma$ cascade and by pair production to the ground state. Higher excited states may also contribute, but those will decay primarily via $\gamma$ cascades to the ground state. 

\begin{figure}[h]  
\centering
\includegraphics[width=0.9\textwidth]{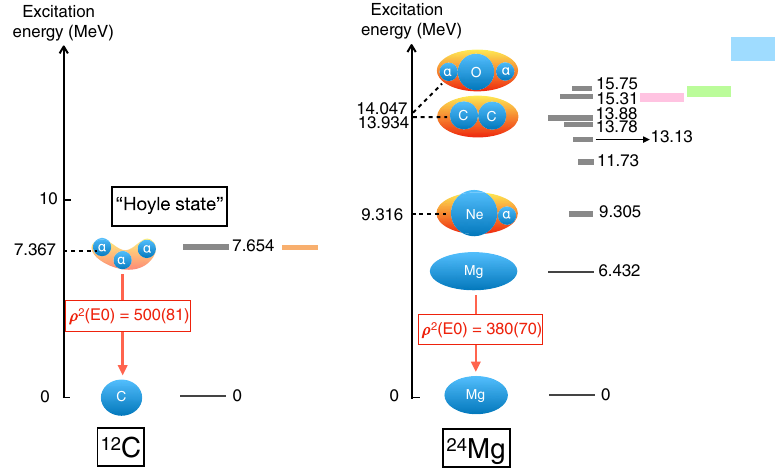}
\caption{The fusion of $\alpha$~particles to Carbon, as well as the fusion of Carbon in Carbon burning, may be facilitates by nuclear states in the compound nucleus, which are characterized by a cluster substructure of the nucleus  (from \citet{Adsley:2022}).}\label{fig_clusterStructures_C-Mg}
\end{figure} 
 
This combination of three $\alpha $ particles has a favorable overlap with a $\alpha$-cluster configuration of the \emph{Hoyle state}, which may thus de-excite into the $^{12}$C ground-state to produce the desired reaction product.  

In first-star environments, exhibiting a high primordial deuterium balance, an alternative multiple-step reaction path of $^4He(^2H,\gamma)^6Li(\alpha,\gamma)^{10}B(\alpha,d)^{12}C$ may operate, feeding $^{12}C$ on a shorter timescale than facilitated through the triple-$\alpha$-process~\citep{Wiescher:2021}.


After the build-up, the subsequent $^{12}$C($\alpha,\gamma$)$^{16}$O reaction in stellar helium burning is often called the \emph{holy grail} reaction because it determines the cosmic abundance ratio between the two isotopes that are key for organic life, $^{12}$C and $^{16}$O. Numerous measurements have been performed to determine the reaction cross section towards lower energies. However, the interpretation of the data for a reliable extrapolation towards the Gamow range around 300 keV remains a challenge. The main difficulty is in the treatment of the pronounced interference effects between the low-energy tails of high-energy resonances, direct capture contributions and the high-energy tails of sub-threshold states.
More specifically, the interpretation of the reaction cross-section~\citep[see][for recent reviews]{de-Boer:2025,deBoer:2017} is based on contributions from a number of broad interfering $1^-$ and $2^+$ resonances in the compound nucleus $^{16}O$ as well as an E2 direct capture to the ground state of $^{16}O$. The $\gamma$ transition therefore shows an a E1 or a E2 multipolarity, with both components adding to the total cross section. Both components are predicted to have comparable strength due to the suppression of the E1 term by the breaking of isospin symmetry by Coulomb interaction as discussed by \citet{deBoer:2017} 
To determine the cross-section data and map the resonance contributions interfering with bound state components extensive angular distribution in $^{12}$C($\alpha,\gamma$)$^{16}$O experiments have been performed over a wide energy range. 

An alternative approach of measuring a nearly time-inverse\footnote{It is not quite the inverse reaction, as the gamma ray beam does not have the multiplicity mix that is obtained in the forward reaction} reaction, the  $^{16}$O($\gamma, \alpha$)$^{12}$C, with a high-energy photon beam at the TUNL HI$\gamma$S facility and at the MAMI facility in Mainz, Germany, is being attempted. Here outgoing particle measurements provide the reaction constraints~\citep{Gai:2023}.  
{This method is complemented by the Coulomb dissociation technique which uses the strong electromagnetic field of a heavy nucleus to simulate the capture of a virtual photon ($\gamma$) by a projectile nucleus $^{16}$O, which then breaks apart into its constituents $^{12}$C and $^4$He\citep{Fleurot:2005,Goebel:2020}. Both methods may help to constrain extrapolation of the $^{12}$C($\alpha,\gamma$)$^{16}$O reaction rate to the stellar energy range (here \about~300~keV).}

A more indirect approach is the study of the \emph{asymptotic normalization coefficient (ANC)} via peripheral nuclear transfer reactions of resonant and subthreshold states~\citep{deBoer:2017,Shen:2023}. The ANCs determination is an indirect method for determining level properties such as the $\alpha$ strength for calculating the contribution to the cross section.

The nucleosynthesis of carbon also plays an important role in quiescent carbon burning in massive stars and explosive carbon burning anticipated of type Ia supernova environments. Like in the case of the Hoyle state for carbon production, $^{12}$C-$^{12}$C cluster states in $^{24}$Mg may also play a role in the fusion of two $^{12}$C isotopes \citep{Adsley:2022}. This analogy is illustrated in Figure~\ref{fig_clusterStructures_C-Mg}.
Such molecular configuration would be reflected in the strength of resonances in the fusion process~
\citep{Wei:2024,Freer:2014,Hafstad:1938}. 
On the theory side, at the same time the \emph{cluster theory} had been further investigated towards potential information on resonances in the $^{24}$Mg product nucleus of the $^{12}$C+$^{12}$C  reaction \citep{Descouvemont:2020,Descouvemont:2023}. 
Such clusters within nuclei may have important implications for nucleosynthesis as a whole \citep{Wiescher:2017,Wiescher:2021}.

While pronounced cluster configurations may enhance the fusion process, a specific \emph{hindrance} suppressing sub-threshold fusion reaction rates towards lower energies had been proposed for heavier-ion fusion reactions \citep{Jiang:2002}. Hindrance is caused by the incompressibility of matter affecting the fusion process of two heavy nuclei~\citep{Misicu:2006}. In the case of carbon fusion, hindrance could have important consequences for several cosmic objects, specifically giant stars, \rd{superbursts in neutron star crusts, and} thermonuclear supernovae~\citep{Gasques:2007}. Therefore, much experimental effort has been devoted to study the carbon fusion reaction and a possible hindrance occurrence for such sub-threshold reactions. 
So far the data from direct and indirect measurements appear to not support the strong drop suggested by hindrance, but measurements towards lower energies are difficult and the critical energy range for hindrance has not been reached yet~\citep{Chieffi:2025}.

{One particular experimental technique to find low-energy resonances in such heavier compound nuclei has been the \emph{Trojan horse} method \citep{Spitaleri:2011,Tumino:2021}. This method had been developed since the 1990$^{ies}$ mainly in the Italian laboratories for nuclear physics in Catania and Legnaro, and also applied at Texas A\&M (US) and RIKEN (Japan).
Herein, a transfer nuclear reaction is initiated above the Coulomb threshold with a target nucleus so that the nucleus of interest breaks up through a quasi-free state into a \emph{spectator nucleus} and its recoil (which is the nucleus of interest), measuring thus the inverse of the production. Analyzing the 3- or 4-particle reaction partners in their momenta and angular distributions, one thus can infer the energy levels of interest at low energies for the nucleus of interest, even though it is not directly measured, because of the low-energy behavior of the components inside the compound nucleus.}
Using the  $^{12}$C($^{14}$N,$\alpha,^{20}$Ne)$^{2}$H reaction with $^{14}$N as target, the Catania collaboration inferred a spectacular result \citep{Tumino:2018}: a number of sub-threshold resonances suggests an increase by two orders of magnitude of the S-factor. 
However, the interpretation of Trojan-horse experimental data is complex; the method does not provide absolute cross-sectional data, and the results therefore need to be normalized to the data of direct reaction studies. This requires sufficient overlap in the energy range studied. The THM analysis is model dependent and depends sensitively on the treatment of the Coulomb barrier~\citep{Beck:2020}. The possibility of hindrance is typically not included in the conversion of THM to cross-sectional data. Therefore, the debate about proper normalizations leaves the $^{12}$C+$^{12}$C rate towards low energies still as an open case.
The impact of such a reaction rate revision on our understanding of the evolution of massive stars during carbon burning has recently been modeled \citep{,Dumont:2024,Monpribat:2023,Chieffi:2021}. But this is controversial because modeling still relies on a variety of assumptions, \rd{as recently summarized by} \citet{Chieffi:2025}.

\begin{figure}[h]  
\centering
\includegraphics[width=0.7\textwidth]{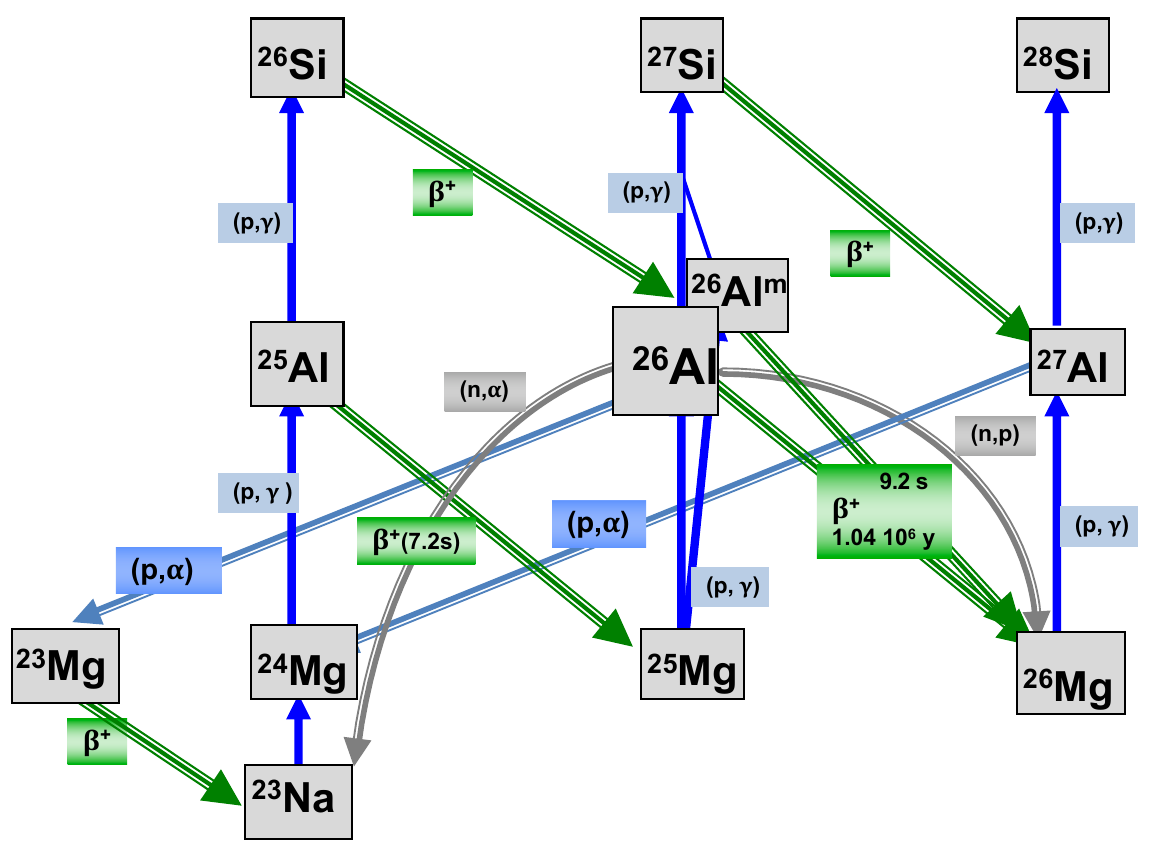}
\caption{The Na-Mg-Al region of the table of isotopes features a cycle of reactions, that processes material from a Na-dominated composition up to heavier isotopes up to Si. (\rd{p},$\alpha$)-reactions return lighter nuclei and thus provide the reverse path compared to the (p,$\gamma$) reactions that characterize the path towards heavier isotopes, thus being properly describes as a cycle. Beyond $^{27}$Al, no such cycles occur, as (p,$\alpha$) reactions are prohibited by the high Coulomb barrier for the $\alpha$~particles with their relatively low energy.}\label{fig_Na-Mg-Al-cycle}
\end{figure} 

\subsection{Photon reactions}\label{sec2.3}
Photon reactions become important as the temperature of reaction sites increases toward a regime that governs stellar neon and silicon burning. These nucleosynthesis phases are dominated by photodissociation releasing a flux of light particles for subsequent nuclear capture reactions. {These photodissociation processes on $^{20}$Ne and $^{28}$Si form $^{16}$O and $^{24}$Mg but also release a large flux of protons and $\alpha$ particles, which are rapidly captured at the high temperature conditions of the burning environment and facilitate a rapid build~up of heavier elements up to the silicon range or nickel iron range, respectively.}

However, photon-induced reactions will also occur at high temperature conditions feeding the so-called p-process \citep{Arnould:2003} or also $\gamma$-process in explosive environments with high photon flux.  
At those conditions, $(\gamma,X)$ reactions on seed nuclei and their thermally excited states will release protons, neutrons, and $\alpha$~particles from nuclei, and thus produce free protons, neutrons, or $\alpha$~particles for further reactions of these with ambient nuclei.
Therefore, binding energies of these particles in nuclei are one of the goals for photon experiments, others being the measurement of inverse exit channels for 
$\gamma$~decays of nuclei of interest. 

Laser backscattering on electron beams from a particle accelerator can provide intense photon beams with desired energies in the MeV regime of nuclear excitation levels, in order to investigate specific nuclear photon reactions. 
Such techniques have been established since the 1960$^{ies}$ in the context of nuclear weapon developments~\citep{Sinars:2020}. 
Compared to broad-band bremsstrahlung photon beams with tagging of the electrons to obtain information on the photon energy, these laser-backscatter facilities provide a much cleaner beam of photons around a specific energy of interest, for specific photo-nuclear cross section experiments.
The HI$\gamma$S facility at TUNL is one such example operated over decades for nuclear experiments \citep{Weller:2015}, another such facility is being set up at the Extreme Light Initiative (ELI) facility near Bucharest in Romania \citep{Balabanski:2023,Gai:2018}.  

More new facilities with very high photon intensities have been developed to study photon reactions of a broad range as well as plasma effects related to screening, thus filling a gap in experimental nuclear astrophysics \citep{Thirolf:2014}. 
Such intense photon beams from lasers with high energy density have been used to implement the wake field acceleration of particles, driven out of a target as the laser pulse deposits a large amount of energy inside a target sheet.
\emph{Inertial confinement} facilities using high-power lasers have also been developed to study how plasma can be energized toward nuclear fusion reactions \citep{Lehe:2025}. 
In this case, the plasma environment of fusion reactions more resembles the situation in cosmic sites, which, e.g., allows to study high-density plasma effects such as screening and Coulomb barrier modifications due to plasma electrons surrounding reaction partners \citep{Cerjan:2018a,Park:2021}.
So, at facilities such as  the Lawrence Livermore Lab's Nuclear Ignition Facility (NIF), the Omega facility at Rochester, variate applications in nuclear astrophysics and corresponding experiments are being set up.
As one such example, experiments could obtain an alternative view on the general physics of \emph{Coulomb explosions}, using the intense photon beam to drive electrons out of a solid \citep{Perrotta:2020}.

\subsection{Neutron reactions}\label{sec2.4}
Reactions involving release or capture of neutrons from a nucleus are important alternatives to charged-particle reactions in cosmic sites, transforming nuclei without the hindering effects of Coulomb repulsion.
Neutron capture rates of individual nuclear species can be measured directly, preparing a neutron beam onto a specific target, or indirectly, measuring photo-neutron production or decay properties of radioactive isotopes with different neutron numbers.
In direct measurements, neutron-producing reactions such as $^7$Li(p,n) are triggered through a proton beam with energy above 2~MeV, and the spectrum of neutrons emerging from the target (here lithium) establishes the neutron source \citep{Reifarth:2009}. 
In this arrangement the energy distribution of neutrons rather closely resembles a Maxwellian thermal distribution of energy 25~keV \citep{Kappeler:2011}. 
This is close to a situation inside a star, in terms of plasma energy, and also because dominant neutron interactions within a stellar plasma are scatterings which establish a thermal energy distribution. 
The direct measurement of neutron captures is realized by identification and counting of de-excitation $\gamma$~rays that reflect the production of a nucleus with an additional neutron captured. 
Alternatively, an initial irradiation with neutrons is followed by measurement of a radioactive-decay in \emph{activation experiments}. 
This second measurement can be arranged off line, or as an in-beam spectroscopic setup, again identifying the product by its characteristic $\gamma$~rays.
Recent developments include constructing a liquid Lithium target to enhance neutron flux \citep{Paul:2020} (at SARAF, Israel), and increasing sensitivity for measuring even smallest amounts of activated radioactive products of neutron capture using accelerator-mass spectrometry \citep{Kutschera:2023}.
Such specific neutron irradiation as well as activation $\gamma$-ray measurements can be optimized through specific target and detector setup arrangements. This allows activations with large neutron fluxes that would otherwise destroy detectors, and, correspondingly, activation measurements in a clean environment that can be sensitive to low intensity $\gamma$~rays. 
Activation measurements are typically used for neutron captures leading to a known radioactive nucleus, while in-beam $\gamma$-ray spectroscopy is used when the neutron capture reaction produces a stable nucleus.
These neutron-capture results from a spectrum of incident neutron energies are useful approximations of the situation in stars. 
Neutrons for such experiments also can be generated from spallation sources as well as from nuclear reactors, and can be used similarly for experiments targeting the integrated neutron capture of specific nuclei.

For measurements of the neutron capture process through specific resonances, a better definition of neutron energies is required. 
When the neutron production is initiated at a specific time through a pulsed particle beam that releases neutrons from a target through a spallation reaction, the time of flight measurement between the production target signal and the detection of a neutron capture signal in the specified target encodes the energy of the reacting neutron \citep{Reifarth:2014,Domingo-Pardo:2025}. 
This is the principle employed at the nToF facility  for neutron experiments at CERN \citep{Domingo-Pardo:2023} and at the LANSCE facility at the Los Alamos National Laboratory in the US \citep{Lisowski:2006}.


\subsection{Beta~decay and neutrino reactions}\label{sec2.5}
{ Weak reactions, i.e. $\beta$~decays and electron captures, are the reactions to change protons into neutrons and vice versa. 
As nuclei become more neutron rich through neutron captures, $\beta^-$~decay transfers the nucleus into a more stable configuration. Similarly, on the proton-rich side, $\beta^+$~decays, or electron captures in this case, produce a more stable configuration, changing a proton into a neutron.
These weak reactions may leave the nucleus spin unchanged, which are called \emph{Fermi} transitions, while spin-changing transitions are named \emph{Gamow-Teller} transitions. 
Inter-shell transitions are called 'forbidden', even though in astrophysical conditions, or in nuclei where shells would be filled completely, such forbidden transitions actually can play important roles.
Theoretical models initially considered nucleons within a nucleus as independent particles, summing up their weak interactions. But correlations among nuclei turned out important, and shell models have become the baseline for weak interaction models, thus accounting for correlations among nucleons \citep{Fuller:1982,Langanke:2000}.
In cosmic environments, the Gamow-Teller transitions provide important reaction channels. This is particular relevant as low-energy reactions dominate, and exit channels with different quantum states provide much of the phase space for the reaction to occur.
Phase space considerations also are important for electron captures, in particular in stellar cores, core-collapse, and neutron stars, i.e. when densities are high so that electrons are degenerate and Fermi energies reach into the MeV region of nuclear reaction Q values.}

Measurements of $\beta$~decay rates for nuclei far away from stability are difficult, since they require the production of a suitable amount of particles.
{Radioactive beams produced by the \emph{ISOL} or spallation technologies can be collected to measure the decay features and lifetime of the radioactive decay process. More recently the use of storage rings has been demonstrated to be very useful to measure the $\beta$-decay rates of specific nuclei, counting nuclei as they get lost by decay from the storage orbit  \citep{Woods:2015}.
This approach helps to test theoretical predictions for the decay processes, which are presently based on QRPA predictions and shell model applications, using Monte Carlo methods are  to explore the transitions within and between different shells \citep{Langanke:2003a}.}
Additionally, charge exchange reactions have proven recently as important experimental tools to study weak reactions rather directly  \citep{Frekers:2018}. 
After pioneering experiments at TRIUMF (Vancouver, Canada), suitable high-resolution beams of light nuclei such as $^3$He were successfully used at KVI (Groningen, Netherlands), and RCNP (Osaka, Japan) to measure (p,n) and (n,p) reactions indirectly through charge exchange from deuterium and $^3$He beams.   

{Reactions of neutrinos with nuclei become important in stars, and in particular near newly-forming neutron stars in the gravitational collapse of massive stars \citep{Balasi:2015,Suzuki:2022,Fischer:2024}.
Similar to the study of $\beta$~decay and electron capture, these reactions are experimentally studied through charge-exchange experiments.
One way to study neutrino reactions directly is the production of $^{12}$C from the rare isotope $^{13}$C using reactor neutrinos.
Theoretical studies, however, are needed for most of the nuclei of interest, and their neutrino interactions, building on the shell model treatments that have proven successful in studies of $\beta$~decays.}

{In this context, the process of neutrino oscillations may alter the neutrino flavors between the neutrino source and the region where interactions with nuclei occur. This has been recognized and studied in recent years \citep[e.g.][]{Ko:2020} \citep[see recent review by][]{Fischer:2024}.}

\subsection{Nuclear theory complements}\label{sec2.6.}

The $\beta$~decay, and even $\beta$-delayed neutron emission rates can be measured in detail for a subset of specific nuclei \citep{Kratz:2001,Moller:2003,Kratz:2017}.
Yet the number of nuclei involved in heavy-isotope reaction paths is large, and many of these nuclei are inaccessible to experiments (see Figure~\ref{fig_ToI-knowledge}).
Also in terms of nuclear properties as needed for neutron binding and $\beta$~decays, there remains considerable uncertainty \citep{Arnould:2007,Lewitowicz:2024}.
Nuclear fission becomes important at the high-mass end, setting a limit to the reaction flow of the r~process beyond masses of \about~260, but difficult to constrain by experiment~\citep{Giuliani:2020}. 
Theoretical models are needed to describe nuclear properties in the wider region of thousands of nuclei that are expected to exist, and to extrapolate beyond measured nuclei \citep{Cowan:2021}. 

\begin{figure}[h]  
\centering
\includegraphics[width=0.6\textwidth]{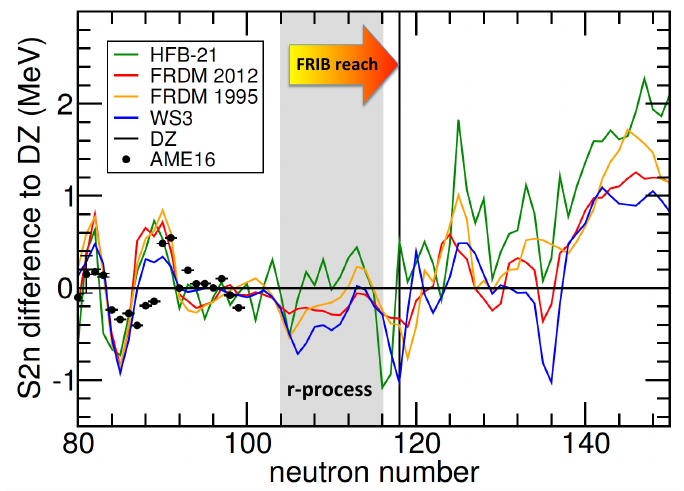}
\caption{Comparison of nuclear binding models away from the region of stable nuclei and towards the neutron drip line. Two-neutron separation energies are shown for Gd isotopes, comparing different mass models (see text) with experimental values as represented by AME2016 (from \citet{Horowitz:2019}). Avoiding spin effects, the 2-neutron separation energy, S$_{2n}$, is determined in measurements along sequences of isobars, and characteristically reveals shell closure effects.}\label{fig_massModels_rProc}
\end{figure} 

The original description of nuclear binding energies (through determining their masses) was made as an ensemble (macroscopic) effect through the liquid-drop model \citep{Weizsacker:1935}.
Since then, there had been an impressive evolution of theories \citep[see][for a personal review]{Moller:2023}.
A bottom-up approach starting from the nucleon-nucleon forces (\emph{microscopic approach}) was pursued, in parallel to efforts aiming at an establishment of an empiric energy density functional (\emph{macroscopic approach}) \citep{Warbinek:2024,Ryssens:2023}. 
Nuclear mass models are a compromise between empirical descriptions of macroscopic (and more-easily measurable) components and microscopic descriptions of nucleon-nucleon interactions deriving from fundamental theory or from empirical prescriptions.   
Nuclear theory attempts to advance models in specific regions of interest, while astrophysical nucleosynthesis models require consistent treatment of neutron captures and $\beta$~decays across the entire path of reactions; therefore, the FRDM baseline from 1995 is still widely used, as $\beta$~decays have been evaluated for a large number of nuclei here.  
Nuclear mass models evolved from the macroscopic \emph{finite range liquid drop} model (FRDM)  \citep{Moller:2016} to   \emph{Hartree-Fock Bogoliubov} (HFB) model versions of different sophistication levels, adding corrections to nucleon-nucleon forces with shell and deformation effects to a Fermi gas model for nuclei, among others \citep[see][for testing model alternatives]{Goriely:2022}. 
Towards the higher range of nuclear masses, nuclear deformations play a large role. The additional degrees of freedom for nuclear rotation and oscillations provide a larger phase space, thus making predictions of effective nuclear binding quite uncertain. 
Experimental results are analyzed and fitted to descriptions of mass models, thus determining model parameters; often models have a large number (\about~30) of parameters determined in this way.  
As a measure, the 2-neutron separation energies $S_{2n}$ between models and measurements (where available) still show an RMS difference of 800~keV (see Figure~\ref{fig_massModels_rProc}), while reliable astrophysical r-process calculations would require a precision of order 100~keV \citep{Horowitz:2019}. 
Weak-interaction rates can be estimated theoretically in the QRPA approach \citep{Sarriguren:2017}. 
But correspondingly, also the weak-interaction processes in atomic nuclei ($\beta$-decay rates, radioactive lifetimes) cannot be reliably calculated in theoretical models, if the details of shells and nuclear structure are not covered; shell models have proven to provide more realistic rates, in better agreement with experimental values \citep[see][for discussions of weak interactions in cosmic environments]{Langanke:2003a,Grawe:2007}. As agreements with experimental values are unsatisfactory, phenomenological models are widely used.
This also extends to the determination of fission barriers, correspondingly \citep{Erler:2012b,Moller:2024}.

Due to the large number of nuclei and nuclear-interaction parameters, also machine learning has been employed. 
For extrapolations into regions of unknown nuclei this may be inappropriate, as the learning algorithms can be trained and thus constrained on available data only \citep{Mumpower:2023}. 
Further advancement of the underlying model and its understanding remains a promising road in any case. 

For a better understanding of nucleon-nucleon interactions and the properties of nuclear matter, also astronomical observations are exploited \citep{Sun:2008,Lattimer:2021} (see discussions of neutron star observations below).

Our limitations of nuclear-physics knowledge incur uncertainties in astrophysical calculations of nucleosynthesis processes whenever nuclear reaction rates need to be extrapolated and estimated from nuclear knowledge rather than from direct experiment, i.e. away from the nuclei near the valley of stability, where it is important for simulating explosive nucleosynthesis, and towards very low energies, where it is critical for modeling quiescent and long-term stellar nucleosynthesis. 
This affects neutron captures and r and i~process, as well as rp~process situations and electron capture and neutrino interactions in the context of core collapses (see Figure~\ref{fig_ToI-reactionpaths} above) \citep{Grawe:2007}.


\section{Cosmic environments for nuclear processes}\label{sec3}   
Cosmic nuclear reaction sites can be grouped into two main categories: 
(I) rather stable environments, { maintaining an equilibrium near a specific temperature, density, and composition values, which evolves moderately in time and space;} 
and 
(II) explosions, with rapid transformations of energy among different degrees of freedom, including nuclear energy.

In case (I) gravity defines the containment of the site, as it collects matter into an object that eventually develops conditions for nuclear reactions, and thus isolates the site from the rest of the universe in terms of matter exchange. We are interested in how nuclear reactions shape the object and its evolution.

In case (II), gravity is one of the agents in the dynamical evolution, {and nuclear energy release is either dominant or at least a key agent for some time. 
In the late stage of a massive star's evolution, electron capture on nuclei triggers the collapse. 
Infalling matter is decomposed into nucleons and $\alpha$s in the shock near the compact newly-forming neutron star, so that re-investment of nuclear binding energy reduces the energy available from gravitational infall. 
Some nuclear binding energy is released as nuclear burning in inner regions reassemble free nucleons and $\alpha$~particles into newly-produced nuclei. The pressure from neutrino reactions in the central regions around the newly-forming neutron star are probably responsible for initiating the explosive dynamics. 
The result of different forms of energy and their interplay remains unclear, the implosion may just end up in a compact remnant, or else produce a supernova explosion.
Other explosions may be initiated when exothermic nuclear reactions are triggered in high-density regions, so that the violent release of nuclear binding energy exceeds gravitational binding and the expansive cooling of the burning region is insufficient to quench nuclear burning. This is the case for thermonuclear supernovae (Type Ia supernovae), where a white dwarf explodes as a result of carbon fusion ignition.
Novae and Type~I X-ray bursts are an intermediate case: Explosive expansion is initiated by nuclear-reaction triggers (hydrogen burning for novae, helium burning for Type~I XRBs). 
But soon after ignition \rd{of a nova}, expansive cooling extinguishes the nuclear reactions, and gravity takes over, gradually settling the material again onto the compact star, slowed down by radioactivity produced in the initial burning. 
\rd{In Type-I X-ray superbursts, strong gravitational confinement restricts such dynamics, and fuel exhaustion causes the explosion to end.}
In neutron star \rd{mergers}, gravity remains dominant and produces a single compact object. But in the outskirts of the violent decomposition of a neutron star, a small part of the total material undergoes nuclear burning and is gravitationally unbound. 
Unusual as such an environment is for nuclear reactions, the ejecta of such implosions have become an important aspect of studying the origins of the r~process.}
\rd{All these explosive situations are addressed in more detail below.}

{In this chapter,} we want to illuminate how nuclear reactions interplay with the other degrees of freedom such as \rd{temperature and density, and} matter and energy flows, as these shape the diversity of stars and of explosions.
We summarize {our current understanding and models for the candidate cosmic sites and environments and their characteristic nuclear reactions with production of new nuclei, and point out achievements and challenges}. 

{Simulations have become a versatile tool to understand and illustrate the details of our current beliefs, theories, and models, mainly due to growing computing capabilities.}
Examples are simulations of stellar structure evolution and of stellar explosions, or even cosmological-model simulations and compositional evolution models of galaxies. 
Ideally, those tools and results can then be used to design informative measurements (see section on Nuclear Astronomy) or experiments (see previous section on Nuclear Physics), in order to learn about potential mistakes or improvements in our interpretations.

\begin{figure} 
\centering
\includegraphics[width=0.8\columnwidth,clip]{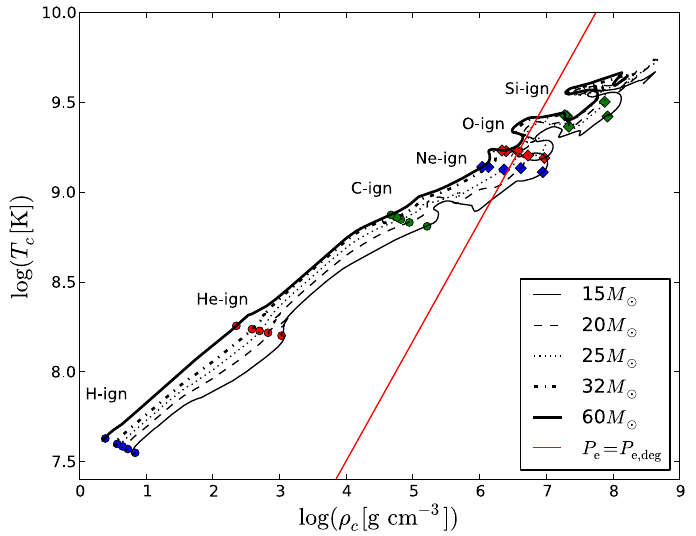}
\caption{The phase diagram for the evolution of stellar cores, through the different burning stages from Hydrogen burning until silicon burning before gravitational collapse. The red line shows the separation between thermal and degenerate plasma (towards bottom right), where in the latter the pressure is dominated by Fermi pressure of electrons in gas that is confined to a small-enough volume. (From \citet{Pignatari:2016}).}
\label{fig_massiveStar_temp_dens}       
\end{figure} 

\subsection{Stellar interiors}\label{sec3.1}   
Gravitational compression implies that stellar cores obtain high temperatures, high enough so that nuclear reactions may occur \citep{Iben:1991}.
Even at slow rates these reactions are important for the fate of stars as cosmic objects.
Thermal energies result in the Gamow peak being located far in the low-energy tail of quantum tunneling through the Coulomb barrier. In the Sun, the burning of hydrogen currently proceeds at a temperature of 15~10$^6$~K \citep{Acharya:2025}, liberating \about~26~MeV of nuclear binding energy per $^4$He formed. This nuclear energy is sufficient to stabilize the star against gravity, for \about~8~10$^9$~years in the case of our Sun and its total hydrogen supply of \about~10$^{57}$ atoms. 
{This phase of hydrogen fusion in the stellar core is called the \emph{main sequence} phase of stellar evolution, as it is the longest-lasting nuclear burning phase. 
It characterizes most of the stars that we can observe, as they are observationally represented in the well-known \emph{Hertzsprung-Russel diagram} of objects characterized by their brightness and temperature; the main sequence is the dominant feature in this diagram, diagonally stretching from low temperature stars such as the Sun in the lower right up to hot massive stars in the upper left.}

Inside stars, energy transport processes and their characteristic times are short compared to the nuclear-burning time scale. 
The star has time to arrange its structure to this given amount of stellar matter providing the gravitational confinement, and to the given nuclear energy liberation in the core, compensating also for energy loss at the surface from radiation and an outflow from a stellar wind. 

Hydrogen burning converts hydrogen nuclei (protons) through a sequence of
proton capture reactions and $\beta$-decays to $^4$He.
For low mass stars,
$M\le 1.5 M_{\odot}$ hydrogen burning is dominated by the
pp-chains~\citep{Salpeter:1952a}, while
for more massive stars, $M\ge 1.5 M_{\odot}$ hydrogen burning
operates through the CNO cycle (discussed above; see Section 2.2), a catalytic reaction
sequence which takes a key role in nuclear astrophysics and its
many stellar environments~\citep{Wiescher:2018a}. Even
for the Sun, for which the CNO cycle contributes only one percent
in energy, the CNO cycle is a source of neutrinos. 
As this is directly associated with the solar metalicity, measurements of neutrinos from the Sun (see section 4 on Nuclear Astronomy) directly {probe} the CNO cycle in solar hydrogen burning \citep{Borexino-Collaboration:2018}.
Since the operation of the cycle relies on the abundance of the
catalytic elements C, N and also O, it cannot contribute to hydrogen
burning in first stars, which by definition have the primordial composition free of CNO catalysts.

\begin{figure} 
\centering
\includegraphics[width=1.0\columnwidth,clip]{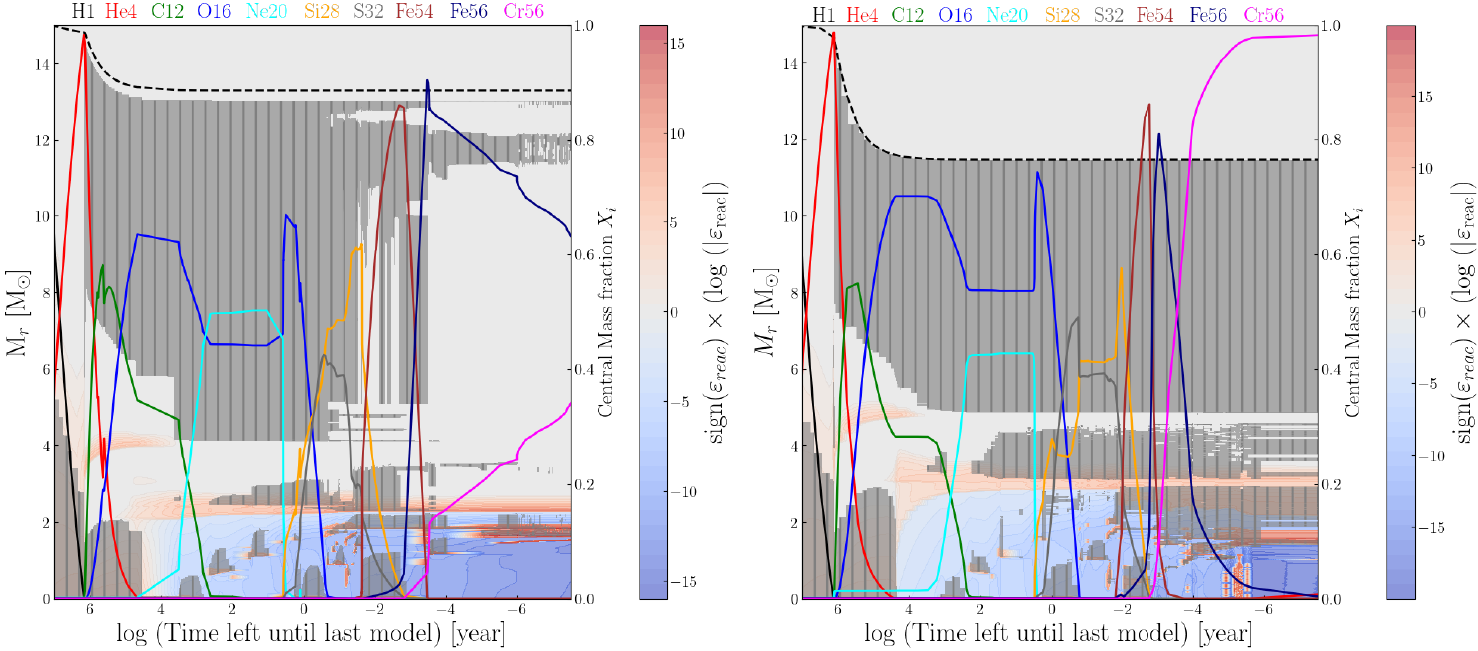}
\caption{The Kippenhahn diagram for the evolution of stellar structure. {The time axis is logarithmic and extends from the creation of the star to its gravitational collapse, emphasizing the late phases through the logarithmic scaling. The vertical axis extends from the center of the star to its surface, and is scaled in units of enclosed stellar mass.} Shown are two different implementations of the physical processes during stellar evolution, from the Geneva code (left) and the public MESA code (right). (From \citet{Griffith:2025}).}
\label{fig_massiveStar_structEvol}       
\end{figure} 

Once hydrogen in the core becomes exhausted, gravitational contraction {results in a temperature increase} until the next nuclear fuel $^4$He can undergo fusion reactions to liberate nuclear binding energy again.
{At the same time,} hydrogen burning from then on proceeds in a shell\footnote{Note that only \about~10\% of the total hydrogen will have been consumed during the main-sequence evolution.}, continuing throughout evolution in parallel to the nuclear processes in the stellar core (see Figure~\ref{fig_massiveStar_structEvol}). 
The radiation pressure caused by nuclear energy release in the shell causes expansion of the outer zones of the star, while the core contracts due to a lack of energy production, removing the required thermal pressure for hydrostatic stability, and thus further expanding the envelope as a reaction to the contraction of the core.
This is called the \emph{giant} phase in stellar evolution \citep[see][for a textbook introduction to stellar evolution]{Clayton:1968}.
Due to the smaller energy gain per $^{12}$C produced (7.3~MeV), the total helium-burning time  is not quite as long as the main-sequence time; for the case of our Sun it is \about~10\% of the main sequence phase. 
This seems sufficiently long for most stars to still achieve and maintain hydrostatic equilibrium, rearranging temperatures and densities to the convective and radiative energy transport processes within the star and the nuclear energy gain in the core balancing radiative and wind losses at the surface.

{So, beyond the main sequence phase, stellar structure becomes increasingly complex. 
Core and shell burnings each liberate nuclear energy {at different locations within the star}. 
As radiative energy transport becomes insufficient to locally maintain hydrostatic stability, convective mixing takes over and thus transports nuclear-burning fuel and ashes within the star. 
A \emph{Kippenhahn diagram} is valuable to illustrate structural evolution~\citep{Kippenhahn:1990} (see Figure~\ref{fig_massiveStar_structEvol}).} 

{The total mass of a star determines its internal gravitational pressure. Therefore more-massive stars with their hotter cores evolve more rapidly due to higher nuclear burning rates, consuming fuel faster. Such a more vigorous internal energy liberation also challenges hydrostatic stability: massive stars in their giant phases experience internal instabilities as convective energy transport occurs, and enhanced radiation pressure results in much stronger stellar winds and episodic mass loss periods.
With core helium burning and continued hydrogen shell burning, two energy liberation regions incur convective instabilities in the giant phase. 
In continued stellar evolution, more shells within the star may experience nuclear burning simultaneously, or intermittently, as illustrated in Figure~\ref{fig_massiveStar_structEvol}. 
As a result, the stellar core proceeds through different stages of nuclear burning, with higher temperatures successively consuming helium, then carbon, \rd{oxygen}, neon, and silicon, {- a sequence of fusion and photo-dissociation processes -} until the composition in the core reaches iron nuclei. 
\rd{Binding energy of nucleons in nuclei reaches a maximum for iron-group nuclei, with $^{56}$Ni as most-bound configuration for symmetric matter of equal numbers of neutrons and protons}, and no further release of nuclear binding energy can occur to stabilize the star against gravitational contraction.
Nuclear burning stages and their characteristic temperatures and densities as shown in Figure~\ref{fig_massiveStar_temp_dens} describe these \emph{gravitationally-confined nuclear reactors}\footnote{quoted from Stan Woosley} rather well. }  

{At higher temperatures of carbon burning and beyond, more of the thermal neutrinos created by plasmon decay, Compton scattering, and bremsstrahlung \citep{Vitagliano:2020} become part of the material mix in nuclear burning zones, in addition to nuclear-reaction neutrinos created in weak reactions. URCA pairs with cycles of electron captures and $\beta$~decays contribute to the neutrino budget as well \citep{Barkat:1990}. This leads to immediate energy losses from the nuclear burning region, because neutrinos escape through the overlying envelope without need for time-consuming energy transport processes. 
As a result, nuclear burning rates must increase to provide the required energy for prevention of collapse, and stellar evolution proceeds more rapidly in these later core-burning phases \citep{Heger:2003,Limongi:2010}. 
With shorter time scales of this nuclear and energetic evolution in their interiors, the higher-mass stars become hydrostatically unstable during  carbon and neon burning. This is reflected in the trajectories in Figure~\ref{fig_massiveStar_temp_dens}, which deviate from a straight diagonal path for these late phases, and reflect the structural rearrangements with their deviations from hydrostatic equilibrium. }   

{Mixing processes are a key agent determining stellar structure and composition, and hence nuclear reactions at various locations within the star.
As radiative energy transport becomes insufficient when nuclear energy generation increases, convection takes over. 
The Schwarzschild criterion in principle defines when gas in a region becomes unstable to convection. But compositional gradients are ignored herein, so that the Ledoux criterion should better be used. 
Convective overturn of materials is complex in detail, due to the momenta carried by the convective flows, and due to the multi-component composition.
Therefore, \emph{extra mixing} has been discussed, sometimes captured by parametrized mixing corrections also termed \emph{overshoot} and \emph{semiconvection}. 
Compositional inversions from such convection may also lead to \emph{thermohaline mixing} when a heavier composition of gas ends up above a lighter composition layer \citep{Karakas:2014}. 
Moreover, the Coriolis force incurs mixing processes from sheer motions, when stars carry internal rotation from the angular momentum transport of material accretion as the star formed \citep{Maeder:2000,Meynet:2013,Meynet:2016}. 
Stellar envelope material also is subject to impacts of magnetic fields initiated by the material flows. 
At the boundaries of convective zones, this results in complex patterns of motion and turbulence.
These lead to intrinsically 3-dimensional mixing processes. 
\rd{\emph{Mixing length theory} had been derived to describe material mixing under hydrostatic conditions within a star, essentially a one-dimensional approach. This theory is clearly inadequate to model such complex 3D processes, or even explosive conditions, yet often applied in lack of a useful alternative.} The empirical adjustment of mixing length scales has been used as an empirical way to approximate these processes within mixing length theory as  commonly applied in all the 1-dimensional stellar models \citep{Lagarde:2012}.  
3D modeling of convective boundaries have recently been undertaken for small regions within a star \citep{Rizzuti:2023}. These then provide a simulation-based calibration of such mixing-length parametrization to account for the many complexities of material mixing within stars. }

{The presupernova evolution of massive stars in their late phases is a challenging topic in astrophysics. 
The {above-discussed processes of mixing  determine how evolution proceeds and modifies} the structure of the star and its hydrostatic balance \rd{ \citep[see][for a recent broad-impact study]{Whitehead:2025}}. 
Many {different scenarios can be inferred from simulations, as mixing prescriptions and other parameters are varied.} 
Examples are: burning shells progressing further inwards (instead of the more-plausible trend to proceed along the availability of unburnt fuel at larger distance from the core) further inwards with time \citep{Jones:2015,Jones:2019}, or merge \citep{Roberti:2025,Issa:2025}, or hot-bottom burning  \citep{Karakas:2014}. 
Even though agreement is reached on the fundamental microscopic processes, their implementation in different numerical codes may provide different results, depending on how these processes are  approximated and numerically stabilized within each code. Figure~\ref{fig_massiveStar_structEvol} illustrates an example; alternative codes for stellar evolution are maintained, e.g., by the Geneva \citep{Schaerer:1993,Ekstrom:2012,Georgy:2013,Meynet:2013,Griffith:2025} and Rome \citep{Chieffi:2010,Chieffi:2013,Limongi:2018} groups. 
Astronomical observations may provide clarifications of such issues concerning stellar interiors (see Section 4 on Nuclear Astronomy below).
}

Hydrostatic evolution with core and shell burning regions characterizes \emph{AGB stars} of the lower mass range (1.5 to about 3~\Msol) on their evolution along the asymptotic giant branch \citep[see][for reviews on AGB star evolution and nucleosynthesis]{Cristallo:2011,Karakas:2014,Lattanzio:2016,Lattanzio:2019}.
Violent overturns of their interiors outside the core region, called \emph{dredge ups}, are an essential outcome of core evolution, and indicate that hydrostatic stability is challenged.
AGB stars are characterized by their unstable outer envelopes.
So-called \emph{carbon pockets} may result from the complex mixing processes discussed above \citep{Jones:2016}, and are held responsible for preparing s-process conditions \citep{Herwig:2005,Karakas:2014,Lugaro:2023}. 
The main s~process as discussed above (section 2.5 on neutron reactions) is believed to occur in these stars. 
It is debated how in detail this occurs. 
The initial assumption that protons from the envelope are mixed into carbon-enriched regions below seems now to be confirmed by 3d model simulations of the hydrogen-helium shell environment \citep{Stephens:2021}. Protons will be captured by the available $^{12}$C abundances to form $^{13}$N and subsequently $^{13}$C by $\beta$-decay. This process activates the neutron-producing $^{13}$C$(\alpha,n)^{16}$O reaction providing the neutron fuel for the s~process.
Total yields then depend critically on the details of such $^{13}$C pocket scenarios {and the impact of so-called neutron poisons, neutron capture reactions on highly abundant light isotopes, e.g. $^{13}$C($n,\gamma$)$^{14}$C, $^{16}$O($n,\gamma$)$^{17}$O, and $^{14}$N($n,p$)$^{14}$C} \citep[see][]{Wiescher:2025b}.
Surface abundances of stars have been found to have larger enrichments of, e.g., carbon than predicted by stellar models. 
Given the intrinsic uncertainties in such stellar mixing \citep[see][for a discussion of issues and challenges]{Lattanzio:2019}, the carbon pocket has become a popular picture for the essence of mixing followed by neutron release and the s~process in AGB stars. 
\rd{But the astrophysical modeling has still to be carried beyond \emph{scenarios} and one-dimensional models, in order to demonstrate that such transient proton ingestion is a well-regulated outcome of the dynamics of shell boundaries in AGB stars.}
Such s-process conditions are also responsible for enriching neutron-rich silicon isotopes. This is one of the key signatures in presolar grains and their attributed origin in AGB stars, \rd{thus providing observational access to the mixing dynamics in AGB stars towards an s~process} {(see section 4 on Nuclear Astronomy below)}. 
Also, { a neutron capture process with neutron irradiation densities in the  intermediate} range of \about~10$^{14}$~cm$^{-3}$  has been attributed to AGB stars, through so-called \emph{proton ingestion events}, similar to, but more effective than the carbon pocket scenario. 
Proton ingestion into convective Helium-burning regions could be a characteristic of  stars in the higher AGB mass range of 5-8~\Msol. 
Alternatively, accreting material from a companion star might destabilize the AGB star envelope \citep{Choplin:2023}.

More massive stars with initial masses beyond 20--25~\Msol evolve into a \emph{Wolf Rayet} phase, characterized by intense mass loss of up to 10$^{-4}$~\Msol per year \citep{Vanbeveren:2009a,Meynet:2009}. 
Despite these dynamic phases of the outer envelope challenging the interior structure, such stars as a whole also still remain close to a hydrostatic structure, within first approximation. 

Very massive stars in excess of 100~\Msol should have been co-created with lower-mass stars in star-forming regions, from the early Galaxy until present. They are copious producers of hydrogen-burning ejecta, according to models, and thus may be important for localized abundance anomalies within a galaxy \citep{Higgins:2023,Higgins:2025}.

{ Stars of initial masses beyond 25~\Msol or even beyond \about~15~\Msol may collapse to a black hole and thus not produce an explosion . 
On one hand, this \emph{explodability} of massive stars upon their collapse is a debated outcome of the explosion physics (see discussion of Stellar Explosions in the next section.)
On the other hand, onset of pair creation as temperatures in the stellar core approach and exceed the pair production threshold of 1.022~MeV result in loss of pressure in the core. This possible trigger of pair instabilities of stellar structure is expected to lead to pulsations and a \emph{pulsating pair instability supernova} (PPISN; in the mass range for He cores of 45 to 65 \Msol) \citep{Fowler:1964,Woosley:2017}, or to disruption of the entire star (PISN; above 65--70~\Msol) \citep{Woosley:2021,Ozel:2010}.
The resulting \emph{black hole mass gap} in this region of stellar masses, however, might be populated by different variants of catastrophic late evolution of stellar cores leading to direct black hole formation \citep{Farmer:2019,Rizzuto:2022,Umeda:2020,van-Son:2020}. Gravitational-wave events where black hole masses in this \emph{gap} range have been deduced are strong evidence of such direct collapses.}

Stars are often formed in groups, and multiple stars may be created as single entities yet close enough to have significant mutual interactions that affects their evolution. 
The impact of a companion star on stellar evolution had been described long ago \citep{Iben:1991}. 
The rather obvious effect of a companion to enhance mass loss and thus stripping the envelope has been modeled \citep{Woosley:1995a,Podsiadlowski:2004,Tauris:2015}. This is understood as a path towards explaining core-collapse supernovae which do not show a hydrogen envelope, and are thus observationally known as supernova types Ib and Ic.
More difficult are successive periods of mass exchange in different directions within a binary system, or phases of a common envelope \citep[e.g.][]{Wei:2024a}.
Thus, binary interactions and their impacts have developed into a major field of stellar evolution \citep{Iben:1991,Vanbeveren:1991,Podsiadlowski:2004,Sana:2012,de-Mink:2013,Izzard:2013,Vanbeveren:2017,Laplace:2021,Marchant:2021,Schneider:2024}. This is mainly addressed by theory and simulations, because it is difficult to obtain and exploit observational constraints on models of binary interactions, as observational phenomena cannot easily by attributed to binary interaction physics as discriminated from instabilities in single-star evolution.

\emph{Compact stars} are stabilized by degeneracy pressure from electrons (the white dwarf stars), and by the complex pressure behavior of dense nuclear matter called its \emph{equation of state} (the neutron stars). 
White dwarfs are the compact remnants of stars after the giant stage and strong mass losses therein \citep[see][for a recent review]{Saumon:2022}. 
Typical mass and size values are 0.16~\Msol  and 0.01~R$_{\odot}$, respectively. 
{Matter is confined to a rather small volume in compact stars. Electrons move freely as a Fermi gas within the charge-neutral star which is bound by gravity that is caused mainly by the nucleons. 
As thermal energies become small and approach zero for the electrons, their kinetic energy level occupancy obeys quantum constrains and the Pauli exclusion principle for identical states. 
Occupancy of energy levels significantly above what would correspond to thermal energies produces a \emph{Fermi pressure} from the electrons. 
This stabilizes the stellar structure for long times, as it is independent of thermodynamic evolution and merely a result of the compactness. 
Radiative energy loss from the surface is modest due to their small surface, and provided by the heat capacity of the stellar matter as dominated by nucleons. 
The nucleons, on the other hand, thus are left at rather cold temperatures, and nuclear interactions do not contribute to energy generation.
The dense nuclear matter of neutron stars is more complex, and the corresponding equation of state reflects the multi-nucleon interactions as they increase with compactness, and phase transitions to quark matter.
Nuclear matter within a neutron star can be considered as `cold' \citep{Baym:2018}, and quark-gluon plasma aspects only arise at extreme compactness (unlike when neutron stars are formed during gravitational collapse of a star; see below).
Radiative losses are modest due to the small surface of neutron stars, with typical radii of 10-15~km at masses of 1-2~\Msol \citep{Ozel:2016}.}

Stellar structure and stability can be calculated from the relativistic description of behavior of such matter and using its equation of state, the Tolman-Oppenheimer-Volkoff equations \citep{Oppenheimer:1939,Tolman:1939}.
{ Neutron stars turn out to be excellent laboratories to study dense nuclear matter at two extremes of parameters that do not exist in nuclei \citep{Lattimer:2021,Sun:2025} 
{(see also our discussion of Extremes of Matter in section 3.3).} 
The neutron richness, or proton poorness, is extreme in neutron stars, the characteristic $Y_e$ variable of net charge and measuring deviations from neutron-proton balance ($Y_e=0.5$).
The symmetry term of nuclear binding, hence, adopts extremes in neutron stars, and its connection to studies of nuclei with different neutron excess is a valuable diagnostic.
The equation of state of neutron star matter describes its pressure dependency on thermodynamic variables of temperature and density. It shapes the compressibility of neutron star matter, which, on the other hand, determines tidal deformability of neutron stars as they collide in neutron star mergers, as well as the rebounce from gravitational collapse of a massive star as it forms a central neutron star, and develops an explosion or not \citep[see review by][in Supernova Handbook]{Piekarewicz:2017}.}
The determination of the equation of state for nuclear matter throughout the range of density and temperature is one of the major challenges of nuclear astrophysics, involving nuclear theory as well as experiments using high-energy collisions of heavy ions, and astronomical observations of neutron star properties \citep{Baym:2018,Lattimer:2021}.

At their high densities beyond 10$^{10}$g~cm$^{-3}$, neutron star interiors, even if relatively cold, provide an environment for nuclear {reactions}.
On one hand, Coulomb barriers are lowered by electron screening and penetrated more easily under this elevated gravitational pressure \citep{Salpeter:1969}.
On the other hand, nuclei are bound in lattice structures and their thermal energies can be neglected. Within these lattice structures the nucleons (neutrons, protons) adjust according to external pressure, as it may increase either through accretion of matter or by sinking deeper into the neutron star  \citep{Haensel:1990}.  
Matter within the crust of neutron stars includes special nuclear phases called \emph{pasta}, as neutron-rich nuclei align in strong gravitational and magnetic fields  \citep{Piekarewicz:2023}.
Nuclear reactions occur in outer regions closer to the neutron star surface, where nuclei are still present \citep{Meisel:2018}.
Within neutron stars, sequences of electron capture on nuclei with successive $\beta$~decay provide \emph{URCA pairs} \citep{Schatz:2014}, and the neutrinos generated within these escape and establish an important source of cooling \citep{Page:1992,Haensel:1994}.

\emph{Pycnonuclear burning} {is expected to occur} somewhat deeper below the surface of the neutron star, at a typical depth of a km, {as a reaction to the pressure increase from accretion.
{Electron captures on heavier neutron deficient isotopes such as the products of earlier rp-process burning in type-I bursts drives the abundance distribution towards the neutron rich side where electron capture induced neutron emission breaks up the heavy element to very neutron rich light isotope species of neon and magnesium~\citep{Lau:2018,Jain:2025}. At near nuclear density conditions these isotopes can undergo fusion reactions, facilitated by screening of the electron gas~\citep{Yakovlev:2006}  }.
This nuclear fusion energy }is not radiated away at the neutron star surface and rather heats the neutron star interiors \citep{Yakovlev:2008} (as opposed to nuclear burning of surface material in bursts as discussed below under explosive nuclear burning).
Nuclear reactions relevant in these high densities deeper into the neutron star crust have been studied extensively \citep{Lau:2018}.
{ This internal source of heat} does not directly determine stellar structure or stability, 
{but} is relevant to properly understand the cooling behavior of neutron stars \citep{Jain:2023,Jain:2025}. {This has been found important when} following surface explosions and their phenomenology to study properties of neutron star matter as it cools.

{Both for white dwarfs and for neutron stars}, accretion of additional matter onto the surface of the compact star leads to an input of gravitational energy and enhancement of gravitational pressure. This may alter the energy balance and heat matter to nuclear-reaction temperatures.
As stability in these cases does not directly relate to thermal pressure, the consequences of releasing nuclear binding energy through nuclear reactions are very different: \rd{As nuclear reactions begin during such accretion}, there is no expansive cooling reaction of the burning site, which otherwise quenches nuclear burning, and due to local high densities conditions the fuel may be processed to release large amounts of energy, until \rd{a runaway occurs, lifting the degeneracy, and }explosion dynamics sets in. This leads to explosions of novae and thermonuclear supernovae in the case of white dwarfs, and of X-ray bursts (called 'type~I') on neutron star surfaces.

\subsection{Cosmic explosions}\label{sec3.2}   
In the case of stellar explosions, the dynamical evolution proceeds much faster than the nuclear-reaction time scales.
An astrophysical description of cosmic explosions commonly adopts several assumptions about the initial pre-explosive state. 
Cosmic explosions that we currently know and study are: the Big Bang, supernovae, novae, kilonovae, and type-I X-ray bursts. 
Definitions may vary, and \emph{impostors} have been discussed, which are transient events that seem to show phenomena of explosions, but have different reasons for their transient behavior (such as interaction with circum-stellar matter, or internal phase transitions/condensations) \citep{Gal-Yam:2019,Gal-Yam:2025}.
In nuclear astrophysics we aim to investigate and understand the release of nuclear energy during explosive nucleosynthesis. 

Beyond the initial states of density, composition, and temperature we face a major constraint of the explosion dynamics from the {evolutionary history and the }gravitational energy that is set by the pre-explosive object.
Boundary conditions for nuclear reactions are different for different types of explosions.
{We distinguish processes within a single object, and processes that characteristically involve a binary companion star or object.}

\subsubsection{Single stars}\label{sec3.2.1}
In the case of \emph{massive stars}, the central energy source from release of nuclear binding energy fades and becomes insufficient to stabilize the star against gravitational collapse \citep{Heger:2003a}.
One cause of this is the exhaustion of nuclear fuel: once the core composition proceeds from neon-silicon to iron, the most-stable nuclear configuration is reached, and further fusion reactions would be endothermic. 

{A second cause is the increase of electron captures as temperatures and densities increase, in particular in degenerate cores, and on abundant $^{20}$Ne and $^{24}$Mg nuclei \citep{Langanke:2003}. 
These continue to occur on various nuclei and proceed throughout the collapse, thus changing the composition and the average charge to neutron-rich material (Y$_e<$0.5) during collapse.
Electron captures are the most-plausible trigger of gravitational collapse in this mass range \citep{Nomoto:1984b,Wanajo:2009,Wang:2024a}, as they eliminate the pressure of the electrons that have been energized through degeneracy in the cores of the low-mass end of massive stars, i.e. 9--13~\Msol.   
The results are called \emph{electron capture supernovae (EC-SNe)}. }

A third cause is the progressive relevance of neutrino components of the matter-radiation mixture (including above electron capture neutrinos), which leads to energy loss { from nuclear burning regions, }instead of building up thermal pressure.

A fourth possibility for very massive stars (if they really are created) in the mass range above \about~100~\Msol is the opening of another lepton-related channel as photons obtain increasing energies with the mass of the star: Photon-photon collisions can lead to pair creation when center-of-mass energies exceed the rest mass of an electron-positron pair of 1.022~MeV, i.e. at temperatures in the core that exceed \about~7~10$^8$~K. 
Pair creation results in substantial loss of energy from the photon field that otherwise stabilizes the star against gravity. At masses above \about~140~\Msol this is expected to lead to \emph{pair instability supernovae} \citep{Fowler:1964,Barkat:1967,Heger:2002,Umeda:2008,Kozyreva:2017} disrupting the entire star.  
Already for massive stars of lower total birth masses from \about~70~\Msol upwards, this pair creation would set in and result in instabilities and pulsations with substantial mass loss, releasing products of nucleosynthesis \citep{Rahman:2022}. 

As a result of these various fading central energy sources towards the end of stellar evolution of stars with masses exceeding \about~8~\Msol, gravitational collapse will take over, and proceed faster with electron capture and hence neutronization increasing in central regions \citep[see][for a recent review]{Jerkstrand:2025}.
{This neutronization of matter near and in the newly-forming neutron star in the center of gravitational collapse results in an additional and very intense burst of electron neutrinos.
This is one of the significant messengers for astronomical measurement (see section 4 on Nuclear Astronomy).}

The collapse is halted when central density reaches values above nuclear saturation density beyond 2.7~10$^{14}$g~cm$^{-3}$, as compression then faces the repulsive part of the strong nuclear potential. 
This is reflected in the equation of state of {matter in the newly-forming} neutron star \citep{Janka:2023}, which describes the pressure reaction of this matter as it cools and becomes more compact. 
The rebound from the sharp pressure increase at the inner region of collapse thus creates a shock {for infalling envelope material} some 50--100~km above the proto-neutron star surface. Infalling nuclei are decomposed in the enhanced density region of this shock and from intense neutrino interactions, thus further consuming nuclear binding energy as infalling heavy nuclei are converted to $\alpha$~particles and free nucleons. A nuclear statistical equilibrium establishes in this region, {while infalling nuclei have become neutron rich from electron captures}.  
Energies are sufficient so that nuclear reactions can shift this equilibrium to include $^{56}$Ni as matter expands and cools \citep{Clayton:1968}. 
{If a supernova explosion can be launched, expansion} of the central region then rapidly reduces temperatures and densities, thus affecting nuclear reactions and cooling properties of the materials during explosion. 
The nuclear statistical equilibrium evolves into freeze-out with increased abundance of $\alpha$~particles. 
The outward-propagating shock is expected to lead to \emph{explosive nucleosynthesis} as it runs through the envelope of the surrounding star that has had no time to fall in \citep{Thielemann:1990}. Explosive nucleosynthesis  is defined as the network of nuclear reactions occurring { in the envelope} as an outward moving shock heats { this matter to nuclear burning temperatures}. 
During expansion, nuclear reactions fade and freeze out characteristic abundances that reflect nucleosynthesis in central regions and the expansion and cooling behavior of matter \citep[see][for pre-supernova and explosive nuclear processing]{Limongi:2018,Limongi:2020}.
\rd{The nucleosynthesis products from this inner region characteristically reflect those conditions \citep[see][for a case study; such measurements are addressed in section 4.1]{Magkotsios:2010}}.

Explosions of collapsing massive stars are not a natural outcome, as rather the formation of a compact black hole { may be an alternative outcome of the gravitational collapse, and already imprinted from earlier stellar evolution \citep{Laplace:2025}:} explodability is a major theme of studies \citep{Smartt:2015,Ertl:2016,Ertl:2020,Muller:2016,Ebinger:2019,Ebinger:2020,Zapartas:2021,Maltsev:2025}.  
The compactness of the core at the time of collapse appears to determine the fate of the collapse, proceeding either towards a central black hole, or towards an explosion of the matter outside the central compact newly-forming neutron star. 
The supernova explosion energy thus covers a wide range, from 10$^{50}$ erg in Type IIP supernovae to 10$^{52}$~erg in hypernovae \citep{Jerkstrand:2025}. 

The expansion { of the ejecta} from a supernova explosion often is not spherically symmetric.
The dynamics of the collapse is strongly affected by the non-spherical accretion and by large-scale motions resulting from previous stellar rotation, mediated by magnetic fields, which develop a key role in controlling angular momentum and energy transport during collapse and explosion \citep{Burrows:2021,Fryer:2000,Muller:2016a}.
Furthermore, the neutronization of matter towards the central object and the intense neutrino interactions near this proto-neutron star are believed to have a decisive role in turning the collapse into a successful supernova explosion.
{ The neutrino mechanism is held responsible for the energy deposits in the gain region of the stalled shock \citep{Jerkstrand:2025,Burrows:2021,Janka:2007}. Turbulent infall as well as magnetic fields are expected to incur deviations from spherical symmetry of the infall and thus of the directionality of the effects from the central neutrino source \citep{Jerkstrand:2025,Sykes:2025}.}
Models for collapsars, magnetic-jet supernovae, and hypernovae plausibly describe such explosive nuclear-reaction environments. 
These {variants} are expected to {each} leave characteristic imprints in the isotopic composition of the nucleosynthesis ejecta from such sources.

The special regions near the proto-neutron star, where electron captures and inverse $\beta$~decays neutronize the nuclear composition, { had also been considered as} plausible locations for the \emph{rapid neutron capture (r-) process} \citep{Woosley:1994a} to occur. 
The early idea of r~process nucleosynthesis in the high-entropy wind behind the expanding supernova shock  \citep{Farouqi:2009} now appears unrealistic, as recent simulations of gravitational collapses rather suggest a proton-rich composition here , i.e., $Y_e \geq 0.5$  \citep{Wanajo:2011b,Cowan:2021}. 
{Consequently, this has led to the suggestion of a \emph{p~process} }occurring in this region.
{On the other hand}, the plasma flows near a magnetically-confined jet may also create the required low-$Y_e$ conditions for a successful r~process in magnetized gravitational-collapse {variants}.
Also, a neutrino-related p~process is being discussed for this inner supernova region. In this context it is interesting that oscillations of neutrinos in the inner supernova may be relevant and affect conditions for heavy-element synthesis \citep{Mori:2025}.
 {The assessment of impacts on nucleosynthesis} is difficult: neutrino transport and their nuclear interactions require sophisticated modeling \citep[see, e.g.][for details]{Janka:2016,Sieverding:2018,Fischer:2024}.

{In the variant of} rapidly-collapsing pair instability supernovae, electron captures {are expected to} be to slow to significantly modify the neutron-to-proton ratio. 
Even-even nuclei as well as the iron group products would be {favored nucleosynthesis products} \citep{Takahashi:2017}. 

Altogether this remains a very active discussion of heavy-element production and exceptional nucleosynthesis occurring in the context of gravitational collapse and black hole formation with its generation of a wide range of extreme conditions energetic enough for nuclear reactions.

\subsubsection{Stars in binary systems}\label{sec3.2.1.2}
Binary evolution provides another path to a stellar explosion through sudden changes in the energy balance of a star. An external trigger {thus  may arise} when the binary separation is close enough {for sufficient transfer of mass}.
The binarity's role could be (i)  mass transfer that either gradually brings a star to a critical state (such as discussed in gravitational collapse initializations above), or the (re)-ignition of nuclear reactions in a stellar core, or (ii) a violent collision of the two binary components, after having gradually lost their orbital momentum.
In either case, the long time scales of mass transfer, or orbital energy loss, imply that such explosions {occur with significant delay after stars have been formed, and they are rare, compared to single star explosions}.
Binary interactions can be quite complex, {once} orbital separations are small enough \citep{Podsiadlowski:2004,Schneider:2024}.
If one of those binary components is a compact star, i.e., a white dwarf or a neutron star, even modest amounts of accreted matter can initiate violent nuclear energy releases {due to the high density in the burning region}.

\begin{figure} 
\centering
\includegraphics[width=1.0\columnwidth,clip]{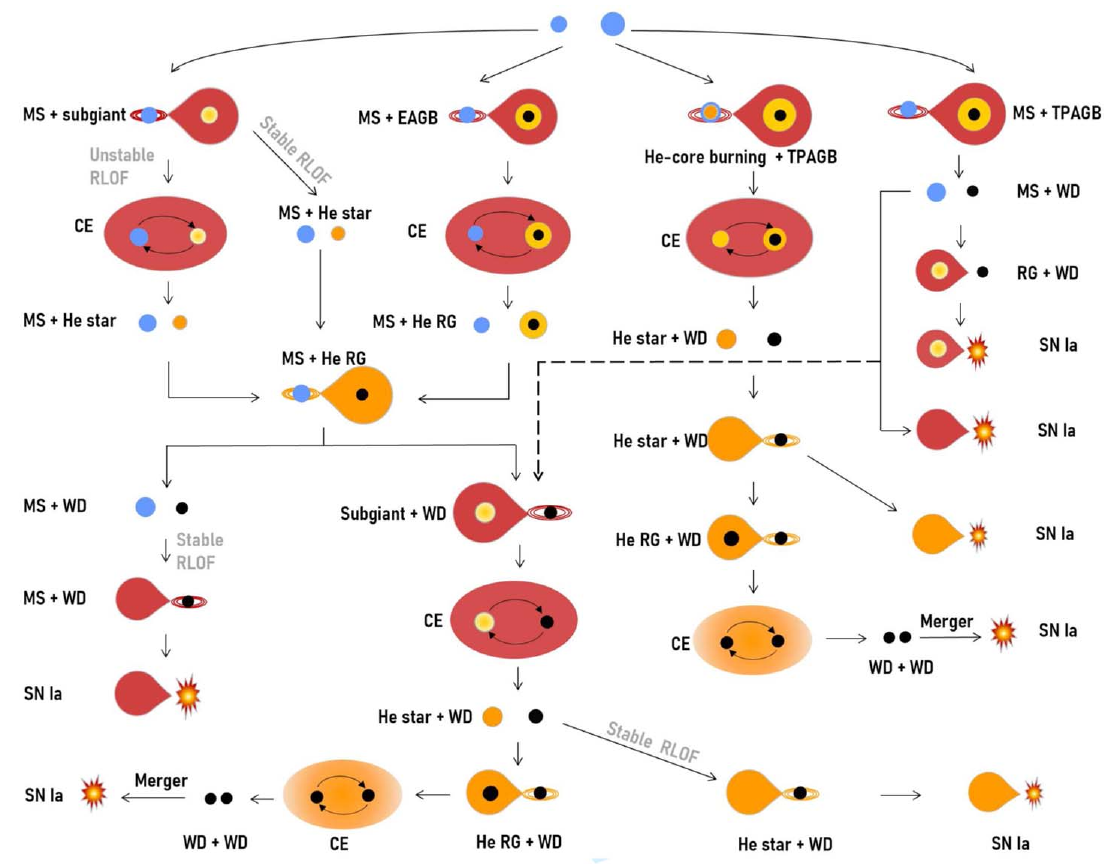}
\caption{The variety of evolutionary paths that may lead to a thermonuclear supernova (SN Ia) from binary-system interactions is large, illustrated in this schematic chart. {Starting from a binary system of stars with different mass (top), evolution that produces a white dwarf and a main-sequence or giant donor of mass with slow and stanble mass transfers will enrich the mass of the white dwarf up to its Chandrasekhar mass limit of stability, where it will undergo thermonuclear explosion. This \emph{Chandrasekhar} scenario had been believed to be the main channel for supernovae of type Ia. Alternatively, several channels lead to helium stars as companions, opening the path towards sub-Chandrasekhar mass explosions. } (From \citet{Liu:2023}).}
\label{fig_SNI-scenarios}       
\end{figure} 

Supernovae from {\emph{white dwarfs} in binary systems, which result from evolution of more-massive stars that went through helium burning and thus produce a carbon-oxygen degenerate remnant} occur as an ignition of central carbon fusion. 
{The steep reaction-rate rise with temperature for this reaction leads to a runaway, as the nuclear energy release at densities near 10$^{13}$g~cm$^{-3}$ does not immediately expand and cool the burning region, due to the degeneracy of white-dwarf matter. This} results in disruption the entire white dwarf. 
These are called \emph{supernovae type Ia}, or \emph{thermonuclear supernovae} \citep[see][for recent reviews]{Liu:2023,Ruiter:2025}.

The nuclear flame  reaches nuclear statistical equilibrium as the initial carbon fusion heating activates many other nuclear reactions. While initially it may have begun as a deflagration front, it may turn into a supersonic detonation in outer, less dense, regions of the star. \rd{The scenarios that parametrize this idea} are called \emph{deflagration-detonation transition (DDT)} explosions \citep{Hoeflich:1996}. 

{Any explosive burning-expansion dynamics} leaves its imprints, as the nuclear reaction flow freezes out: 
major \rd{expected} nucleosynthesis products are $^{56}$Ni from central regions, and incompletely-burned products or unburnt carbon and oxygen from further-out regions \citep[see][for a review of SN Ia nucleosynthesis]{Seitenzahl:2017}.
Intermediate-mass nuclei \rd{are expected to} show a characteristic abundance distribution with peaks for $\alpha$~multiples.
The ratio of intermediate-mass isotopes to those of the iron group provides a diagnostic of the completeness of nuclear burning across the white dwarf  \citep{Mazzali:2006a}. 

{In general, from nuclear burning on surfaces of compact stars, we expect that }the progress of reaction paths up the nuclear chart is incomplete, and the contribution of heavier versus less-heavy ashes shows up in the radioactive heating of the expanding envelopes {of compact stars}.

{The evolution of a white dwarf in a binary system towards a thermonuclear supernova, however, may be possible in a diversity of pathways, shown in Figure~\ref{fig_SNI-scenarios}.
Several of the possible paths result in a steady growth of the white dwarf mass through mass transfer from the primary, which ends catastrophically as the Chandrasekhar mass limit for stability of a white dwarf is reached. 
But as shown in Figure \ref{fig_SNI-scenarios}, many alternative pathways lead to either helium-star companions or white-dwarf binaries. These constitute a \emph{sub-Chandrasekhar} or \emph{double-degenerate} path towards thermonuclear supernovae. 
It is currently unclear which of these scenarios provides the majority of supernovae of type Ia \citep{Liu:2023}. Even the exploitation of binarity studies that reveal a variety of astronomical signatures could not yet resolve this \emph{SN~Ia progenitor puzzle}. }

Accumulation of hydrogen from a companion {may occur in different ways: Accreting at a small-enough rate of 10$^{-8}$ to 10$^{-10}$~\Msol,  hydrogen ignition already during accretion can be avoided (see below for the resulting nova phenomenon). At somewhat higher accretion rates, hydrogen burning during accretion would occur slow enough to \rd{end up as steady-state burning during accretion, and } lead to the phenomenon of the \emph{supersoft X-ray sources} \citep{Kahabka:1997,Podsiadlowski:2010} as hydrogen is converted to helium. }Now helium would end up to be deposited on the surface of the white dwarf. 
{The ignition of helium burning (the 3-$\alpha$~process, see section 2)} is expected above a critical mass of order 10$^{-4}$~\Msol. 
{It would occur in a flash, sending a shock into the underlying white dwarf star. This shock may be sufficient to trigger the nuclear carbon fusion reaction} in the central regions of the white dwarf, thus again leading to the thermonuclear explosion of the entire white dwarf. 
This is one of the \emph{sub-Chandrasekhar} scenarios discussed to trigger supernovae of type Ia, complementing the above-discussed Chandrasekhar limiting mass scenario.
Other such triggering precursors of thermonuclear supernovae could be accretion of helium from a companion star that already has lost its hydrogen envelope, or the {more or less-violent} collision of two white dwarfs. In each such scenario, the nuclear ignition and flash of helium burning provides the initial trigger for the explosion of the white dwarf, launching its violent carbon fusion reactions in central regions.
It is a current topic of research to identify the relevance and detailed evolution of each of these{ evolutionary pathways} \citep[see, e.g.][for a detailed recent discussion]{Liu:2023}.

If hydrogen{ can be} accumulated from accretion {on the white-dwarf surface}, {after accumulation of a critical mass of hydrogen, ignition will occur on the white dwarf surface from gravitational compression and heating; due to lower density, non-degeneracy, and slower reaction rates or hydrogen burning (see section 2),} the result is less violent: 
A \emph{nova explosion} results \citep{Jose:2020,Jose:2016a}. 
Ignition at the surface of the white dwarf occurs at densities of a few 10$^3$g~cm$^{-3}$. 
As the {underlying} degenerate matter does not respond to the nuclear energy input by expansion that would cool the burning region, temperatures of a few~10$^8$K can be reached.
{The thermonuclear runaway of this surface burning is primarily driven by hydrogen burning along the CNO cycle 
 (see section 2.2 on Carbon Reaction Examples above) and proceeds along the proton-rich side of the valley of stability. 
 Bypassing $\beta$~decay of $^{13}$N by proton capture, this is called the \emph{hot CNO cycle} \citep{Lazareff:1979,Wiescher:2010}.
This cycle limits the energy production. {If the origin of the white dwarf is from a higher-mass star that included oxygen burning, a \emph{neon nova} then would be} driven by a continuous reaction flow from neon upwards along and near nuclear stability  up to the closed-shell nucleus of $^{40}$Ca.}
A wide variety of p~capture reactions processes matter, as envelope expansion lifts the degeneracy, and convection further cools the burning region and terminates the nuclear reactions \citep{Jose:2024}.  
Plausibly, convection and sheer flows \rd{generate a complex three-dimensional gas dynamics \citep{Casanova:2011} which} mixes some material from the underlying white dwarf with its heavier elements, predominantly C and O, and up to Ne for the more-massive white dwarfs, into the burning region, thus producing heavier isotopes in the nucleosynthesis chain.
It is unclear how much of this processed material will finally be ejected in the nova, before the expanded envelope settles down. Values of 10$^{-7}$ to 10$^{-4}$~\Msol have been argued \citep{Jose:2007,Jose:2020}.

For \emph{neutron stars},  a binary companion star and its mass transfer also may trigger { explosive} nuclear burning.
But due to the large gravitational binding, it is difficult to release sufficient energy to liberate neutron star matter or its ashes against these \about~200~MeV per proton of gravitational energy. 
\rd{Although some modest material release (of order 0.1-1\%~of the entire neutron-star envelope, i.e. 10$^{-14}$\Msol) has been predicted  \citep{Herrera:2023,Woosley:2004} through, e.g., radiation-driven winds,} such events are probably irrelevant for enriching interstellar gas with nucleosynthesis products.
{But the resulting} \emph{type-I X-ray bursts} {observationally} signify explosions on or near the the surface of a neutron star \citep[see][for a review]{Galloway:2021}. 
High densities of the gravitationally-compressed matter enable proton and $\alpha$~capture reactions to proceed beyond the hot CNO cycle reactions, and further hydrogen-burning reactions proceed along the nuclear chart up to Sn in the \emph{rapid proton capture (rp) process} \citep{Schatz:1998,Woosley:2004,Fisker:2008,Jose:2010}.
Ignition of remaining \rd{ashes from this rp~process, and in particular abundant} carbon ashes somewhat deeper in the crust result in {\emph{superbursts}, {more vigorous} as the nuclear-energy liberation rate is higher}. 
The light curve of such bursts is characterized by an initial rise that reflects ignition of nuclear burning, {expansion and thus cooling of the} burning site {stops the rise, giving way to} a gradual decline that is characterized by the radioactive afterglow of the mix of nuclei that resulted from nuclear burning \citep{int-Zand:2017}. 
Complexities arise, e.g., from neutron star spin \citep{Galloway:2018}, and from cooling properties including neutrino emission \citep{Meisel:2022}, and interfere with a straightforward interpretation of the lightcurve tail in terms of radioactive content of the burning ashes \citep{Goodwin:2019}. 
{Nevertheless, measuring lightcurves and their} variations at high time resolution provides { information on the explosive burning, and details from oscillations} can be related to material dynamics during the bursts \citep{Watts:2012}.

We may also speak about a `stellar explosion' of a neutron star in cases of collisions of two neutron stars that make up a binary system: Sufficient energy is provided by the collision, so that significant amounts of matter (although very likely far less than a percent, \rd{i.e. 10$^{-2}$~\Msol or much less}) may be ejected during such a binary neutron star collision. This can be viewed as an extreme case of `binary mass transfer'.
\rd{Similarly, compact white dwarfs may collide among each other or with main sequence stars. For main-sequence stars this might lead to some nuclear burning and to  mass ejections, although mainly from the main-sequence star that does not survive such a collision. This has been studied in the context of stellar dynamics in dense star clusters \citep{Van-der-Merwe:2024}, but implications for the interstellar isotopic abundance composition have not been pursued so far.
White-dwarf-white-dwarf collisions, on the other hand, may be more or less violent mergers, triggering an explosion; these are also discussed among the pathways to thermonuclear supernovae \citep{Liu:2023,Ruiter:2025,Pakmor:2010,Pakmor:2012} (see above).}

On the other hand, \emph{neutron star mergers} had long been proposed to be important sources for heavy nuclei beyond the iron group through r process nucleosynthesis \citep{Argast:2004,Hotokezaka:2013}.
From the spectacular kilonova GRB20170817/AT2017gfo observed in 2017 \citep{Abbott:2017} {(see section 4 on Nuclear Astronomy)}, an ejecta mass in a range up to 10$^{-2}$~\Msol has been inferred from the observations of the kilonova afterglow of the binary neutron star collision \citep{Abbott:2017a}.

In the collision, one of the colliding neutron stars is expected to be tidally decomposed by gravitational interaction, resulting in \emph{dynamical} ejecta. After merging, explosive nucleosynthesis may result in production of heavy elements and release of nuclear energy through radioactive isotopes, providing an energy source for the kilonova phenomenon. The later formation of a black hole is expected to occur through a transient \emph{accretion disk} with a lifetime of typically 100~ms. Within and from this accretion disk, {a substantial amount of matter} may be released {in the form of} \emph{disk wind} ejecta.  
{These simulations and their} studies have {resulted in} interesting {and very detailed} insights into nucleosynthesis under such violent and dynamic circumstances. 
{Collision geometries and the viewing aspect of these are critical for observed phenomena, and different types of ejecta are discussed; this results in a large parameter space of possibilities with corresponding uncertainties on the observed or inferred nucleosynthesis outcomes of this class of neutron star merger (kilonova) events} \citep{Arcones:2023}.

\subsection{Extremes of matter and energy}\label{sec3.3}
The gradual release of nuclear binding energy during hydrogen and helium burning in stellar interiors is key to stabilize stars against gravity. 
Temperatures of stellar plasma in these cases are relatively low, an order of magnitude between the GK regime where nuclear reactions occur in explosive conditions, and densities fall into a range of between one and below 1000~g~cm$^{-3}$. This is regarded as a normal state of interacting plasma in astrophysics. 
But cosmic environments where nuclear reactions occur can be more extreme.
The criterion we use to tag a site as 'extreme' often is a major deviation from what we consider as 'normal'. 
In normal situations, as above, we are used to describe matter as composed of free particles, exchanging energy through collisions among themselves or with the photon field. Temperature is a characteristic environmental characteristic, and any value of momentum and energy can be adopted, as required by temperature.
Particles interact exchanging energies, and if they otherwise store or liberate energy, then we can fully address this tracking such internal energies globally, e.g. through thermal energy and conservation of energy.

We already encountered a non-normal state of matter above: degenerate matter in stellar cores or white dwarfs. Here, not all degrees of freedom for thermal energy exchanges are available, due to compression of matter in a smaller volume of phase space and the Pauli exclusion principle for fermions. The occupation of states follows availability. Temperature ceases to be a measure of particle energies. 
In cosmic extremes, the particle energies and their connection to other system variables of energy lose their familiar connections.

Extreme cosmic sites are characterized often by a much larger number of degrees of freedom being available to share energy in different quantum states, phase transitions including systematic changes in availability of particle states. Moreover, the coupling of particles to the photon field may change, leading to differences in temperatures of photons and particles.

A prominent example is the \emph{Big Bang}: Here, during expansive cooling, there is a time window where particle collisions which are sufficiently energetic for nuclear reactions are frequent, so that nuclear transformations lead to compositional changes. This period of Big Bang nucleosynthesis fades as matter expands, reducing collision frequencies and relative particle energies.
{ Nucleosynthesis calculations for this case rely on a network of reactions among light nuclei, which mostly have been determined experimentally in underground laboratories \citep{Coc:2012} Therefore, the agreement of \emph{big bang nucleosynthesis} with observations of primordial gas abundances of $^4$He, $^3$He, D, and $^7$Li at a baryon-to-photon ratio as inferred from measurement of the cosmic microwave background radiation by the Planck satellite is one of the pilars of the Big Bang model and the commonly-agreed standard model of cosmology \citep[see][for a review of Big Bang nucleosynthesis]{Coc:2012}. The tension reported earlier for $^7$Li, the \emph{cosmological Li problem}, has recently been {interpreted} as due to systematics in observations of $^7$Li in stars: atmospheric processes lead to mixing of the outer layers of these stars, and selection of different subsets of stars revealed systematic effects that can explain observational abundance biases \citep{Korn:2024,Miranda:2025}.}
In the Big Bang, by contrast to explosions of compact stars, the total energy density is dominated by the initial thermal energy, i.e., by the photon content of the early universe \citep[see][for recent reviews]{Cyburt:2016,Coc:2017,Pitrou:2018,Grohs:2022}. 
The initial material composition is just composed of protons and neutrons after freeze-out of neutrino reactions, and the lack of stable nuclei of masses 5 and 8 provides a bottleneck for nuclear reactions producing heavier elements, and beyond the synthesis of helium not much has time to happen before temperature and density freeze out nuclear reactions \citep{Pitrou:2018}. 
There are, however, discussions on variants of the early evolution of the universe. One of the ideas relevant to nucleosynthesis is the suggestion of an inhomogeneous big bang that also could produce heavy elements with an r-process pattern, thus solving the issues related to determination of the source of the surprisingly homogeneous r-process ingestion from the early to the current universe \rd{ \citep[see][for two alternative scenarios]{Roepke:2025,Rauscher:1994}}.

Another example is the case of \emph{supernova explosions}{, as described above.}

A third example of nuclear-reaction environmental extremes is \emph{neutron star matter} \citep{Lattimer:2021}. 
\rd{Within the dense, neutron-dominated hadronic matter of a neutron star, nuclear processes are expected. These include \emph{kilonovae} \citep{Metzger:2020} and the phenomenological variants of ignitions near the surface, \emph{type-I X-ray bursts} and \emph{superbursts} \citep{Galloway:2008}, as discussed in section 3.1 above.} 
 

In a dynamical situation such as gravitational collapse, the properties of the inner compressed matter { in the newly-forming neutron star} has an important role for the progress of collapse, and its eventual reversal into an explosion.
\emph{Compressibility} of { \emph{matter beyond nuclear densities} in neutron stars} is characterized by the equation of state. This describes the pressure increase with increasing compression of matter \citep{Lattimer:2021}. 
Herein, the repulsive characteristics of the strong nuclear force are decisive at the extreme end of high densities that are reached in gravitational collapse, and nuclear phase space aspects in less extreme parts of the parameter space.
Measurements of compressibility have indirectly been inferred from the merging of neutron stars in kilonovae \citep[see][for a review]{Metzger:2017}. Although few events only have been measured, the gravitational radiation signature allows an analysis of compressibility \citep[see][for its discussion]{Lattimer:2021}.
On the other hand, measurements on heavy nuclei in the laboratory provide an independent check and assessment of nuclear-matter properties, such as through measurements of heavy-ion collision aspects and giant resonances \citep{Garg:2018}.

\rd {The \emph{pair instability} for very massive stars (as discussed above in sections 3.1. on Stellar Evolution and  3.2 on Stellar Explosions) also is an example of an extreme of matter.}  
Pair-instability supernovae are most-likely characterized by violent nuclear burning in their central regions.
While the initial study \citep{Barkat:1967} suggests ejection of mainly oxygen-burning products, later studies suggest high amounts of $^{56}$Ni as a characteristic result of such nucleosynthesis \citep{Kawashimo:2024}.   



\begin{figure}[ht] 
\centering
\includegraphics[width=\columnwidth,clip]{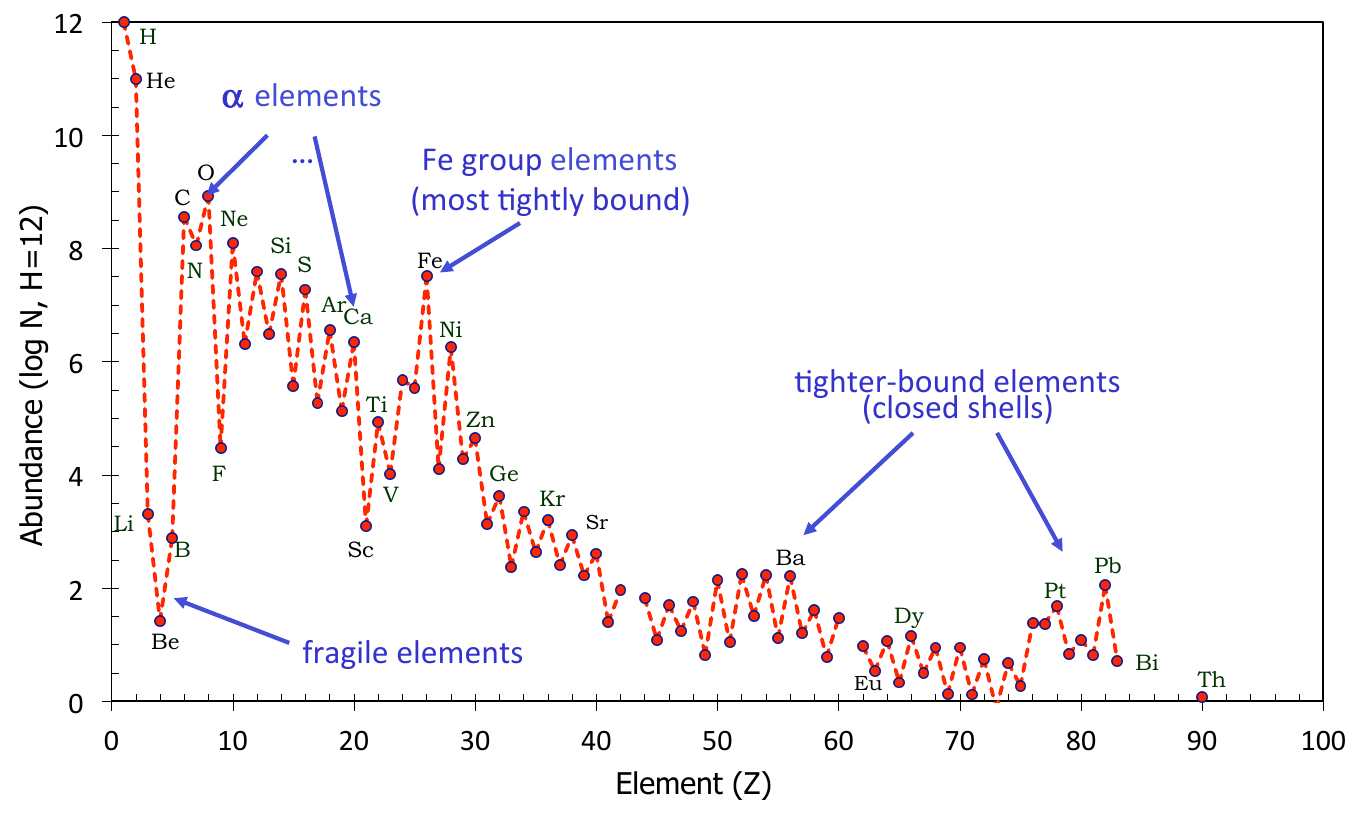}
\caption{The elemental abundances {(references see text)} are characterized by a dynamic range extending over twelve orders of magnitude, with a dominance of Hydrogen and helium, a prominent broad peak around iron-group nuclei with zig-zag abundance patterns bridging the region from light elements to iron, and two characteristic double-peaks across the heavy-element region beyond iron. This plausibly is the result of nuclear-stability physics, as convolved with nuclear-reaction environments for the different cosmic sites where they can occur.}
\label{fig_abundances}       
\end{figure} 

\section{Nuclear astronomy}\label{sec4} 
{Figure~\ref{fig_abundances} shows the observed cosmic elemental abundances. These} have been derived from photospheric and meteoritic data {(discussed below in more detail) }.
{ The elemental abundances} show signatures that are reminiscent of nuclear fusion reaction characteristics.
The light and intermediate-mass elements show \rd{local enhancements} in abundance that appear to correlate with nuclear binding: Local maxima are seen for $\alpha$~multiples such as carbon, oxygen, magnesium, silicon, and then generally of the iron group elements; the broad maximum around the Fe group suggests that cosmic nuclear reactions often reach \rd{high-temperature conditions that allow to obtain maximum nucleon binding}. 
A significant abundance drop beyond the iron peak {indicates} that different nuclear reaction processes dominate below and above the iron peak. This abundance drop suggests that the {production of} heavier elements {occurs in less-frequent events and/or} in rarer environments. 
Among these heavy element abundance patterns, the secondary maxima of abundances at adjacent element groups again are reminiscent of nuclear binding properties, here the \emph{magic numbers} that favor stability of nuclei with specific nucleon numbers (8, 20, 28, ...). 
The {local abundance maxima} at these magic numbers, hence, can be associated with matter processed in long-time stable stellar interiors by the s~process, whereas the maxima next to these and towards lower mass may be thought of as matter that had been driven \rd{far off stability} to these more-stable magic number isotopes and $\beta$-decayed thereafter, hence r~processed matter { with abundance maxima at smaller mass numbers}. 
Thus, the elemental abundance distribution  suggests {a dominance of } near-equilibrium nucleosynthesis in hot and dense environments (such as stars and their explosions), and less-frequent and/or more extreme nucleosynthesis sites contributing {to create the} rarer nuclei.
\rd{This was recognized by the 1950ies and led to the pioneering publications about nuclear reactions in cosmic site \citep{Urey:1952a,Suess:1956,Burbidge:1957,Cameron:1957a} (see section 1 (Introduction) above), interpreting the abundance distribution shown in Figure~\ref{fig_abundances}.  Nuclear astronomy attempts to measure which sites are responsible for which products, using above-discussed nuclear physics and astrophysics knowledge.}

{The elemental abundances (Figure~\ref{fig_abundances}) represent} the original and first astronomical result about nuclear processes in the universe, {the motivations from nuclear physics and astrophysics have been discussed in chapters 2 and 3 }above. 
{This abundance inventory} is the cumulative result of nuclear reactions during the evolution of the universe{, which enrich the gas between stars, and thus the gas that will form new stars subsequently; observing the gas composition in stellar photospheres, we thus obtain a record of the composition of \emph{cosmic gas} at the times those stars were formed.}
The elemental abundance pattern is an indirect measurement. 
{It is the result of} a complex convolution with contributions from sources in different regions of the solar neighborhood as mediated by interstellar transport processes {(see section 4.4. on Cumulative Nucleosynthesis)}. 
{Nucleosynthesis in very nearby regions shaped the composition of} the interstellar gas as it ended up in the pre-solar nebula 4.567~Gy ago. 
Nevertheless, this abundance measurement is informative on nuclear processes in cosmic sources. 

Abundances are best determined for the Sun. 
The result shown in Figure~\ref{fig_abundances}  for our solar system is determined after combining spectroscopy of solar radiation \citep{Asplund:2005, Asplund:2021} with isotopic analyses of meteoritic samples \citep{Lodders:2009,Lodders:2020}. 
{(For a discussion of the methods and accuracy of stellar photospheric abundance determinations see reviews in \citep{Allende-Prieto:2016,Jofre:2019}, and an assessment of solar abundances from photosphere, meteoritics, solar wind, and helioseismology \citep{Bergemann:2026}.}
It is unlikely to obtain such a precise measurement for other regions even in the nearby universe. But a multitude of messengers can be exploited to study nuclear physics in cosmic environments.
{Nevertheless, this pattern} of elemental abundances appears to be largely the same throughout the current universe and its star-forming galaxies \citep[see][for $\alpha$ elements in star-forming regions]{Esteban:2025} \citep[see][for coupling of Galactic abundances]{Mead:2025} \citep[see][for a review of universality of the r process pattern]{Sneden:2008}. 
Even galaxies of the early universe with low metallicities show surprisingly similar nucleosynthesis signatures of the processes discussed above , although a special role for first-generation stars appears prominent \citep{Rossi:2024}. 
Note that different star formation histories and galaxy collisions, however, make such analyses uncertain, and an evolution of different enrichment histories for the various nucleosynthesis sources is likely, and part of chemical-evolution modeling (see Cumulative Nucleosynthesis below).

Measuring direct, more specific and detailed signals from nuclear processes and reactions involved in cosmic nucleosynthesis is a challenge. 
Main obstacles are that (i) these reactions occur in environments that are surrounded by major layers of matter {that absorb or distort} most such signals, and (ii) the duration of signal emissions often is short, i.e. order of seconds.

\emph{Astronomy} is the method to obtain information from phenomena and messengers of physical processes that occur in cosmic objects throughout the universe. 
A \emph{nuclear astronomy} must develop a focus on messengers of nuclear processes and nuclear reactions.
As in the above case of nuclear reaction experiments, we again distinguish between \emph{direct} and \emph{indirect} astronomical tools.
For an extended discussion of the variety of nucleosynthesis messengers see \citet{Diehl:2022}.

\begin{figure}[ht] 
\centering
\includegraphics[width=0.9\columnwidth,clip]{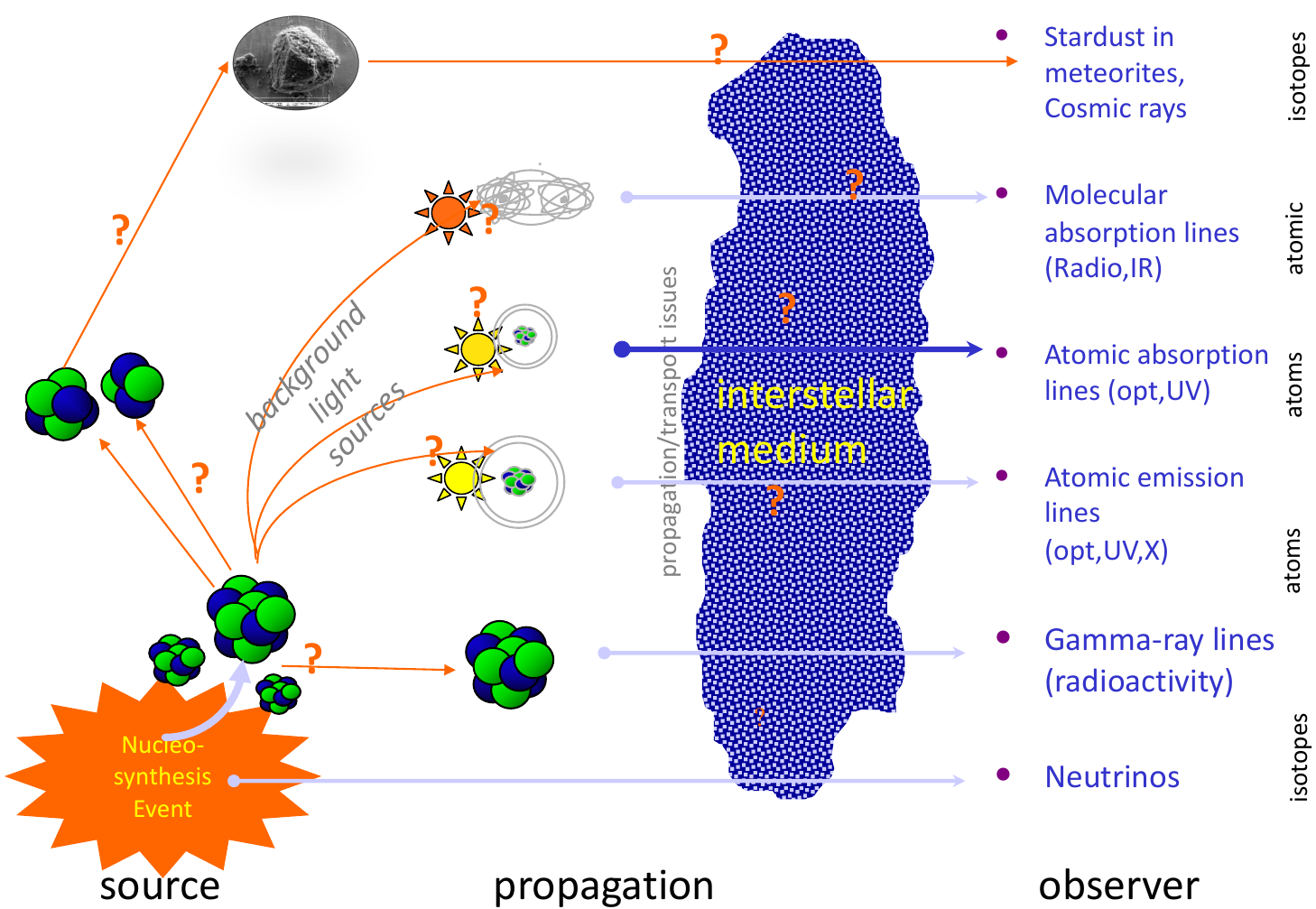}
\caption{Different methods obtain information about the abundances of specific elements and isotopes of ejecta from a nucleosynthesis event. Although we may call these all \emph{nuclear} astronomical methods, most of these are only indirectly related to the nucleosynthesis event of interest. Gamma-ray lines from decay of radioactive by-products of nucleosynthesis may be considered most directly relate to the nucleosynthesis ejecta. Even more directly, neutrinos can escape directly from the dense environments of nucleosynthesis and directly relate to the nuclear reactions that form nucleosynthesis within the event. See text for discussion of the different nuclear messengers. }
\label{fig_nucleosynAstronomies}       
\end{figure} 

\subsection{Nucleosynthesis events and their ejecta}\label{sec4.1} 
Most directly, observations target specific objects and events where nucleosynthesis occurs, and its nuclear-reaction phenomena and ejecta can be measured. 
Figure~\ref{fig_nucleosynAstronomies} illustrates different methods that have been established to observe signals that inform us about cosmic nucleosynthesis events.

\begin{figure} 
\centering
\includegraphics[width=0.7\columnwidth,clip]{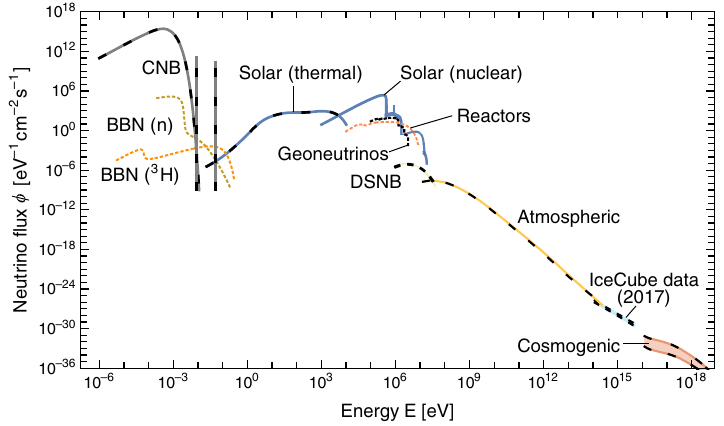}\\
\includegraphics[width=0.7\columnwidth,clip]{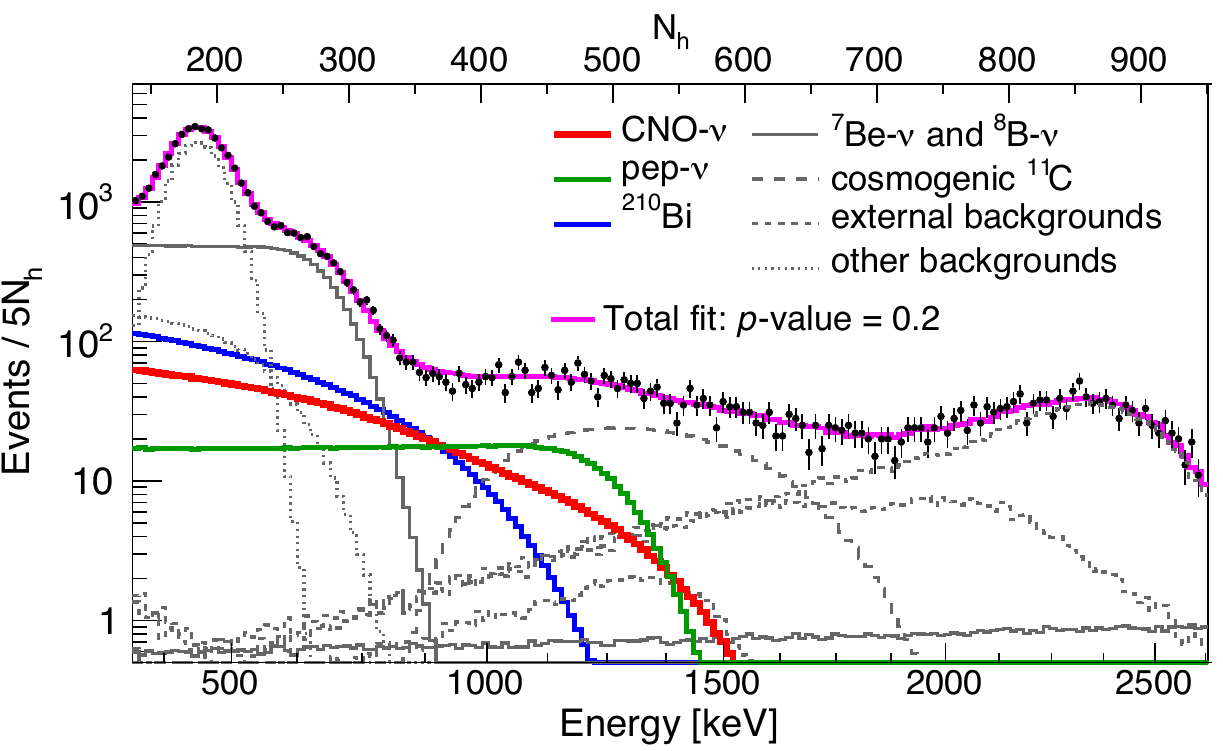}
\caption{{\it Top:} The combined spectrum of neutrino luminosities from the variety of sources shows all different components, with the prominent nuclear components from the Sun, with its dominating proton-chain continua and the $^7$Be and $^8$B lines, with continua from the CNO reactions adding spectral modulation \citep[from][]{Vitagliano:2020}.
{\it Bottom:} The CNO component of solar neutrinos as measured with Borexino. 
The red line shows the CNO neutrino component as fitted in the analysis, the purple line marks the total fit including all background types. The best signal to noise is obtained in the yellow-marked region (see text).
(Updated by \citep{Appel:2022} from \citep{Borexino-Collaboration:2020}).} 
\label{fig_solarNeutrinos}       
\end{figure} 

\emph{Neutrinos} are a direct product of nuclear reactions, with characteristic neutrino {spectra} reflecting specific reactions. Due to their penetrability, they escape from the dense nuclear reaction site even through massive envelopes, and thus can reach a remote observer.
{Cosmic sources are expected to create neutrinos through a variety of processes, each with a characteristic energy spectrum. Shown in Figure~\ref{fig_solarNeutrinos} (left) are the cosmological neutrinos as they decoupled from the plasma of the early universe (lowest energies), then the solar neutrinos from nuclear reactions (MeV energies) \citep{Vitagliano:2020,Ianni:2026}. 
Towards higher energies, supernova neutrinos from the cooling proto-neutron star with a continuum spectrum centered at 10--30~MeV is expected from theory, consistent with the few events recorded from SN1987A \citep{Loredo:2002}. 
Hadronic interactions of cosmic rays in the Earth atmosphere form powerlaw-like spectra towards higher energies that have been measured in detail by, e.g., IceCube and Kamiokande experiments \citep{Abbasi:2023,Takhistov:2021}. 
Probably the jets created in black hole formation or accretion such as in gamma-ray bursts and active galactic nuclei, respectively, will create similar neutrino spectra at even higher energies, corresponding to the most-energetic cosmic rays \citep{Murase:2023}.}

{For the Sun, characteristic energies of neutrinos from nuclear reactions are around the MeV scale, emitted by reactions along the hydrogen burning sequences \citep{Raffelt:1996,Adelberger:2011}. These are dominated by neutrinos from the pp~chain reactions, which are mostly of energies below 0.5~MeV and therefore hard to detect, as neutrino cross sections steeply rise with energies. The pep~neutrinos from the initial deuterium formation in the pp chain however are mono-energetic at 1.44~MeV, and thus a target of measurements. 
Solar neutrinos include CNO cycle neutrinos with energies up to 1.7~MeV, from $\beta$~decays of $^{13}$N, $^{15}$O, and $^{17}$F, and also $^7$B neutrinos with two lines at energies 0.86 and 0.36~MeV, that are most intense and thus easier to detect \citep{Fowler:1958}.} 
This is a challenging energy range, because many processes of nuclear physics in the detector and its surroundings lead to similar signals, and it is difficult to recognize such background  against the few desired signatures.
Moreover, spectroscopic resolution at the 0.1~MeV level or better is required to attribute the neutrinos to specific nuclear reactions.
{The discovery of neutrinos from the Sun and assessment of its measurement in terms of modeling the Sun and understanding the nature of neutrinos has been a major triumph of nuclear astrophysics.
An underground experiment based on conversion of \rd{chlorine} to argon by neutrinos reported a flux limit that lay a factor 7 below the flux of (the most energetic) $^8$B neutrinos predicted by models of the Sun \citep{Davis:1968} \citep[also see the corresponding Nobel prize lecture by][]{Davis:2003}). 
The SAGE and GALLEX experiments similarly used chemical extraction methods to measure solar neutrinos, and confirmed the data recorded by the Homestad experiment. 
But more convincing appeared the direct neutrino event detection in the Kamiokande imaging water Cherenkov experiment, which later verified this discrepancy and established the \emph{solar neutrino problem}, at the same time discussing its potential solution through neutrino flavor oscillations \citep{Koshiba:1986}. 
\rd{The heavy-water based SNO experiment operated in Canada at the Sudbury Neutrino Laboratory then added the capability to measure the neutral-current neutrino reactions through weak-interaction neutron release from deuterium (producing a characteristic neutron capture $\gamma$-ray line at 6.25~MeV), in addition to their measurements of the charged-current $\nu_e$ interactions, thus providing experimental proof of neutrino flavor oscillations \citep{Aguilar-Arevalo:2002}.}
The theoretical predictions and experimental confirmations of neutrino flavor oscillations were assessed as the explanation of the measured deficit: electron neutrinos emitted in the nuclear reactions change flavor along their trajectories, so that a lower electron neutrino flux interacts with detectors on Earth.
Including these neutrino properties, current solar-model predictions now are in general agreement with measured values \citep{Ianni:2026}.}

The pp reaction chains result in a nuclear-reaction neutrino flux from the Sun of 1.6~10$^{10}$~cm$^{-2}$s$^{-1}$, while the detection efficiency of a typical instrument such as Borexino is about 10$^{-18}$. 
\rd{Solar neutrinos are not only produced by reactions associated with the pp-chains, but also by the CNO cycle \citep{Adelberger:2011}. This produces less than roughly 1\% of the solar energy, but is of particular importance for determining the metalicity of the solar core \citep{Cerdeno:2018}.}
Borexino in the past has reported solar neutrino detections of the pep \citep{Borexino-Collaboration:2020}  and CNO neutrinos  (Figure~\ref{fig_solarNeutrinos}, right)  \citep{Appel:2022,Basilico:2023}. 
Future improvement of these results will be expected from JUNO {in China} after 2025 \citep{An:2016}, but also instruments with partially-overlapping capabilities such as Panda-X or DUNE \citep{Abi:2021}, {which aim at determining fundamental neutrino properties, and  supernova-neutrino aimed neutrino experiments such as SNO+ or }Superkamiokande with its Gd loading are expected to contribute to higher-energy solar neutrino results (the $^8$B part above \about~4~MeV). 
As can be seen (Figure~\ref{fig_solarNeutrinos}, right), the challenge is formidable, due to the penetrating nature of neutrinos and the large detector backgrounds, even in the case of the Sun; it is unlikely that diagnostic neutrino spectra can be obtained from other even nearby nuclear reaction sites. 

Through \emph{gamma~rays} from nuclear de-excitations, direct and penetrating characteristic radiation from nuclear reactions can be measured. 
The spectral signature of nuclear gamma ray emission allows direct attribution to a specific isotope.
A prominent example is the discovery of the decay of $^{26}$Al  by the HEAO-C spacecraft experiment \citep{Mahoney:1984}. Identifying this isotope with its 1~My lifetime through the characteristic 1809~keV line from the $^{26}$Mg nuclear de-excitation, it was clear that nucleosynthesis towards $^{26}$Al  production and ejection into interstellar space is ongoing in the current (My time scale) Galaxy.

\begin{figure}[h] 
\centering
\includegraphics[width=0.9\textwidth]{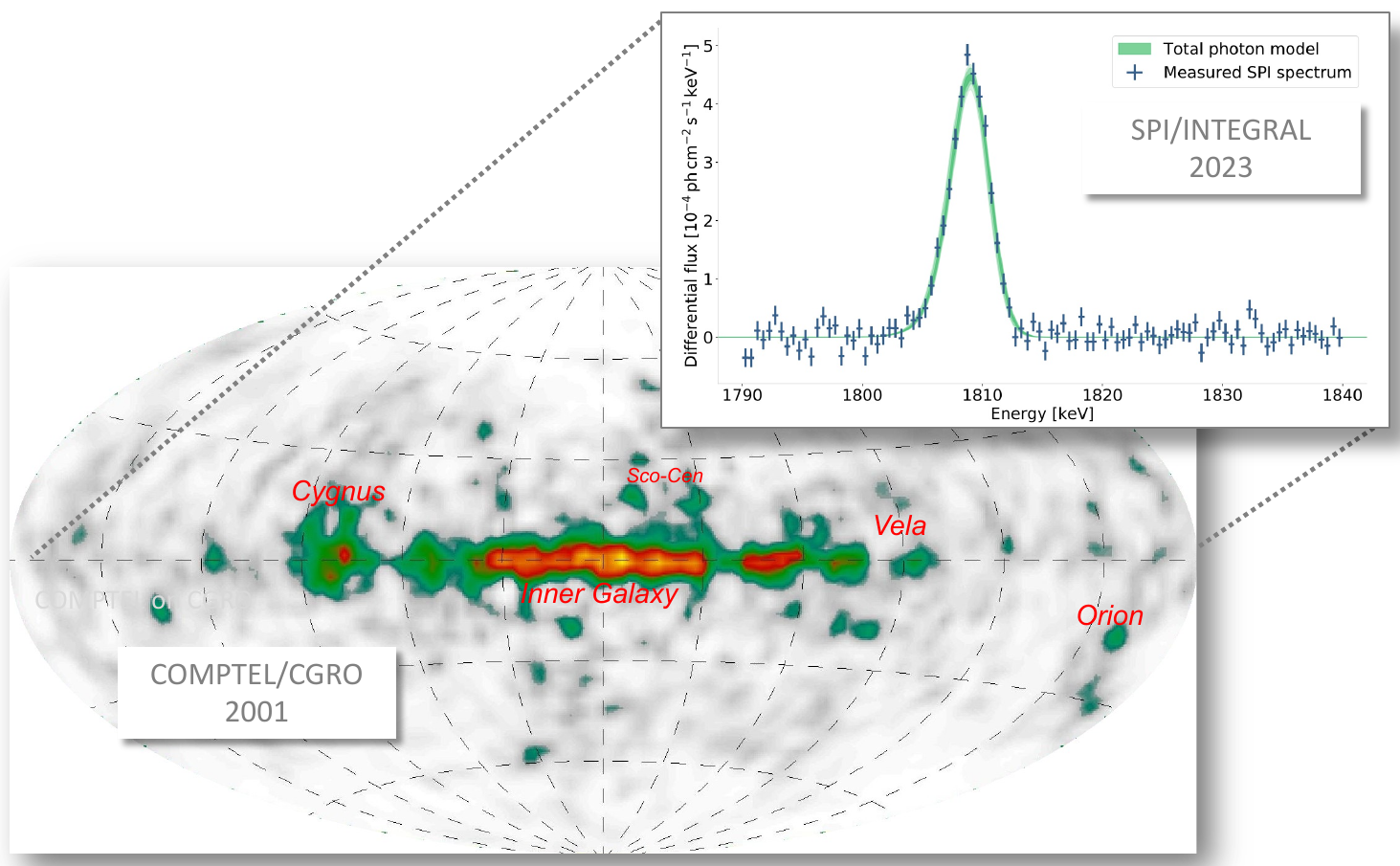}
\caption{Cumulative $\gamma$-ray emission from the decay of radioactive $^{26}$Al ($\tau \sim$1~My) reflects the afterglow of nucleosynthesis in massive stars and their supernovae. This image (COMPTEL, \citet{Pluschke:2001c}) shows nucleosynthesis source contributions along large regions within the Galaxy. Spectroscopy with SPI \citep{Pleintinger:2023} revealed large-scale cavities surrounding sources, and bulk motion of ejecta \citep{Diehl:2021}. }
\label{fig_26Al-image}
\end{figure} 

The Compton Gamma-Ray Observatory mission (1991-2000) and its COMPTEL instrument later allowed imaging of this $^{26}$Al emission at \about~3~degree resolution \citep{Diehl:1995g,Pluschke:2001c}. The emission extends over large parts of our Galaxy, and highlights regions which are known for their massive-star population. Thus, a dominant $^{26}$Al  origin from massive stars and their core-collapse supernovae was concluded \citep{Prantzos:1996a}.
The INTEGRAL space mission of ESA (2002-2025) with its gamma-ray spectrometer telescope SPI supported measurement at 3~keV spectral resolution, which provided interesting kinematic information about the $^{26}$Al nuclei in interstellar space \citep{Kretschmer:2013}. From this, we learned that ejections typically seem to occur in large interstellar cavities, the \emph{superbubbles} blown by stellar winds and supernovae around a cluster of massive stars \citep{Krause:2015}.
Meanwhile, a multitude of interesting studies could be made about Galactic $^{26}$Al and several specific parental massive star clusters, as summarized in \citet{Diehl:2021}.
This summary also discusses diffuse $\gamma$-ray emission from another isotope, $^{60}$Fe. With a lifetime of 3.8~My, it is suitable to provide a complementary view on the same massive-star sources: Its origin is plausibly related to core-collapse supernovae, which eject the nuclear ashes produced deep in the pre-supernova star, specifically its helium and carbon shell regions. Neutron emitting reactions enable there an s-process-like successive neutron capture on $^{54,56}$Fe seed nuclei to produce radioactive $^{60}$Fe.
Specific nucleosynthesis reactions of $^{26}$Al and $^{60}$Fe relate to these observations, as discussed in detail in \citet{Diehl:2021b,Laird:2023}. 
The context of the creation, evolution, and final dispersal of massive star clusters is an active topic of current astrophysics \citep{Krumholz:2019,Krause:2020}.  
Individual stars and supernovae cannot be separately detected within such clusters, as ejections accumulate during a radioactive lifetime of $^{26}$Al, \rd{and result in diffuse $\gamma$-ray afterglow from all these individual sources.}

\begin{figure} 
\centering 
\includegraphics[width=0.45\columnwidth]{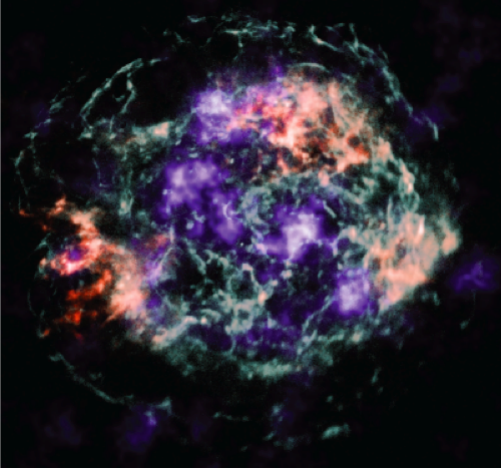}
\includegraphics[width=0.54\columnwidth]{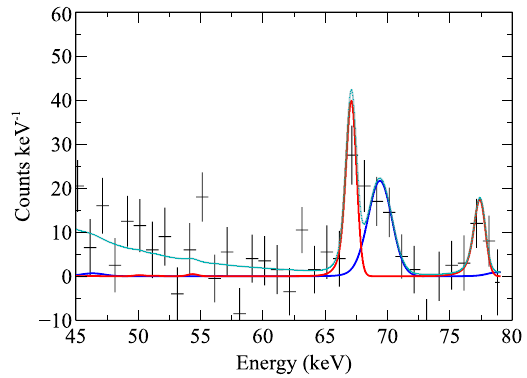}
\caption{{\it Left:} Image of the Cas~A supernova remnant from different messengers:  Characteristic lines from $^{44}$Ti decay (blue) reveal the location of inner ejecta through this radioactive nucleosynthesis product, while X-ray line emissions from Iron (red) and Silicon (green) atoms, that also are emitted from those inner ejecta, show a somewhat different brightness distribution, due to ionization emphasizing the parts of ejecta that have been shocked within the remnant. (From \citet{Grefenstette:2014}).
{\it Right:} The spectrum for a specific region within the remnant shows a narrow red-shifted component of the ejecta (where both the 68 and 78~keV line are within NuSTAR's range) as well as a blue-shifted component (only 68~keV line). (From \citet{Grefenstette:2017}).}
\label{fig_CasAgammas}
\end{figure} 
 
Characteristic gamma rays also were measured from specific and individual nucleosynthesis events, {specifically from} supernova explosions, through more-shortlived isotopes.
$^{44}$Ti, with a 89~year {radioactive} lifetime, provides a long radioactive afterglow of such supernova explosions, with {emission of } characteristic $\gamma$-ray lines at 68, 78, and 1156~keV. 
It was discovered with COMPTEL from the \about~380-year old supernova remnant Cas A \citep{Iyudin:1994}. With more measurements, also with the instruments on the INTEGRAL satellite \citep{Grebenev:2012,Siegert:2015}, and in particular with the hard X-ray imaging telescope of the NuSTAR mission \citep{Grefenstette:2014,Grefenstette:2017} (see Figure~\ref{fig_CasAgammas}), it was clearly established that the core-collapse supernova that created the Cas A remnant was \rd{not} a symmetric explosion, ejecting $^{44}$Ti from its inner region in a few clumps. A predominant redshift was seen both in SPI/INTEGRAL \citep{Weinberger:2020}  and NuSTAR \citep{Grefenstette:2017} data.
{This indicates that the bulk of these ejecta move away from our observational viewpoint.}
The observation of Cas~A and its $^{44}$Ti emission morphology provided an important observational ingredient towards our current picture of core collapses as being complex, a-spherical explosions resulting from the interior neutrino interactions and hydrodynamical processes that turn collapse into explosion {(see discussion in section 3.2 on Stellar Explosions above)}. $^{44}$Ti gamma rays directly reflect inner-region nucleosynthesis products and their kinematics.
Detecting these characteristic $^{44}$Ti gamma rays from supernova SN1987A \citep{Boggs:2015} further supports this role of nuclear observations.

Also for thermonuclear supernovae, a first successful case of observations of nuclear gamma rays on nearby SN2014J turns out to be quite informative.
Even though it is expected that centrally-ignited carbon burning and its main radioactive product of $^{56}$Ni should be occulted by the massive white-dwarf envelope for about three months, characteristic $\gamma$~rays from $^{56}$Ni decay (with a radioactive lifetime $\tau$ of 9~days) at 158 and 812~keV could be seen \citep{Diehl:2014}.
This illustrates that either the envelope and central burning ashes adopt a more-complex morphology, or else some  $^{56}$Ni had been synthesized near the white-dwarf surface, possibly from helium accreted from a companion, in a non-Chandrasekhar scenario { (explained in section 3.2 on Stellar Explosions)}  \citep[see also discussion in][]{Diehl:2014}.
From this same SN2014J event, the characteristic $\gamma$-ray lines from the decay of  $^{56}$Ni-decay daughter $^{56}$Co (with radioactive lifetime $\tau$ of 111~days) were measured by both INTEGRAL gamma-ray telescopes  \citep{Churazov:2014}. 
They appear characteristically Doppler~broadened from the explosion. {But beyond,} SPI $\gamma$-ray spectra in high-resolution detail show indications of a non-spherical explosion, i.e. individual bulk-velocity spikes rather than a broad and smooth Doppler broadening at all times  \citep{Diehl:2015}.
Observations of supernova light  is likely to diffuse and probably not show such information, as bolometric absorption and down-scattering in energy within a massive envelope is required for such radiation conversion.  

\rd{Other radioactive by-products of nucleosynthesis might be detected  in their characteristic $\gamma$-ray lines \citep[e.g.][for science perspectives with a typical next-generation telescope]{de-Angelis:2018}. 
Candidates are $^{22}$Na (radioactive lifetime $\tau\simeq$3.8~y), expected to be produced in novae \citep{Jose:1998} or in massive-star core collapses \citep{Woosley:1995,Pignatari:2025}, or $^{126}$Sn (radioactive lifetime $\tau\simeq$3.3~10$^5$~y), predicted to be produced in kilonovae \citep{Liu:2025}.}
{For novae, none of the candidate lines \rd{from positron annihilation of shortlived $\beta^+$~nuclei (511~keV line), $^7$Be, or $^{22}$Na,} could be detected \citep{Siegert:2018}. 
An interesting \rd{tentative} detection of $^7$Be from INTEGRAL data \citep{Izzo:2025} appears promising for future instruments.}
 
Observations of nuclear gamma rays have been difficult, as cosmic-ray bombardment of spacecraft and instruments in space create local radioactivity and thus lead to a high instrumental background. 
Different techniques have been employed to cope with this and provide imaging as well as adequate spectral resolution.
The COMPTEL Compton telescope \citep{Schoenfelder:1993} aboard NASA's Compton gamma-ray observatory \citep{Gehrels:1993a} featured a characteristic two-detector coincidence arrangement to measure incident $\gamma$-rays through a Compton scatter, followed by absorbing the scattered $\gamma$-ray photon in a second detector.
{Even for this large instrument, a} modest effective area of \about~18--20~cm$^{-2}$ for nuclear lines is the price from this coincidence setup. 
The SPI instrument \citep{Vedrenne:2003} on ESA's INTEGRAL satellite \citep{Winkler:2003} implements a coded-mask camera, { as an alternative method. Its} aperture of \about~250~cm$^{-2}$ results in an effective area that was {somewhat} larger (by a factor of a few). 
But {mainly} its high spectral resolution from using Ge detectors (3 keV, versus 200 keV for COMPTEL's scintillation detectors) provided important spectroscopic advances. 
A next planned gamma-ray mission is COSI \citep{Tomsick:2021} (launch by NASA/US in 2027), a Compton telescope based on Ge detectors. With a surface area of just \about ~250~cm$^{-2}$, its sensitivity will be in the same range, depending on the efficiency of their {more-compact} coincidence detector {setup}. 
{Next-generation} instruments also should be capable to detect specific single sources, such as Wolf-Rayet stars or novae.

\emph{X~ray spectroscopy} from hot plasma in young supernova remnants reveals atomic transitions, that can be attributed to specific chemical elements as only few ionization stages are populated under these circumstances. Figure~\ref{fig_CasAgammas} (left) shows images of the Cas A supernova remnant in such lines of silicon and iron. 
As this image demonstrates, interpretations must account for population of suitable ionization levels: In fully-ionized plasma, those lines will not occur, and hence only the radioactivity emission from $^{44}$Ti correctly represents the inner ejecta, while iron, also plausibly co-located with $^{44}$Ti , does not shine in X~rays in the inner region.
Many supernova remnants have been studied through X-ray lines, making use of high-resolution spectroscopy, of the Chandra and XMM missions, and most-recently with the XRISM Resolve instrument \citep{XRISM-Collaboration:2024}.These have been providing important constraints on nucleosynthesis as well as on morphology and dynamics of ejecta \citep{Vink:2012,Mori:2018,Seitenzahl:2019,Gronow:2021a,Bamba:2025,Vink:2025}.
Also massive stars can be hot enough to produce higher ionization levels of gas in their upper photosphere and chromosphere. This results in characteristic emission lines that reflect abundances {in such outer regions}. 

Stellar photospheric \emph{spectroscopy in the optical window} provides for more-easily accessible abundance measurements, such as the absorption line spectroscopy of sunlight that led to the results shown in Figure~ \ref{fig_abundances}; this has been the classical backbone of cosmic abundance data. 
The underlying star (here our Sun) provides the background light source that illuminates photospheric gas from below. 
In the early years of the 19th century, Joseph Fraunhofer and John Wollaston had noticed and studied striking absorption features in optical spectra (about 400-700~nm).  
With Robert Bunsen and Gustav Kirchhoff's work about thermal radiation and its properties in heated solids, and Wilhelm Hittorf's and Julius Pl\"ucker's application to gas in discharge devices, this led to the spectroscope instrument (1859), a tool for then consolidating the interpretation of absorption lines representing abundances of specific elements in the solar photosphere.
Since then, starlight has been used to infer elemental abundances across the table of elements, using telescopes across all countries and in space.
In contrast to the nuclear gamma-ray lines, spectral line signatures at lower energies arise from electronic transitions in atomic or molecular shells. 
Gas temperature and ionization states determine the line features, and need to be accounted for in conversion into elemental or molecular abundances.
Spectral resolution has improved substantially from technological advances, and recently could resolve the minor offsets in spectral lines that occur for different isotopes of a few elements in exceptionally-favorable cases
Limitations occur from systematic effects, as the underlying radiation and the state of absorbing gas must be modeled for proper interpretation of line features.
The existence or deviations from local thermodynamic equilibrium plays a role, then motions of gas in the photosphere, and the atmospheric pressure needs to be modeled properly. Opacities are complex due to the many atomic and molecular transitions that may be involved. 
The radiation transport through a stellar atmosphere is a complex problem at the microscopic level, and solutions used in stellar spectroscopy depend on a mix of  laboratory spectroscopy, theory including \emph{ab-initio} quantum-mechanical calculations, and empirical prescriptions \citep[e.g.][for a recent discussion]{Barklem:2016}. 
Classical methods such as equivalent-width measurements are often too imprecise and depend on questionable assumptions \citep[see][for a discussion of such limitations and current stellar modeling with corresponding improvements]{Bergemann:2019,Deal:2020,Booth:2020}. 
A wide range of techniques have been developed { for the connections to stellar parameters, formulating complex models to iteratively derive abundances  \citep[e.g.][]{Ness:2015,Ting:2019,Fabbro:2018}, also }including machine learning  \citep[e.g.][]{Buder:2018} as well as hierarchy concepts \citep{Jofre:2019}. 

Despite these caveats, elemental photospheric abundances have been {successfully} used to constrain cosmic nuclear reactions:
(i) for elements that only have radioactive isotopes; the prominent example is Tc, the abundance of which was measured in AGB stars as a first proof of currently-ongoing nucleosynthesis \citep{Merrill:1952};
or, as another example, (ii) for elements which have isotopes that preferentially can be attributed to either the r-process or the s process.
Exploiting magic number peaks of  nucleosynthesis in the s-process \citep[see discussion in review by][]{Lugaro:2023}, $^{88}$Sr, $^{89}$Y, $^{90}$Zr (first s-process abundance peak), with 82, 100, and 51\% of fractional isotope abundance for each element, can usefully be assigned to s process. Similarly,  $^{139}$La and $^{140}$Ce (99.9 and 88\%, respectively)  (2$^{nd}$ peak), and $^{208}$Pb (62\%) (3$^{rd}$ peak) are key isotopes where elemental abundances reflect s-process origin.
Correspondingly,{ abundance peaks for  Se, Xe, and Pt result after} $\beta$~decays modify  production of magic-number nuclei by the r~process and its high neutron flux exposures, although always significant s-process contributions also are to be expected in these cases.  
Beyond the peaks resulting from magic number nuclei, elements may have specific isotopes that are shielded from the $\beta$-decay path (s-only), or separated from the valley of stability by one or more unstable isotopes along the neutron capture path (r-only), with significant relative abundance within the element to allow some conclusions on the respective process.
Prominent cases are $^{96,94}$Zr (17\%) versus $^{90-92}$Zr, or  $^{76}$Ge (7\%) versus $^{70-74}$Ge.
{As lifetimes of unstable nuclei above Pb and Bi are short compared to s-process time scales, all elements above Pb and Bi should have r~process origins exclusively, and can be attributed unambiguously. }

The link of stellar-photospheric data to {their parental} nucleosynthesis event is somewhat indirect: the ejecta from such an event must have found their way into the material from which this particular star has been formed; {moreover}, processes during stellar evolution should have had no impact on photospheric gas composition.

Abundance signals from absorption  lines can be obtained similarly with other background light sources.
{Of particular importance are} background stars beyond interstellar gas clouds, or quasars or gamma-ray bursts as background sources illuminating gas from behind.
At redshifts beyond z=2 the hydrogen Lyman-$\alpha$ line (121.6~nm at rest) becomes accessible to optical spectroscopy. 
Spectra from high-redshift quasars and $\gamma$-ray bursts with absorption lines shortward of the Lyman~$\alpha$ line from the source may come from absorbing gas clouds on the near side of these sources, allowing a study of \emph{gas composition in the early universe} \citep[see][for a review]{Peroux:2020}.
In particular for determination of primordial abundances and their interpretations with respect to big-bang nucleosynthesis, this has been important for measuring helium isotopes and deuterium in gas clouds against high-redshift quasar background light \citep{Pitrou:2018,Pitrou:2021a}.

Emission lines may be generated as stellar gas is {heated by} high luminosity in UV emission \rd{from}  very massive stars. Interstellar gas {may be heated} through shocks from winds or explosions; these convey compositions of gas in such energized gas. 
Shocks from winds and explosions can provide transient {energy inputs to} envelope or circumstellar gas,  thus {providing additional} information about wind or explosion ejecta. This is similar to the situation in supernova remnants. Similar caveats, therefore, apply {about biases from the thermodynamical state of gas}.
 
In all these cases, external energy sources {are the }cause of abundance signatures. We must be aware that the particular {circumstances} of the observed gas { and its relation to the external energy source} provide a bias, and {it is not easy to obtain} typical cosmic-gas abundances,  or even less to obtain abundances {attributed to} specific nucleosynthesis events.

Molecular lines provide an opportunity to measure abundances in \emph{interstellar gas}, where temperatures are low enough to not thermally destroy molecules.
Spectroscopy at sub-mm wavelengths , e.g. with the ALMA observatory \citep{Belloche:2015}, had led to a report 
of rotational lines of $^{26}$AlF from the nova-like source CK Vul \citep{Kaminski:2018}.
This radioactivity detection through molecules in the gas surrounding a specific source presents a new quality of a nucleosynthesis product measurement.
The production of such molecules in or near the source is a precondition, and adds another element to an astrophysical model of the source. 
Generally, nucleosynthesis ashes will not often develop molecular appearances, and the relevance of such bias needs careful assessment.
{In the case of $^{26}$AlF, }general conclusions on $^{26}$Al sources can not be derived in light of such uncertainty, but {the proof of} presence of nucleosynthesis products near such specific sources presents another constraint for modeling production and transport of, in this case, $^{26}$Al in its cosmic variety.
A similar detection of fluorine through the HF molecule in a galaxy at a redshift of 4.42 \citep{Franco:2021} points to early nucleosynthesis of fluorine due to very massive stars, while in the current universe, AGB stars are established as dominating the fluorine production.

\subsection{Sampling cosmic matter}\label{sec4.2} 
{The capture and analysis of cosmic materials in laboratories can measure isotopic abundances in great detail, as a fundamentally-different method of astronomy (that also contributed to abundances shown in Figure~\ref{fig_abundances}). }
The link to nucleosynthesis events is indirect, still: We mostly do not know the trajectories that such samples have taken before they reach our detector.

\begin{figure} 
\centering 
\includegraphics[width=0.95\columnwidth]{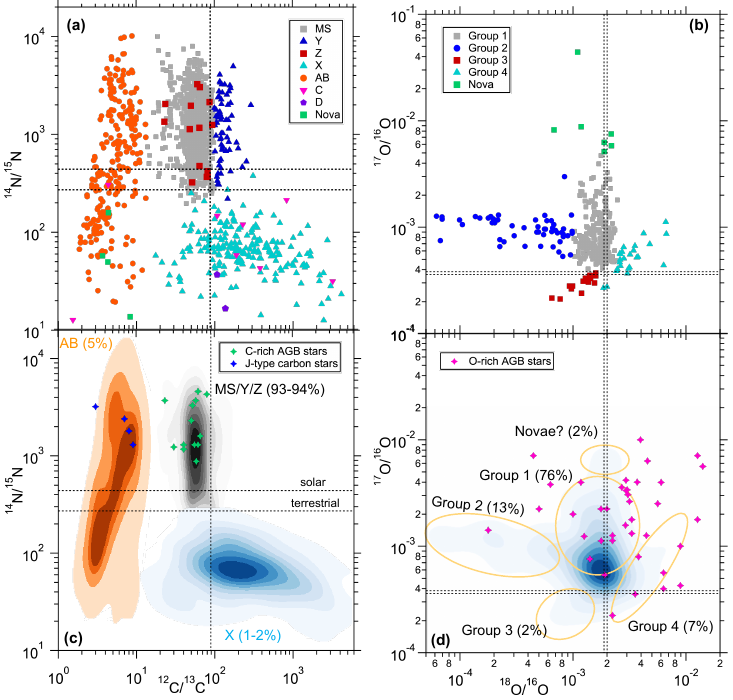}
\caption{Isotope analysis of stardust samples reveal abundance patterns that allow identification of potential cosmic sources. {\it Left column:} silicon carbide stardust, with measurements ({\it top}) and attributed sources ({\it bottom}). {\it Right column:} similar for oxide stardust grains. Carbon and Oxygen spectroscopic results from stars (dots in c,d) can be compared to the classification probabilities shown in the lower two graphs.
 (From \citet{Liu:2024a}).}
\label{fig_stardust}
\end{figure} 

\emph{Stardust} particles of sub-micron sizes can be isolated from meteoritic materials. 
These stardust grains have origins that clearly reach beyond the immediate vicinity of the solar system and its material. They have, however, been ingested into meteorites, and have been identified as meteoritic material was analyzed in the laboratory.
Therefore, those stardust grains are probes of material created in Galactic nucleosynthesis preceding formation of the solar system at 4.6~Gy before present. Considering the age of the Galaxy and its evolution, we may still consider this as representative for the current Galaxy (see Section Cumulative Nucleosynthesis below).
\emph{Meteorites} largely consist of the material that decoupled from interstellar material to form the solar system with the Sun and its planets, and meteorites which are a part of the solar system occasionally find their way onto Earth. 
They can be identified as non-terrestrial material samples only if they are found under suitable circumstances: In deserts or on antarctic snow, samples of rock stand out, and sometimes a meteorite fall is observed by its spectacular light track as it breaks up in the atmosphere, starting a dedicated search for unusual rocks. Sample bias is substantial.
Most of the meteorites are attributed to remains of asteroids after their collisional break-ups \citep{Ciesla:2006a}. 
A subset of these includes \emph{chondrules}, roundish condensations that.  are believed to have formed very early in the presolar nebula, and have been preserved since then \citep{Ciesla:2010}. 
Considering known systematics from sample type, chemistry, and elemental volatility \citep{Dauphas:2016}, their isotopic abundances can be precisely determined, and thus complement those from solar spectroscopy \citep{Lodders:2020}, to compose the well-determined solar-system composition.

Measuring isotopic abundances within such samples of cosmic material, their extra-solar origin has been evident from extremely unusual isotopic ratios in many abundant elements \citep{Zinner:2008,Nittler:2016}.
Using an ion beam to release atoms from the dust grain, secondary-isotope analysis (nano-SIMS) has been established since the first identifications of stardust in 1987 \citep{Zinner:1998}, now achieving nm~precision to control the ion beam, hence being capable to analyze smallest grains of stardust.
Modern instruments achieve excellent isotopic identifications, sometimes lasers are used to excite resonances and thus select specific species (RIMS) \citep{Stephan:2016}.
Recently, a database has been established for access to measurements of different samples and by different laboratories \citep{Stephan:2024}.
This provides a useful tool to study correlations in isotopic ratio patterns.
Comparing carbon and nitrogen isotope signatures between stellar observations and stardust grain analyses (see Figure~\ref{fig_stardust}, the apparent groups in these isotopic patterns can be attributed to stars in the intermediate-mass region mainly \citep{Liu:2024a}, with deviant groups possibly attributed to novae \citep{Jose:2004} or supernovae \citep{Nittler:1996,Liu:2024}.

The critical unknown in interpreting these precise data lies in our lack of knowledge of how dust is formed in the vicinity of nucleosynthesis sites \citep{Ivlev:2024}. 
There is general consensus that graphite grains as well as silicates are major constituents, and organic molecules may be major constituents of the smallest grains \citep{Hensley:2023,Weingartner:2001}.
It is unclear, in particular for explosive sites, where and how dust may be formed \citep{Liu:2024a}. 
The nucleation of gas and condensation into small grains crucially depends on availability of refractory elements, then on the ambient photon field and gas density and temperature \citep{Ivlev:2024}.  
Many sites of nucleosynthesis are violent energy sources, typically heating surrounding gas to MK temperatures, and thus evaporating ambient dust as well as preventing its immediate formation. 
Furthermore, the interstellar medium is quite dynamic and energized by shock fronts from stellar winds and supernovae. These lead to destruction of ambient dust, or at least to spallation and re-formation of dust, altering the size distribution. In quiet and cold interstellar medium, dust grains grow ice mantles, thus increasing substantially in size. Water ices are very common, as detected first through molecular lines including deuterium indirectly, and recently in infrared spectroscopy with JWST \citep{Decleir:2025}. 
The nanoparticles that have been extracted from meteorites and identified as stardust apparently are main constituents of current dust models, larger dust grains being coagulations of these, plus icy mantles \citep{Hensley:2023}. 
\rd{Within the solar system, the enrichments of meteorites with apparently s-process enriched material has been found to be inhomogeneous across the solar system, in a way that reflects thermal destruction of infalling dust within a protostellar disk \citep{Ek:2020}. This is another hint that dust particles are important carriers of nucleosynthetic ashes through interstellar space.}

Irrespective of these uncertain dust production aspects, the isotopic detail available in stardust is a valuable asset from such astronomical measurement towards astrophysical constraints \citep{Clayton:2004}.
The stardust data of \emph{mainstream silicon dust} could be associated most clearly to low-mass AGB stars of near-solar metallicity \citep{Liu:2022,Clayton:2004}. 
This association of SiC stardust with AGB stars is confirmed by observations of circumstellar environments of AGB stars showing the characteristic SiC features in infrared spectra \citep{Speck:2009,Marini:2021}. 
Analyzing SiC stardust grain compositions, there is a clear pattern of Silicon isotope variations $^{28,29,30}$Si seen in mainstream grains.
This pattern apparently is at odds with AGB yield variations or uncertainties and their coupling to Galactic chemical evolution models  \citep{Fok:2024}. The latter may trace stellar migration within the Galaxy, one of the possible explanations of such linear variation of silicon isotope ratios.

Other important constraints {on specific nucleosynthesis sources other than AGB stars} have been derived from analysis  of isotopic correlations including isotopes that are typical for burning sites.
\rd{Exotic isotopes that are attributed to the p~process thus have been constrained from meteoritic isotopic abundances \citep{Rauscher:2013}.}
{Furthermore, starting from the pattern of Si isotopic ratios and selecting for $^{28}$Si enrichments to point to equilibrium burning without neutron excess,} this includes $^{44}$Ti. {This} isotope than cannot be synthesized in stellar burning, but is expected to be produced in interiors of core-collapse supernovae {(see discussion in above sections 2.1. and 2.2. on Nuclear Reactions and 3.2. on Stellar Explosions)}.
Grains with excess $^{44}$Ti thus have been attributed to supernovae, and have been used to constrain supernova nucleosynthesis \citep{Nittler:1996,Liu:2024b}.

\begin{figure} 
\centering
	\includegraphics[width=0.9\columnwidth]{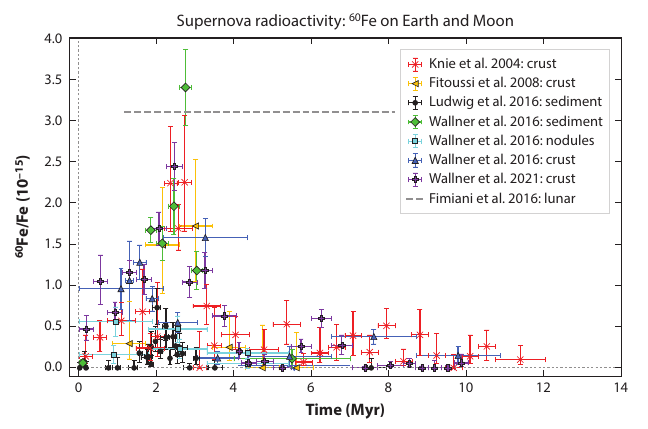}
	\caption{$^{60}$Fe detections in samples of Moon and sediments on Earth, versus sediment age \citep{Fields:2023}. Enhanced deposition occurs near 3~Myr ago, and possibly earlier near 6--8~My. }
	\label{fig_60Fe-sediments}
\end{figure}

Stardust also may directly enter the Earth atmosphere and reach surface. As in the case of meteorites, it is only under special circumstances that we can identify \emph{stardust deposits} on Earth or Moon.  Moreover, the transport of interstellar dust is mediated by the magnetic fields of Sun and Earth, because interstellar dust grains are charged; it is difficult to calculate the magnitude of such shielding and transport \rd{efficiency} \citep{Fields:2005,Fry:2020}.
A main idea was to search for some kinds of un-natural radioactivity, such as from the longer-lived actinides $^{244}$Pu (80.6 My) and $^{247}$Cm (15.7 My) \citep{Fields:1970}. A search in ice cores and deep-sea sediments for deposited dust grains from nearby nucleosynthesis events was motivated as a side activity to measure in this way the history of increased production of cosmogenic nuclides (e.g. $^{10}$Be) in the Earth atmosphere, as expected from cosmic ray intensity variations. 
Target isotopes were $^{10}$Be, $^{26}$Al, $^{36}$Cl, $^{53}$Mn, $^{60}$Fe, and $^{59}$Ni, and also the longer-lived $^{129}$I, $^{146}$Sm, and $^{244}$Pu \citep{Ellis:1996}, and the way to find very small admixtures in sediments was Accelerator Mass Spectrometry, reaching a 10$^{-17}$ fraction of contaminations \citep[see][for a review of the AMS technique]{Kutschera:2023}.
Analyses of samples of ocean crust obtained from deep ocean explorations far from contaminations of civilization, and also from Antarctic snow as well as lunar samples, have been investigated 
\citep{Knie:1999,Fitoussi:2008,Wallner:2015,Wallner:2016,Ludwig:2016,Feige:2018,Korschinek:2020,Koll:2020,Wallner:2021}.
The detections of $^{60}$Fe can be dated using a combination of $^{10}$Be deposits and other dating methods, to display a deposition history shown in Figure~\ref{fig_60Fe-sediments}. Evidently, in the period around 2--3~My ago there is an enhanced deposition \citep{Wallner:2016,Fields:2023}. The temporal width of this deposition appears incompatible with a debris wave from a single supernova event, and it is debated if this signature may be from multiple supernovae or from relative motion between the boundary of the Local Bubble cavity and the solar system. 
In any case, this graph demonstrates that deposition of rather fresh ejecta material from nearby nucleosynthesis activity occurs, and can be measured.
Furthermore, a search for $^{244}$Pu ($\tau=$80~My) also recently has been successful  \citep{Wallner:2015,Wallner:2021,Fields:2023}, and indicates a correlation with the 3~My deposition peak; a rather constant influx of this radioactive species, relative to the peak-like signature from $^{60}$Fe, can however not be excluded, considering uncertainties \citep{Wallner:2021}.
Possible interpretations have been discussed using simulations of interstellar dynamics \citep{Schulreich:2023}. An origin of this $^{60}$Fe deposition from stars and supernovae of the Scorpius-Centaurus stellar group appears plausible.
 
At higher particle energies, \emph{cosmic rays} reach interplanetary space from the nearby Galactic interstellar medium, and can be captured by suitable experiments on balloons and spacecraft. Mass spectrometry then allows to determine elemental compositions over a wide range of energies.
Cosmic ray energies reach from the MeV region up to 10$^{21}$~eV. At low energies, solar modulation again occurs from the heliospheric and terrestrial magnetic fields. 
Abundance analysis reveals that at lower energies the composition is reminiscent of the solar-system pattern. Major deviations are due to spallation in interstellar space, filling the abundance minima for Li-Be-B nuclei as they are produced from decomposition of carbon. 
From detailed analyses of the CRIS instrument aboard the ACE satellite, more details on such interstellar spallation has been used to constrain cosmic-ray transport, and determine that in general cosmic rays pervade the entire Galaxy including halo regions.
However, a recent detection of $^{60}$Fe nuclei in cosmic rays \citep{Binns:2016} demonstrates that nucleosynthesis ejecta also find their way into cosmic-ray acceleration, { and follow interstellar trajectories that reach detectors in the Solar System } \citep{de-Sereville:2024}. Note that $^{60}$Fe cannot be produced through collisions and spallation in interstellar space in significant quantities, and hence is {clearly} attributed to nucleosynthesis.

\begin{figure*}[ht] 
\centering
\includegraphics[width=\columnwidth,clip]{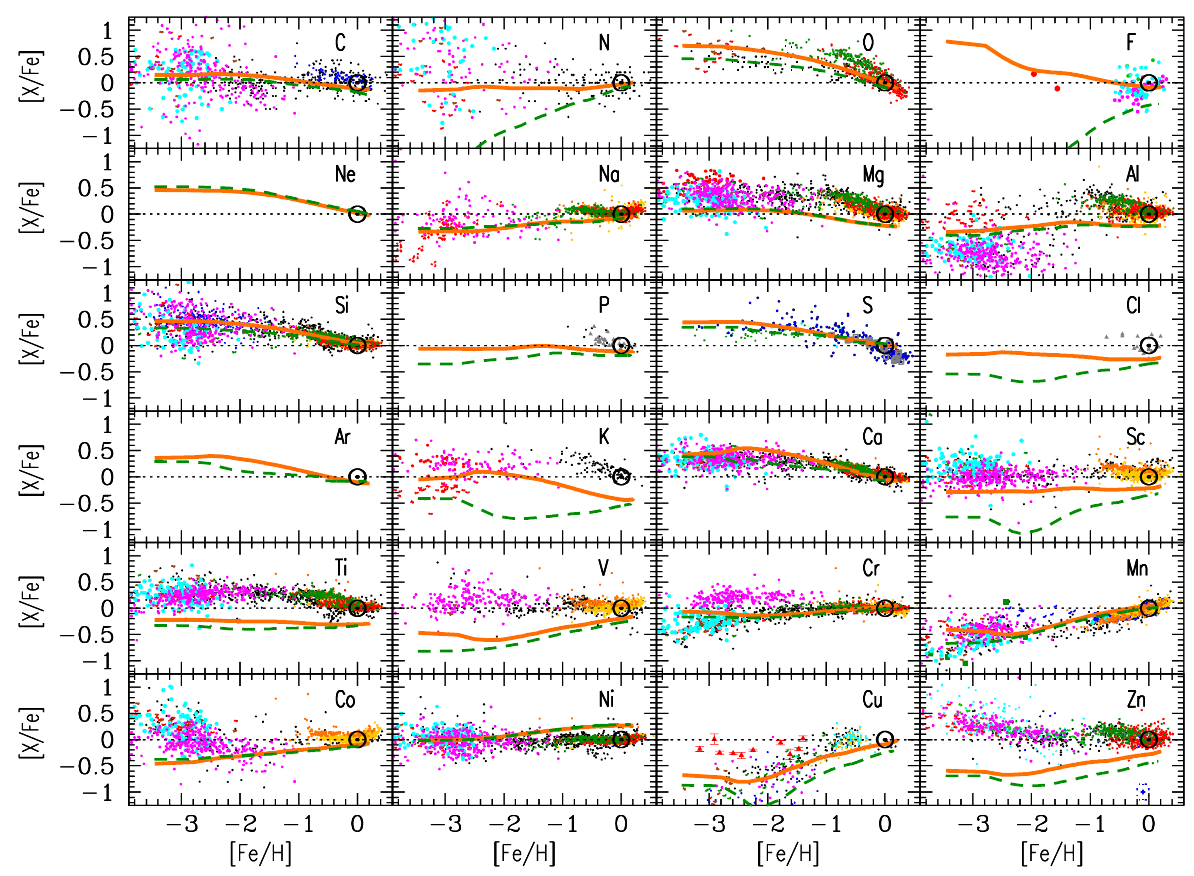}
\caption{The elemental abundances as measured in stellar photospheres for elements up to Fe (dots), are mostly in agreement with  current models of Galactic compositional evolution (lines). {The solar symbol shows the location of solar-system abundances. Solid lines show the baseline analytical model using stellar yields including the effects of stellar rotation, dashed lines show the same model using stellar yields without rotational effects.} (From \citet{Prantzos:2018}).  }
\label{fig_abundanceGal}       
\end{figure*} 

\subsection{Cumulative nucleosynthesis}\label{sec4.3} 
We noted above that abundances measured in stellar photospheres reflect the composition of stellar material at the time this particular star formed from interstellar gas. This composition thus reflects the history of nucleosynthesis and the transport of ejecta in the vicinity of the star that has been measured. 
Measuring many stars throughout the Galaxy, a database has been assembled of stellar abundances, which allows studies of the Galactic history of nucleosynthesis. 
This field is called \emph{Galactic archeology}, { as it is capable of addressing } the early Galaxy and nucleosynthesis at low metallicity.

Millions of stellar spectra have been obtained from spectroscopic surveys (SDSS/APOGEE \citep{Abdurrouf:2022}, GALAH \citep{Buder:2025}, Gaia-ESO \citep{Gilmore:2022}, LAMOST \citep{Luo:2015,Yan:2022}, and RAVE \citep{Steinmetz:2020}) with multi-fiber medium and high-resolution spectrographs \citep[e.g.][]{Donor:2020,Kos:2025}.
Analysis often is conducted through automatic pipelines, using empirical abundance fitting engines such as 'The Cannon' \citep{Casey:2016,Ness:2018} and benchmark stars for calibrations and then deriving abundance results.
From these, it has been concluded that the Galaxy's inner region and bulge, as well as the halo, is mostly composed of old stars that were created in the early phase of the Galaxy's evolution, while the disk and spiral-arm regions include stars of all ages, and show the result of ongoing star formation over the Galactic history.

A description of such compositional evolution has been developed from the early work of \citet{Tinsley:1972}. 
{This work expanded into various models of \emph{Cosmic (or Galactic) Chemical\footnote{We note that 'chemical' here refers to the chemical elements; no chemical reactions are addressed in CCE/GCE models.} Evolution} describing} the increase of metal abundances in terms of the star formation rate and the yields of the variety of nucleosynthesis sources \citep{Clayton:1985b,Prantzos:1999,Gibson:2003,Matteucci:2003a,Kobayashi:2020a,Matteucci:2021}. 
{The sources that are included here are mainly stars and their supernovae of all types I and II, but also rarer sources from binary systems such as neutron-star mergers, as well as rare supernova species such as magnetic-jet supernovae and hypernovae, have been included.
Herein, the delay times between star formation and the ejection of fresh nucleosynthesis material from the respective sources provide an important ingredient, and rely on astrophysical theory or modeling. }
{Figure~\ref{fig_abundanceGal} shows the performance of a one-zone model for the solar neighborhood, with infall at a characteristic time scale of 10~Gy, comparing it to a large assembly of photospheric abundance data \citep[for details see][]{Prantzos:2018}.} 

{An important ingredient in all models is the mass exchange between the extragalactic medium and a galaxy, the \emph{inflow and outflow}. This is taken from galaxy evolution modelings or as an analytical prescription.}
Generally, underlying to all descriptions is an understanding and model of galaxy structure and evolution, characterized by components such as disk, bulge, and halo, with their respective star formation histories and interstellar-medium properties. The latter encode the efficiency of mixing nucleosynthesis ejecta over larger spatial regions and into new star-forming regions. 
{Thus, advances in our understanding of galaxy formation and evolution and the star forming regions and rates herein would ideally be reflected in revisions of chemical-evolution descriptions.}

Chemical-evolution models have advanced from one-zone models to 2- and 3-dimensional. 
{Although not based on fundamental-physics modelings, these descriptions provide a framework, in which the different components of nuclear astrophysics can be linked together to provide a bridge to astronomical observations \citep[see reviews by][]{Matteucci:2021,Kobayashi:2023,Diehl:2023}.} 
{Analytical descriptions are complemented by numerical approaches such as \emph{smooth-particle hydrodynamics}, which enable also inhomogeneous chemical evolution prescriptions. In this may, links appear to simulations of galactic evolution, and stimulate cross comparison between the evolution of galactic structures and of compositional evolution.} 
The determination of stellar ages \citep[see][for a recent review]{Soderblom:2010} provides one of the major uncertainties in the data themselves, comparisons across selection biases are important. 
Since the precision kinematic measurements of Hipparcos \citep{de-Zeeuw:1999} and in particular the Gaia mission \citep{Gaia-Collaboration:2022}, abundances can be compared with kinematic data, and a chemical tagging of kinematic groups obtains new structural information \citep[e.g.][]{Cantat-Gaudin:2024,Eilers:2022}.
This revealed that the Galaxy's history included collisions with smaller galaxies, disturbing the gradual build-up of abundances beyond primordial composition from nucleosynthesis in stars and their explosions, thus confirming earlier detections of stellar streams \citep{Helmi:2020}. 

{An interesting \emph{data driven modeling of compositional evolution} has been proposed recently, representing the stellar abundance data and their trends in a limited number of 'processes', each of these characterized by a specific yield pattern and delay time behavior \citep{Weinberg:2024}. As a first result, spectroscopic data could be represented by two processes, with characteristically different evolutions of their elemental abundances with respect to the abundance of magnesium; thus, magnesium has been recognized as a good tracer of the evolutionary clock, and two types of contributors appear sufficient to explain the spectroscopic data of photospheric abundances \citep{Weinberg:2024,Weinberg:2022}.}

Such models can be tested and verified, using the abundances measured from stars across the Galaxy and all ages.
In general, elemental abundances can be understood for many elements in terms of these models. But for some elements, puzzles and discrepancies remain, and are used to re-assess nucleosynthesis yields, abundance measurements, or transport models.
Future surveys such as 4MOST \citep{Mainieri:2023}, SDSS future surveys, and WEAVE \citep{Jin:2024} will provide the next steps in such massive data collection, allowing discovery of subtle features in abundance evolution and its interpretations.

\begin{figure*}[ht] 
\centering
\includegraphics[width=\columnwidth,clip]{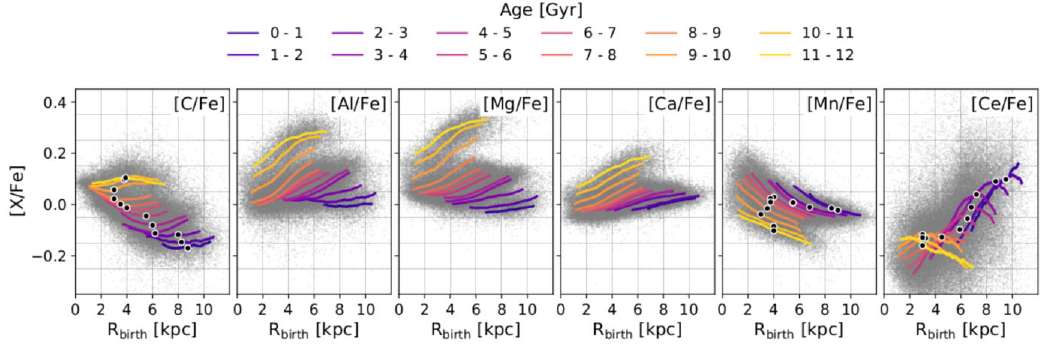}
\caption{The method of chemical tagging allows to identify stellar populations and their evolution in Galactic history. Here the galactocentric birth radii identify groups, and their individual and different evolutions combine to the broad pattern that earlier did not reflect much of a galactocentric-radius effect of metallicity evolution. (From \citet{Ratcliffe:2023}).  }
\label{fig_abundanceGal_chemTag}       
\end{figure*} 

\rd{A completely independent way to study the cumulative effects of nucleosynthesis is the field of \emph{cosmochemistry} \citep{Zinner:1998,Clayton:2004,Nittler:2006}. Here, material samples  are analyzed and interpreted in terms of cumulative nucleosynthesis.
The inclusions of pre-solar stardust grains and their interpretation with respect to specific sources of nucleosynthesis has been discussed above (section 4.2); this was based on the plausible assumption of such stardust identified by strikingly-different isotopic signatures having condensed in the vicinity of such a sources, thus freezing-in and preserving its characteristic ejecta \citep[see, however,][for cosmic evolution effects in stardust]{Nittler:2017}.
The sample of interstellar gas that was isolated as the solar system formed is preserved by all matter attributed to the solar system, including meteorites (their bulk material, beyond the specific stardust inclusions); it carries the complete nucleosynthetic history of the past, mediated by material flows connected to it. 
As described above, the \emph{solar abundances} shown in Figure~\ref{fig_abundances} are a composite result of spectroscopy of solar radiation and of mass spectrometry of meteoritic material, the latter essentially contribution all isotopic abundance information \citep{Lodders:2020,Lodders:2021}.
On Earth, abundances of elements and isotopes have been altered through a combination of physical, chemical, and antropomorphic processes; the abundances in the crust of our planet deviate significantly from the distribution shown in Figure~\ref{fig_abundances}, with depletion of gases and enrichments of refractory elements.
The microscopic fractionation processes that may have altered meteorite compositions over their history of 4.6~Gy are complex, but mostly rather well understood \citep{Dauphas:2016}. 
The interactions and dynamics caused by the planets within the solar system lead to macroscopic effects of some second-order heterogeneity across the meteoritic samples \citep{Kleine:2020}. 
Nevertheless, meteorites are a valuable repository for cosmochemistry, as discussed in the review by \citet{Nittler:2006}. 
As in the case of stardust, the key is that mass spectrometry in terrestrial laboratories provides access to isotopic information for the variety of even rarest elements.
}

\rd{Clear signatures of s-process material are evident in meteoritic abundances, and in particular the isotopic abundances can discriminate s-abundance patterns from r-process contributions \citep{Arnould:2007}.
The sequences of Ba isotopes obtained a prominent role herein, as $^{134}$Ba and $^{136}$Ba can only be a result of an s process, being shielded against the $\beta$-decay chain from r-process isotopes.
Similarly, molybdenum isotopes include the r-only isotope $^{100}$Mo, in addition to the other Mo isotopes of potentially mixed origins, with particular demand for p-process contributions for the light Mo isotopes. In addition to providing an alternate s- versus r-process origin discriminator, Mo (as well as Ru) isotopic abundances are valuable tests for our understanding of how abundances were built up during the evolution of the Galaxy and its nucleosynthesis sources \citep{Dauphas:2002,Stephan:2025}.
Two other isotopes on the proton-rich side of the valley of stability and thus attributed to a p-process origin  are $^{92}$Nb and $^{146}$Sm, with well-determined meteoritic abundances \citep{Lugaro:2016}. 
The interpretation of solar-system isotopic abundances in terms of cumulative nucleosynthesis and models of abundance evolution with a constraint dated 4.6~Gy before present thus is a tool similar to as described above for photospheric abundances. 
Herein, meteoritic isotopic data can provide strong constraints for the source components that describe the s and r~process, respectively, i.e. AGB stars, massive stars, and neutron star mergers.
}

\rd{Traces of radioactive material in samples from the early solar system provide an additional source of information, as they include the temporal signatures of radioactive decay, and thus the ingestion history of material into the mix of early solar system matter \citep{Lugaro:2022}.
In total 19 isotopic species have been analyzed with respect to their relative enrichment, comparing to abundance fractions as expected from the variety of nucleosynthesis processes \citep{Davis:2022}.
Thus, a clear enrichment of $^{26}$Al by about a factor 10 over the large-scale interstellar abundance is interpreted as a late ingestion of some specific nucleosynthesis ashes. Candidates could be a nearby AGB or Wolf-Rayet star \citep{Wasserburg:1977,Boss:2010a,Dwarkadas:2017b,Vescovi:2018}, being prominent sources of $^{26}$Al, and presumably less destructive to a protostellar system than a supernova might be. But among the many scenarios discussed, it remains difficult to weigh the different plausibilities for ingestion of fresh nucleosynthesis ejecta into a protostellar system \citep{Lugaro:2022}.
Another example is the constraint on a candidate source of r-process ejecta in the vicinity of the solar system near its formation time, derived from the combined $^{129}$I and $^{247}$Cm isotopic abundances inferred for the early solar system \citep{Cote:2021}.
}

\subsection{Constraining stellar matter in compact stars}\label{sec4.4} 
The characteristics of nuclear binding and nucleon interactions shape the macroscopic behavior of matter, which is observed in particular in compact stars that are not stabilized by the above-discussed release of nuclear energy through fusion reactions. 
This shapes astronomical phenomena from such stars, which in turn can be exploited to constrain nuclear binding at a macroscopic level. 

\emph{Gravitational waves} have come available as an astronomical tool in the last two decades, with LIGO and VIRGO experiments in the US and Italy, respectively. 
{The discovery of the kilonova AT2017gfo \citep{Smartt:2017} associated with gravitational-wave event GW170817 \citep{Abbott:2017} and with gamma-ray burst GRB170817A \citep{Abbott:2017g} provides one of the great recent successes in nuclear astrophysics. It confirms earlier predictions \citep{Li:1998,Rosswog:2005,Metzger:2010} of such an event to exist, and also confirms its expected signatures to first order in an unprecedented broad range of astronomical messengers, now also including gravitational waves.}
\rd {Gravitational waves have revealed characteristic signals within the Hz to kHz regime \citep{Abbott:2017} as expected from neutron star collisions \citep{Bauswein:2012}, as well as from neutron star / black-hole collisions. From these signals, constraints have been derived on the elasticity of neutron star matter \citep{Lattimer:2021}, as one of the neutron stars is tidally destroyed in the late phase of the inspiral (or early phase of the collision). }
{ The Japanese KAGRA interferometer joined LIGO and VIRGO in 2021 for collaborative measurements in the 'O4' run, and is the first detector for gravitational waves installed underground for better noise suppression \citep{Akutsu:2021}.}
{ Space-based interferometer instrumentation will enhance the range of period that can be measured in gravitational waves beyond the sub-second periods of ground-based instruments from tens of seconds to years, thus possibly accessing the in-spiral phase of binary systems and ringing from compact-star instabilities.         }
Multi-messenger signals from binary neutron stars thus will contribute constraints on the equation of state of nuclear matter \citep{Margalit:2019}.  
Gravitational waves provide a kind of astero-seismology for neutrons stars and their crustal nuclear composition \citep{Shchechilin:2025}, as well as the nuclear modeling of the behavior of neutron star matter \citep{Grams:2025,Grams:2025a}. 

\emph{X-ray timing} measurements of neutron star rotation and its changes as matter is accreted or as nuclear explosions expand the outer envelopes have become more precise after the pioneering RXTE mission with the NICER instrument on the space station.
Oscillations observed herein have been understood as an independent and precise measure of the neutron star radius, constraining the innermost stable orbit of accreting matter \citep{Miller:2021}. Complementing this, mass measurements in binary systems including a neutron star are very precise.
Comparing mass-radius constraints to predictions of models with different behavior of neutron star matter then is informative on nuclear characteristics of neutron star models.

Oscillation frequencies measured at stellar surfaces provide tools \rd{to indirectly} constrain the interiors of a star \citep[see][for a recent review of \emph{Astero-seismology}]{Aerts:2021}.
In the case of the Sun, one of the parameters driving stellar oscillations is the depth of the convective zone, which sensitively depends on the composition of matter, specifically their metallicity, or abundances of main elements Carbon, Nitrogen, and Oxygen \citep{Asplund:2009,Vinyoles:2017}. 
Helioseismology has obtained a precision of 0.5\% and thus enabled detailed constraints on the depth of the convection zone in the Sun. This can be compared to solar model predictions, evaluating uncertainties in solar composition, and how this depends on nuclear reaction rates \citep{Acharya:2025}. 
{ Various revisions of the solar metallicity since the value of 0.0187 from the early standard reference \citep{Anders:1989} have been made \citep{Asplund:2005,Asplund:2009,Asplund:2021}, and currently converge on a metallicity value of 0.0177 \citep{Magg:2022}, alleviating the earlier reports of a major reduction to values near 0.014, and now consistent with helioseismic constraints.}
With the Kepler and TESS missions, astero-seismology has become a tool to better constrain stellar ages \citep{Denis:2024}, as internal structure changes with consumption of fuel in nuclear burning.
Similarly, the pulsations of a white dwarf star reflects the internal composition. In this way,  specifically the central Oxygen content, could be derived \citep{Giammichele:2018}.
Thus astero-seismology had an important impact on studies of Galactic archeology. 
Improving age estimates for stellar groups \citep{Spitoni:2019,Spitoni:2020,Silva-Aguirre:2018}, 
this is a key step to disentangle stellar groups of different origins, as the degeneracy between age effects and different star forming histories is lifted.

\section{Summary and Conclusion}\label{sec5}
Nuclear astrophysics research connects the fields of nuclear physics and of astrophysics, and addresses challenging extremes in each of these fields. 
In cosmic sites, {charged-particle} nuclear reactions mostly occur at low energies, where tunneling probabilities through the Coulomb barrier drive the reaction rates. Resonances play key roles, in particular those which are at energies lower than can be measured directly.  
Conventional nuclear-reaction experiments at charged particle beams are in use, and in underground facilities with their reduced backgrounds.
\rd{Nuclear-structure effects, derived mainly from theory, complement our knowledge of nuclear binding. }
Radioactive beam facilities and fragment separators following spallation are important to also address reactions away from the domain of stable nuclei.
Indirect techniques such as the Trojan-horse method usefully complement direct reaction measurements, but face complexities of normalization to infer reaction rates at the low energies of stellar environments for the reactions of interest.  
Newly developed photon facilities based on high-power lasers allow different types of experiments, probing nuclear levels and specific exit channels {important for the $\gamma$~process and for $\beta$~decays}. \rd{Inertially-confined high temperature plasmas} will address important topics such as screening and the relevance of plasma conditions for nuclear reaction rates. 
Neutron facilities with enhanced neutron fluxes and facilities with better neutron energy definitions enable measurements for the wide range of heavier nuclei that are involved in heavy-element production through neutron captures and $\beta$~decays.
Similarly, radioactive ion beam facilities and storage rings \rd{might} fill a niche in measuring long radioactive lifetimes, \rd{and are being developed for} $\beta$~decays, and low neutron capture rates.
{Weak reactions such as $\beta$~decay and electron capture occur at rather high energies in the regimes of reaction Q~values (MeV scale), driven by degeneracy of the electron gas in compact stellar cores.
Their determination mostly relies on theories such as QRPA calculations and shell model evaluations including excited nuclear states, evaluating Fermi and Gamow-Teller transitions and also including forbidden transitions.}

Evaluating measurements in such laboratories towards cosmic-site reaction rates requires accounting for experimental systematics and uncertainties, as well as theories to extrapolate measurements into the domains not accessible to experiments. 
R~matrix theory and Bayesian methods combined with Monte-Carlo methods prove valuable in dealing with experimental uncertainties.
Statistical methods are required to deal with reaction rates where large numbers of nuclear levels are involved, and often combine with nuclear models to represent the nuclear level structures. 
Here, \rd{first-principle,} or microscopic, models have been used, and are key when extrapolations far  away from measurements are needed; empirical or macroscopic models are still more-precise, though, in such regimes, and when \rd{supporting data} from experiments are available.

Communication between nuclear physicists and astrophysicists on reactions that are relevant in cosmic environments has made use of the concept of \emph{processes}. These encompass networks of nuclear reactions that share properties such as reaction equilibria or dominance of specific reaction types. 
With refined observations and modelings, it has however become clear that, e.g., neutron captures in cosmic sites include more than the classical s~process or r~process situations, as neutron fluxes in cosmic sources cover a broader range, and 'heavy' and 'light' and 'intermediate' neutron capture situations are being discussed. The r~process appeared early on as a clear case where a unique source was thought to operate from the early Galaxy times until present. {Recent insights reveal a diversity of candidate sources that is challenging}, from neutron star collisions through many different (and probably rare) supernova variants. 
{Studies of these sources} are important {even though} physics and interactions get more challenging, as it the case for the various neutrino processes and the 'p' or $\gamma$' process. With more detail, recognizing 'key reactions' then will provide a bridge towards specific nuclear experiments or modelings.

Cosmic environments range from stellar interiors to explosions of and on stars.  {A good first-order} understanding has been achieved for what makes and stabilizes a star, and how explosions of these and on these may be triggered. {But we are far from having understood stellar interiors and explosions }in sufficient detail to {realistically model} the nuclear-reaction environments {and outcomes, a diversity and range of results has been found}. 
Stellar models {have recently established a better} understanding of internal mixing processes {from complex dynamics of stellar rotations and turbulent convection in stars, in order to improve the approximations in 1D models of structural evolution of stars}.
Explosion models {are now able to} describe a variety of {supernova and explosion} scenarios in great detail. Their validity and relevance for cosmic nucleosynthesis across cosmic ages {has been demonstrated. The next steps of validation and constraints} by adequate data for a quantitative assessment {are ahead, and depend on opportunities of a sufficient number and  sufficiently nearby events}.

The discovery of a neutron star collision and its kilonova signature in {2017} was mediated by gravitational-wave detection in combination with a gamma-ray burst event, {thus establishing a highlight of multi-messenger astrophysics. 
The discovery of a kilonova representing a characteristic and predicted type of nucleosynthesis event } has led to unprecedented detail in astronomical data for such an extremely rare type of nucleosynthesis events.
{Great} enthusiasm from this validation of important predictions from kilonova theory has led to {intense} activities in theoretical work as well as a boost for facilities exploring the domain of r-process reaction paths. Nuclear-reaction networks and astrophysical scenarios can be combined to describe the multi-messenger observations of this kilonova.
But many other { geometrical and dynamical} scenarios and nuclear-reaction configurations {in these models} provide similarly close descriptions of {the astronomical} data. Uncertainties will remain large, because from the low event rates and difficult astronomical challenges {we cannot} expect observational constraints for {a significant number of} other possible collision and neutron star parameters, {considering} our varying aspect angle of the event {and its critical role in the appearance of a kilonova}. 

\begin{figure}
  \centering
  \includegraphics[width=\columnwidth]{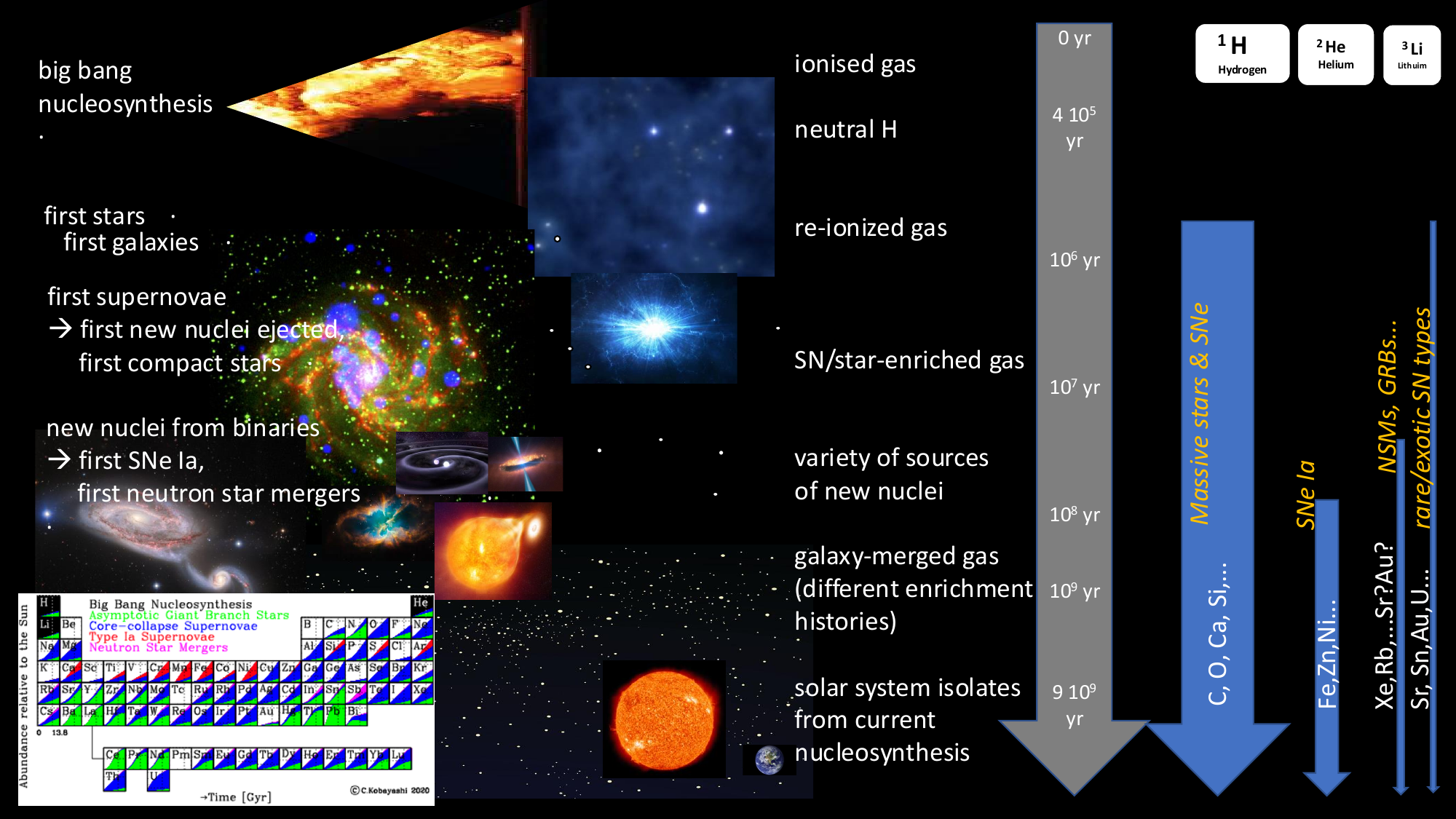}
   \caption{The cosmic enrichment of matter with new nuclei from nucleosynthesis occurs from a variety of sources, and at different times. The y axis of this graph is cosmic time, with the Big Bang origin at the top, and current time at the bottom. Arrows indicate the epochs where different sources contribute, with arrow width indicating their relative  importance. \emph{Nuclear Astrophysics} is the study of how the composition of cosmic gas evolved from primordial abundances (shown on top right) today's elemental and isotopic composition (shown on the bottom left as an elemental table with inferred evolutions in time \citep{Kobayashi:2020}).}
  \label{fig_cosmic_evolution_metallicity}
\end{figure}

Nuclear astronomy must rest on a broad range of mostly-indirect measurements of nuclear reaction outcomes. 
Specific nuclear reactions in cosmic sources can only be constrained with neutrino and $\gamma$-ray spectra, both being challenged by difficult detection techniques and high instrumental backgrounds.
More-accessible astronomical methods are biased from physical processes within the sources, to generate phenomena that relate to elemental abundances and their radiation. Also dust grains and other material samples that reach near-Earth detection methods carry an unknown creation and transport history that complicates interpretations, but in this case unique isotopic abundance detail provides valuable fingerprints of nucleosynthesis. .
Therefore, observations of atomic lines from X~rays through optical and into the sub-mm and radio regime need to be interpreted within the frameworks of astrophysical source models, such as supernova remnant expansion and shocks, or stellar photosphere thermodynamics and circumstellar medium processes, or even interstellar transport models. The classical astronomical backbone of cosmic nucleosynthesis, which is the observation of elemental abundances in stellar photospheres, has reached a culmination with modern spectrographs and the addition of kinematics for millions of stars. Yet, here an even more complex model of recycling of nucleosynthesis ejecta from its various sources into star-forming regions and their ingestion into stars of a great variety needs to be considered.
This may be less important for frequently-occurring sources related to massive stars, but could be crucial for a proper interpretation of nucleosynthesis from rarely occurring sources, in particular those that relate to binary systems and thus include uncertain delay times.
Models of cosmic chemical evolution now have become quite detailed, and include the various astrophysical processes involved in this cycle of matter as good as they have been constrained by various astronomical data and astrophysical models. 
Galactic-evolution models have shown that neutron star collisions and their nucleosynthesis appear not to be dominant sources of r-elements except for the current epoch. {Supernova} variants are being explored {as a promising alternative}.
{All these modelings depend, however, on a realistic description of the processes of star formation with their unknown rates and of ejecta mixing within the galaxy; a link to our understanding of galaxy evolution has become recognized as important in recent years.}
Data, models and simulations, and theoretical tools combine to  descriptions of the cosmic cycle of matter as illustrated in Figure~\ref{fig_cosmic_evolution_metallicity}, which takes the cosmic compositional evolution from primordial few elements through a variety of star-forming gas in galaxies and their nucleosynthesis sources to the variety of elements and isotopes that characterize the current universe and its material composition.

Nuclear astrophysics as a science thus {involves a} detailed discussion of nuclear laboratory experiments on specific reactions, { to obtain} proper ways to interpret their relevance to the cosmic environments in sources of nucleosynthesis. {Using models of those cosmic sources, nuclear astrophysics research tries to understand astronomical observations of a great variety that include nuclear, elemental,  and isotopic information}. Improving the agreements between such observations and the models, we learn about nuclear processes in cosmic environments.
 
\backmatter

\bmhead{Acknowledgements}

We are grateful for the \rd{constructive comments of the two referees}, which helped greatly to improve this review. 
RD acknowledges support from the Max Planck Institut f\"ur extraterrestrische Physik and from the 'Origins' cluster of excellence, both Garching, Germany, and from a Pifi fellowship of the Chinese Academy of Sciences.
MW acknowledges support by the National Science Foundation, USA, Grant No. PHY-2310059.



\end{document}